\documentclass{egpubl}
\usepackage{egsgp2025}
 
\SpecialIssuePaper         

\CGFccby

\usepackage[T1]{fontenc}
\usepackage{dfadobe}  

\usepackage{cite}  
\BibtexOrBiblatex
\electronicVersion
\PrintedOrElectronic
\ifpdf \usepackage[pdftex]{graphicx} \pdfcompresslevel=9
\else \usepackage[dvips]{graphicx} \fi

\usepackage{egweblnk} 

\graphicspath{ {./figures/} }
\usepackage{subcaption}
\usepackage{dblfloatfix}

\usepackage{amsmath}
\usepackage{amsfonts}
\usepackage{booktabs}
\usepackage[linesnumbered,ruled,vlined,norelsize]{algorithm2e}

\newcommand{\surf}{\mathcal{M}}
\newcommand{\mpar}[1]{\vspace{0.5em}\noindent{\bfseries#1.}\quad}
\newtheorem{definition}{Definition}
\newtheorem{theorem}{Theorem}

\makeatletter
\def\ps@titlepage{%
  \let\@oddhead\@empty
  \let\@evenhead\@empty
  \def\@oddfoot{\scriptsize{Copyright \textcopyright\ 2025 The Authors.\hfill}}%
  \def\@evenfoot{\hfill\scriptsize{Copyright \textcopyright\ 2025 The Authors.}}%
}
\def\ps@plain{%
  \def\@oddfoot{\scriptsize{Copyright \textcopyright\ 2025 The Authors.\hfill}}%
  \def\@evenfoot{\hfill\scriptsize{Copyright \textcopyright\ 2025 The Authors.}}%
}
\makeatother
\title[Robust Construction of Polycube Segmentations via Dual Loops]%
      {Robust Construction of Polycube Segmentations via Dual Loops}

\author[Maxim Snoep, Bettina Speckmann, and Kevin Verbeek]{
\parbox{\textwidth}{
    \centering 
    Maxim Snoep,
    Bettina Speckmann,
    and Kevin Verbeek
}
\\
{\parbox{\textwidth}{
    \centering
    TU Eindhoven, The Netherlands
}}
}

\begin{document}

\teaser{
    \centering

    \subcaptionbox{Dual loops}{%
        \includegraphics[width=0.31\linewidth]{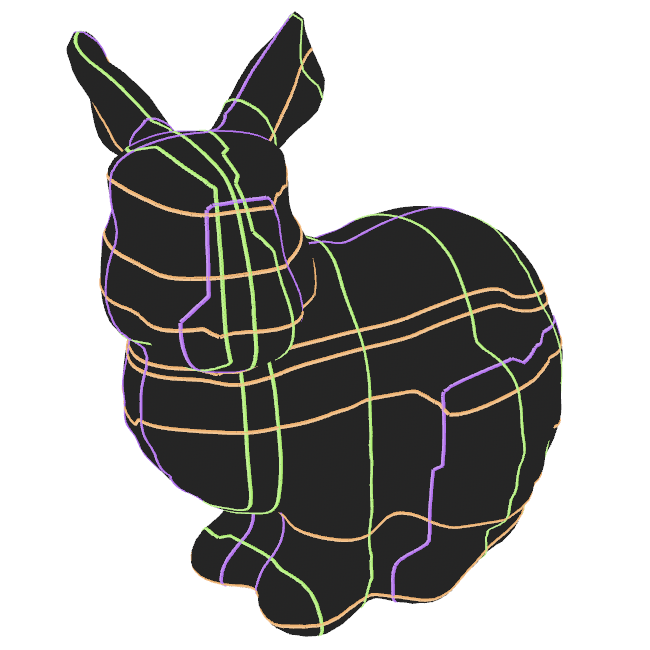}
    }
    \hspace{0.1in}
    \subcaptionbox{Polycube}{%
        \includegraphics[width=0.31\linewidth]{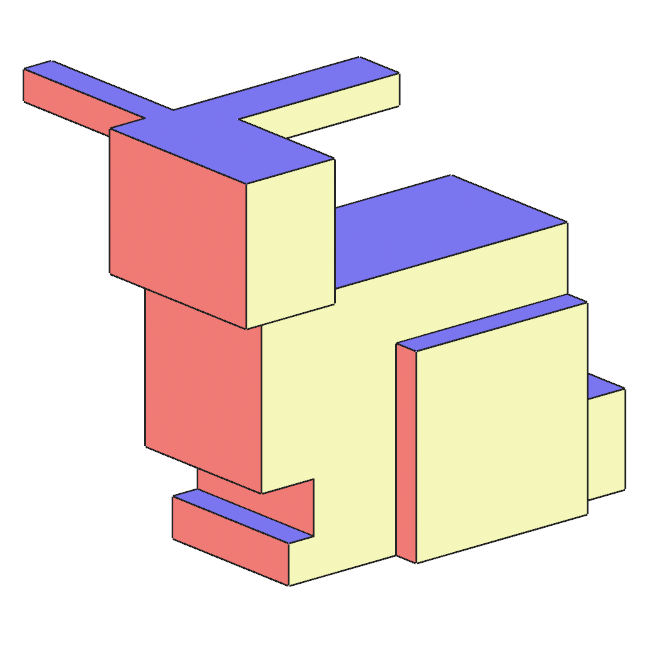}
    }
    \hspace{0.1in}
    \subcaptionbox{Polycube segmentation}{%
        \includegraphics[width=0.31\linewidth]{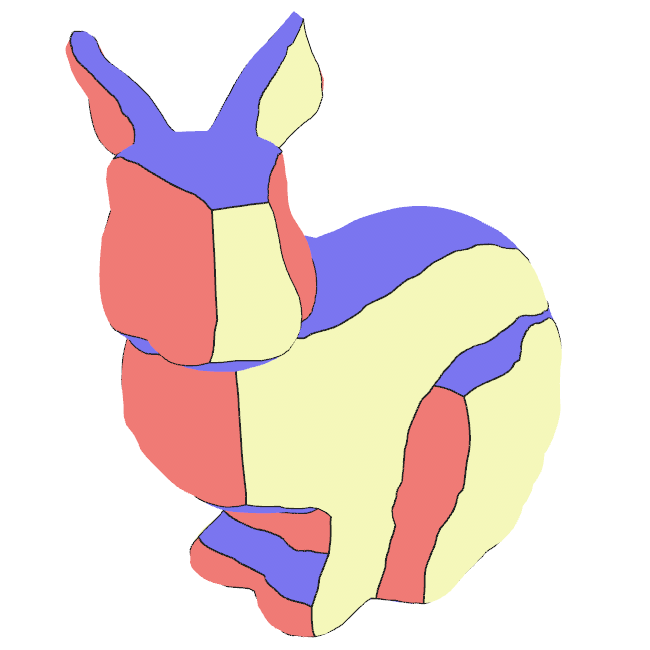}
    }
    
    \caption{The dual approach for the construction of polycube segmentations.}
    \label{fig:cover}
}

\maketitle
\begin{abstract}
Polycube segmentations for 3D models effectively support a wide variety of applications such as seamless texture mapping, spline fitting, structured multi-block grid generation, and hexahedral mesh construction. However, the automated construction of valid polycube segmentations suffers from robustness issues: state-of-the-art methods are not guaranteed to find a valid solution. In this paper we present \emph{DualCube}: an iterative algorithm which is guaranteed to return a valid polycube segmentation for 3D models of any genus. Our algorithm is based on a dual representation of polycubes. Starting from an initial simple polycube of the correct genus, together with the corresponding dual loop structure and polycube segmentation, we iteratively refine the polycube, loop structure, and segmentation, while maintaining the correctness of the solution.
DualCube is robust by construction: at any point during the iterative process the current segmentation is valid. Its iterative nature furthermore facilitates a seamless trade-off between quality and complexity of the solution. 
DualCube can be implemented using comparatively simple algorithmic building blocks; our experimental evaluation establishes that the quality of our polycube segmentations is on par with, or exceeding, the state-of-the-art. 
\end{abstract}

\section{Introduction}
\label{sec:introduction}

Polycubes, or orthogonal polyhedra with axis-aligned quadrilateral faces, enable efficient approaches to various challenging geometric problems. \emph{Polycube maps} are bijective mappings from general shapes to polycubes. They allow for the transfer of efficient solutions computed on polycubes to those general shapes.
As such, polycube maps are used to solve texture mapping~\cite{tarini2004polycube}, spline fitting~\cite{wang2007polycube}, multi-block grid generation~\cite{boelens2009f16}, and hexahedral meshing~\cite{pietroni2022hex}. 

Formally, a polycube map $f$ is a continuous function from a polycube $Q$ of genus $g$ to a closed 2-dimensional surface $\surf$ with the same genus. The edges of $Q$ mapped onto $\surf$ define a segmentation of \emph{patches}, where each patch corresponds to face of $Q$. This segmentation is called a \emph{polycube segmentation} (see Figure~\ref{fig:cover}). Each patch has a corresponding \emph{label} $\{+X, -X, +Y, -Y, +Z, -Z\}$, indicating the direction of the normal vector of the mapped face. The construction of a valid polycube segmentation is a common prerequisite for generating high-quality polycube maps. However, current methods fail to ensure the validity of the segmentation.

Since their introduction in 2004~\cite{tarini2004polycube}, numerous methods have been proposed for constructing polycube maps. The quality of a polycube map is determined by two competing factors: the complexity of the polycube and the distortion of the mapping. Methods that construct polycube maps must balance these conflicting factors. The state-of-the-art achieves a good trade-off between low distortion and low complexity~\cite{gregson2011all, livesu2013polycut, dumery2022evocube}. 

A common approach for constructing polycube maps is to first construct a polycube segmentation. Developing a polycube segmentation into a full (surface or volume) polycube map can be done via various established methods~\cite{gregson2011all, protais2022robust}. Most current approaches construct a polycube segmentation \emph{directly}~\cite{dumery2022evocube, fu2016efficient, gregson2011all, hu2016centroidal, hu2017surface, livesu2013polycut, zhao2017robust}, we call this the \emph{primal} approach. These construction methods suffer from various robustness issues~\cite{mestrallet2023limits, protais2022robust, sokolov2015fixing, zhao2019polycube}. The methods incorrectly identify invalid segmentations as valid, and vice-versa, since they can not guarantee that the constructed polycube segmentations are valid.

The methods rely on a \emph{characterization} of orthogonal polyhedra~\cite{eppstein2010steinitz}, used for checking whether a segmentation corresponds to a valid polycube. This characterization~\cite{eppstein2010steinitz}, and its extensions~\cite{he2024expanding, zhao2019polycube} are neither necessary nor sufficient~\cite{mestrallet2023limits, sokolov2016modelisation}. As a result, the primal methods unnecessarily restrict their solution space or generate invalid solutions. We recently introduced a complete characterization of polycubes of arbitrary genus via dual loops~\cite{snoep2025polycubes} by extending work of Biedl and Genc~\cite{biedl2004when} and Baumeister and Kobelt~\cite{baumeister2023how}. Currently, no algorithm exists that utilizes the dual characterization of polycubes and the accompanying loop structure to construct high-quality polycube segmentations.

In this paper we describe \emph{DualCube}: an iterative robust algorithm to construct polycube segmentations based on our characterization of polycubes via their dual loop structure~\cite{snoep2025polycubes}. This characterization is complete, but characterizes a specific subset of polycubes: those with strictly quadrilateral faces. As such this characterization cannot be used directly with the primal methods, since they construct polycube segmentations with arbitrarily complex patches. Note that we can turn quad-only polycube segmentations into arbitrarily-complex polycube segmentations by simply ignoring the so-called \emph{flat edges}, edges between two patches with the same label. 

\begin{figure}[t]
    \centering
    \includegraphics[width=0.25\linewidth]{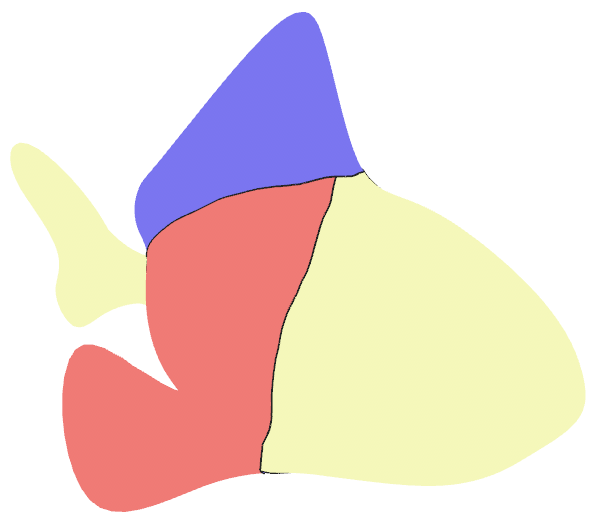}
    \hspace{0.2in}
    \includegraphics[width=0.25\linewidth]{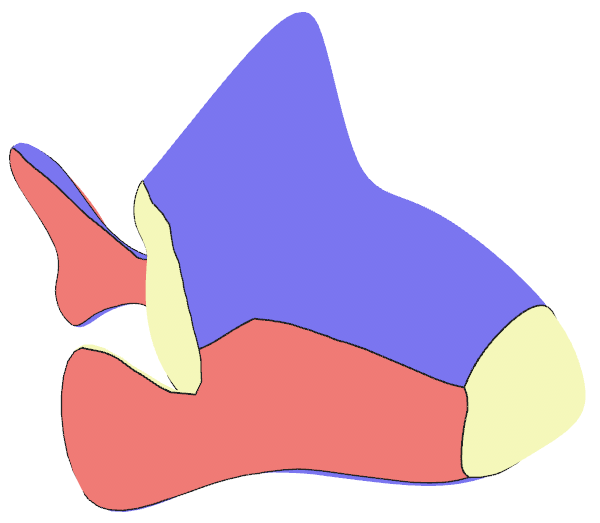}
    \hspace{0.2in}
    \includegraphics[width=0.25\linewidth]{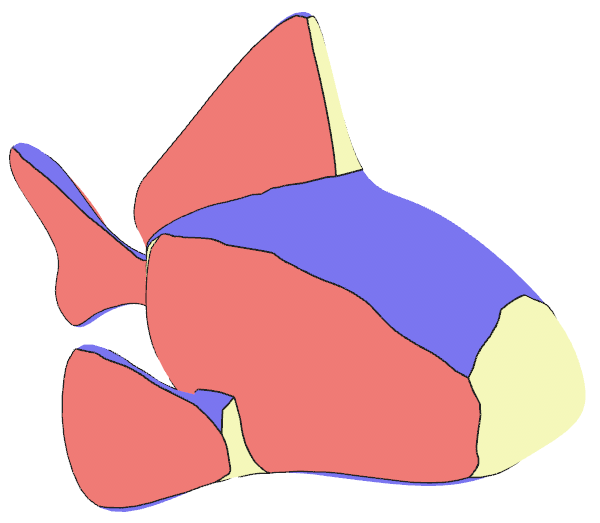}

    \includegraphics[width=0.25\linewidth]{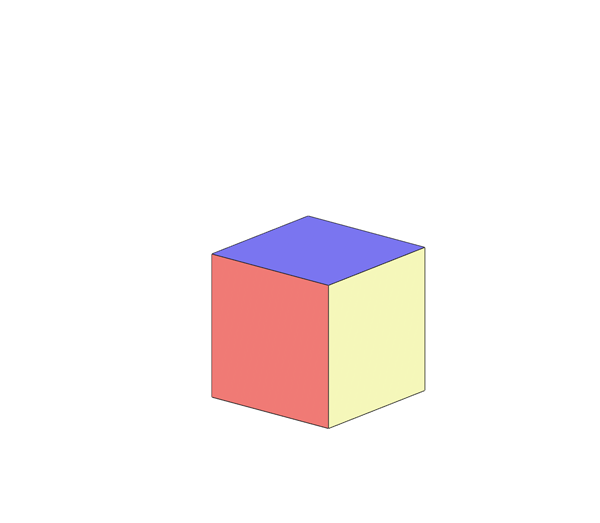}
    \hspace{0.2in}
    \includegraphics[width=0.25\linewidth]{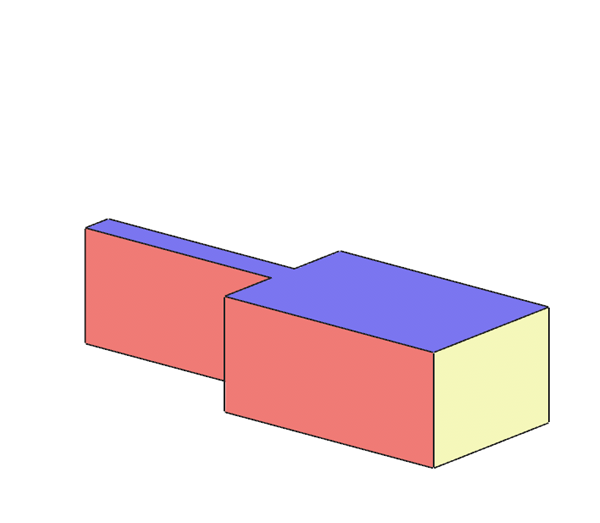}
    \hspace{0.2in}
    \includegraphics[width=0.25\linewidth]{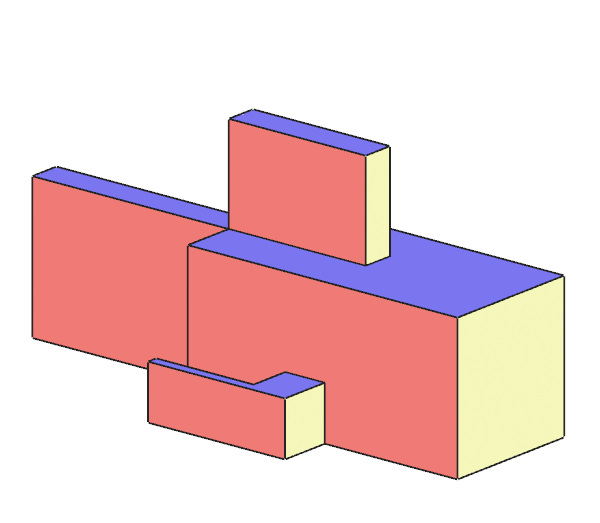}
    
    \caption{Iterative construction of a polycube segmentation, akin to the development of an egg to a guppy to an adult fish.}
    \label{fig:iterative-construction}
\end{figure}

The iterative nature of our algorithm allows us to explicitly control the trade-off between the complexity of the polycube and the distortion of the mapping, see Figure~\ref{fig:iterative-construction}. First, we embed a simple loop structure that corresponds to a polycube that has the same genus as the input surface $\surf$. Then we iteratively \emph{add} or \emph{remove} loops to refine or simplify the loop structure in a way that preserves validity. The embedded loop structure can be transformed into a polycube segmentation through a primalization step.

We review related work in Section~\ref{sec:related}. We discuss the necessary mathematical background, including the dual representation of polycubes, in Section~\ref{sec:prelim}. We describe our algorithm, including all technical details, in Section~\ref{sec:algorithm}. We evaluate DualCube experimentally and showcase the results in Section~\ref{sec:results}. Finally, we conclude with avenues for future work in Section~\ref{sec:conclusion}.

\section{Related Work}
\label{sec:related}

In 2004 \cite{tarini2004polycube} introduced polycube maps as an extension of cube maps. Since polycube maps proved to be an effective tool to solve challenging problems such as hexahedral meshing, many methods have been proposed to automate their construction. The first methods by \cite{tarini2004polycube} and \cite{wang2008polycube} constructed polycube maps from manually constructed polycubes. The first fully automatic construction methods had significant drawbacks, as they produced polycubes that were either very coarse \cite{lin2008automatic} or very detailed \cite{he2009divide}.

The most prominent methods for polycube map construction are based on the deformation approach by \cite{gregson2011all} which was later improved or extended by many~\cite{cherchi2016polycube, dumery2022evocube, fang2016all, fu2016efficient, guo2020cut, huang2014l1, livesu2013polycut, mandad2022intrinsic, yang2019computing, yu2014optimizing, zhao2019polycube}. The deformation-based approaches find an initial polycube segmentation in various ways, none of which guarantees that the segmentation corresponds to a valid polycube. Then, the input shape is gradually deformed until the surface faces are oriented toward their assigned target label. The initial polycube labeling may be adjusted during the deformation process. These methods use the characterization of orthogonal polyhedra by \cite{eppstein2010steinitz}, which is neither sufficient nor necessary for valid polycubes~\cite{mestrallet2023limits, sokolov2015fixing}. 

Another set of methods is based on voxelization~\cite{wan2011topology,yang2019computing,yu2014optimizing}. In general, a polycube map can be obtained by projecting a voxelized version of $\surf$ back onto $\surf$. Although voxelizing $\surf$ yields a valid polycube, the projection cannot be done robustly and may fail to preserve the topology. In addition, voxelization often leads to overly complex polycubes, which is not desirable for most downstream applications.

Finally, since polycube maps are popular for solving the hexahedral meshing problem, some methods focus on optimizing specifically for hexahedral meshing. These methods transform the input surfaces into more manageable shapes that result in higher quality hexahedral meshes by introducing cuts or deforming the input surface~\cite{fang2016all,guo2020cut,mandad2022intrinsic}. While these techniques are effective for interior volumetric meshing, such as hexahedral meshing, they may heavily distort the original surface or apply topological simplifications such as the cutting of handles. These modifications may preserve the volume but alter the original surface. In contrast, exterior volumetric meshing applications, such as multi-block grid generation, impose stricter constraints: the output must remain topologically and geometrically consistent with the input.

A different line of work started by Biedl and Genc~\cite{biedl2004when, biedl2009cauchy} and Baumeister and Kobbelt~\cite{baumeister2023how} takes a \emph{dual} approach, characterizing polycubes through their dual structures consisting of loops traversing the quadrilateral faces of a polycube. We recently introduced a complete characterization of polycubes of arbitrary genus via these dual loops~\cite{snoep2025polycubes}. In the same paper, we briefly outlined a proof-of-concept algorithm that first constructs a valid dual structure, which is then \emph{primalized} into a guaranteed valid polycube segmentation, inspired by the dual approach used for quad meshing~\cite{campen2012dual}. However, our prior work focused on theoretical foundations and did not include technical details, algorithmic analysis, and experimental evaluation against the state-of-the-art.

\section{Polycubes via Dual Loops}
\label{sec:prelim}

Our algorithm builds on our characterization~\cite{snoep2025polycubes} of polycubes via dual loops: loops that correspond to axis-aligned strips of quadrilateral faces. In the following we review the necessary mathematical background and the salient parts of our characterization.

\begin{figure}[b]
    \centering
    \includegraphics[scale=0.7]{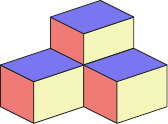}
    \hspace{5pt}
    \includegraphics[scale=0.7]{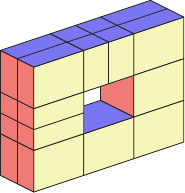}
    \hspace{5pt}
    \includegraphics[scale=0.7]{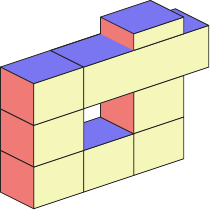}
    \caption{The variety of polycubes that satisfy Definition~\ref{def:polycube}. Note that polycubes are in principle allowed to globally self-intersect.}
    \label{fig:definition}
\end{figure}

A \emph{quadrilateral mesh (quad mesh)} consists of vertices, edges, and quadrilateral faces. Each vertex is adjacent to at least one edge. Each edge is adjacent to one or two faces. Each face consists of four vertices and four edges. A quad mesh is \emph{closed} if each edge is adjacent to exactly two faces. A quad mesh is \emph{orientable} if a consistent circular ordering of vertices can be assigned to each face, such that edge-adjacent faces have opposite vertex orders along their common edge. A quad mesh is \emph{connected} if every vertex can be reached from any other vertex by traversing edges. We can now define a polycube (see Definition~\ref{def:polycube} and Figure~\ref{fig:definition}).

\begin{definition}\label{def:polycube}
    A \emph{polycube} $Q$ is a closed, connected, orientable quad mesh with vertices $V(Q)$ such that: \begin{enumerate}
        \item Each vertex $v \in V(Q)$ has a position $p(v)$ in $\mathbb{Z}^3$,
        \item Each vertex has degree at least $3$,
        \item Positions of adjacent vertices differ in exactly one coordinate,
        \item Edges incident to the same vertex cannot overlap.
    \end{enumerate}
\end{definition}

Each polycube defines three partial orders on its vertices, corresponding to the three principal axes ($X$, $Y$, and $Z$). The partial order for the $X$-axis is defined as follows: for two vertices $v$ and $w$, we say that $v \leq_X w$ if the $x$-coordinate of $v$ is less than or equal to the $x$-coordinate of $w$, and there is an edge between $v$ and $w$. The partial orders for the $Y$-axis and $Z$-axis are defined similarly.

\begin{definition}
\label{def:order-equivalent}
    Two polycubes $Q_1$ and $Q_2$ are \emph{order-equivalent} if there exists an isomorphism $f\colon V(Q_1) \rightarrow V(Q_2)$ between the quad meshes of $Q_1$ and $Q_2$ such that, for all $v, w \in V(Q_1)$ and $\Delta \in \{X,Y,Z\}$, we have that $v \leq_\Delta w$ if and only if $f(v) \leq_\Delta f(w)$.
\end{definition}

\subsection{Dual structure}

\begin{figure}[t]
    \vspace{10pt}
    \centering
    \includegraphics[]{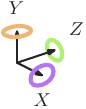}
    \hspace{5pt}  
    \includegraphics[scale=0.7]{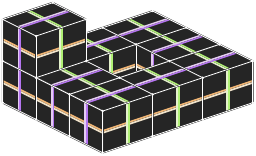}
    \caption{A polycube and its oriented $X$-, $Y$-, and $Z$-loops.}
    \label{fig:dualloops}
\end{figure}

Polycubes exhibit a dual loop structure, where each loop represents a strip of quadrilateral faces whose center points share a single coordinate~\cite{biedl2004when}; see Figure~\ref{fig:dualloops}. This structure forms a system of intersecting loops. The places where two loops intersect are called \emph{loop intersections}. The parts of a loop bounded by two intersection points (but containing no intersection points) are called \emph{loop segments}. The areas bounded by loop segments are called \emph{loop regions}. Note that each loop region of a polycube corresponds to exactly one corner of the polycube.

If you consider the normals of the quadrilateral faces of a single loop, they always correspond to exactly two axes ($\{X,Y,Z\}$). As such, the loops can be labeled as either an $X$-, $Y$-, or $Z$-loop as follows: an $X$-loop traverses faces with normals aligned to $Y$- and $Z$-axes. The $Y$- and $Z$-loops are similarly defined. 

Furthermore, each loop is oriented with a defined \emph{positive} and \emph{negative} side. Crossing a loop on the polycube from its negative to positive side corresponds to an increase in coordinates on the polycube vertices. Crossing in the opposite direction results in a decrease. In our figures, we use the colors purple, lighter purple, orange, lighter orange, green, and lighter green for $+X$, $-X$, $+Y$, $-Y$, $+Z$, and $-Z$, respectively. See Figure~\ref{fig:dualloops}.

We can use the full set of $X$-loops of an oriented loop structure to partition the underlying space (surface or polycube) into regions. We refer to these regions as \emph{$X$-zones}. We can use these zones to capture the global structure of the polycube (with regard to the X-axis) in a directed graph. Specifically, the \emph{$X$-graph} has a vertex for each $X$-zone, and a directed edge $(u,v)$ for every $X$-loop with $u$ and $v$ corresponding to the $X$-zones on the negative and positive side of the loop, respectively. We can define zones and graph for $Y$ and $Z$ similarly. The $X$-, $Y$-, and $Z$-graph are also called the level graphs.

\subsection{Characterization}
Consider a loop structure embedded on some arbitrary surface (not necessarily a polycube). In the remainder of this paper we implicitly assume that each loop in a loop structure is both oriented (defining negative and positive sides) and labeled with either $X$, $Y$, or $Z$, unless stated otherwise. There exists a complete set of rules that characterize which of these loop structures correspond to \emph{polycube loop structures}; those that are dual to a polycube.

\begin{definition}
\label{def:dual}
    A loop structure is a \emph{polycube loop structure} if:
    \begin{enumerate}
        \item No three loops intersect at a single point.
        \item Each loop region is bounded by at least three loop segments.
        \item Within each loop region boundary, no two loop segments have the same axis label \emph{and} side label.
        \item Each loop region has the topology of a disk.
        \item The level graphs are acyclic.
    \end{enumerate}
\end{definition}

\noindent We previously proved the following theorem:

\begin{theorem}[Snoep, Speckmann, and Verbeek~\cite{snoep2025polycubes}]
\label{thm:dual}
    \newline
    For every polycube $Q$ there exists a polycube loop structure $\mathcal{L}$ that forms the dual of $Q$. Furthermore, given a polycube loop structure $\mathcal{L}$, there exists exactly one polycube $Q$ (up to order-equivalence) that corresponds to $\mathcal{L}$.
\end{theorem}

\subsection{Valid modifications}

\begin{figure}[b]
    \centering
    \includegraphics[scale=0.7]{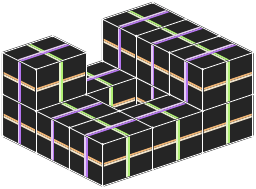}
    \hspace{15pt}   
    \includegraphics[scale=0.7]{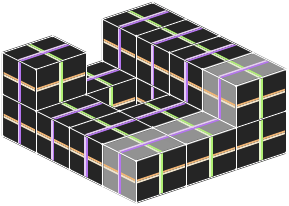}
    \caption{Loop additions (left to right) in the dual of a polycube. Loop removals can be seen as the reverse of addition (right to left).}
    \label{fig:modifications}
\end{figure}

Our dual characterization of polycubes~\cite{snoep2025polycubes} supports two basic valid operations that locally transform one valid structure into another: \emph{loop addition} and \emph{loop removal}. See Figure~\ref{fig:modifications}. These operations are sufficient for constructing many, but not all, polycube configurations. It remains unclear whether they suffice for reachability between all genus-zero polycube configurations~\cite{snoep2025polycubes}. For higher-genus cases, counterexamples exist: for example, adding or removing a single loop cannot change the orientation of a hole, making configurations with differing hole orientations unreachable from one another using only (single) loop additions and removals.

\mpar{Loop addition} 
Loop addition involves inserting a new loop into an existing structure such that the modified configuration remains a valid polycube loop structure. To do this, the authors define three directed graphs that encode permissible transitions between loop segments based on local constraints of Definition~\ref{def:dual}.

Given a loop structure $\mathcal{L}$, a base graph $G_L$ is constructed with one vertex per loop segment and edges between segments that share a loop region. From this, three filtered graphs $G_V^X$, $G_V^Y$, and $G_V^Z$ are derived, corresponding to candidate loops with labels $X$, $Y$, or $Z$, respectively. Each filtered graph retains only the edges whose insertion into the loop structure results in valid loop regions, according to Conditions 1–4 of Definition~\ref{def:dual}. A candidate loop is considered valid if and only if it forms a non-trivial, simple (visits a loop region at most once), directed cycle in its corresponding graph $G_\Delta^V$, where $\Delta \in \{X, Y, Z\}$. In~\cite{snoep2025polycubes} we show that such loop additions do not introduce cycles in the level graphs, thus also preserving Condition~5.

\mpar{Loop removal} 
Loop removal is the inverse operation, in which an existing loop $\lambda$ is removed from the loop structure. Let $\lambda$ be an $X$-loop, its removal merges each pair of loop regions separated by loop $\lambda$. Conditions 1–4 of Definition~\ref{def:dual} can be verified locally at all loop regions adjacent to $\lambda$. To ensure that Condition 5 is preserved, the resulting $X$-graph must be checked for cycles.

\section{Algorithm}
\label{sec:algorithm}
In this section we present DualCube: our algorithm  for computing polycube segmentations of triangulated 3D models. We first give a general overview, before we dive deeper into the technical details for modifying the polycube loop structure, constructing the polycube segmentation from a polycube loop structure, and optimizing the polycube loop structure. 

The input to DualCube is a 3D model represented as a triangulated surface mesh $\surf = (V, T)$, where $V$ is the set of vertices and $T$ is the set of triangular faces, each defined by a triplet $(v_i, v_j, v_k)$ of vertices. The edges of $\surf$ are implicitly defined by the faces. We assume that $\surf$ is a well-formed, orientable, manifold surface of arbitrary genus that bounds a single volume. Additionally, $\surf$ is embedded in $\mathbb{R}^3$, meaning that every vertex $v \in V$ has an associated position $p(v) = (x(v), y(v), z(v)) \in \mathbb{R}^3$, and each triangle $t \in T$ has a corresponding normal vector $n(t)$.

Our goal is to compute a \emph{polycube segmentation} $S(\surf) = (C, \mathcal{P})$ of $\surf$. Here, $C$ is a set of points on $\surf$ representing \emph{corners}, and $\mathcal{P}$ is a collection of \emph{paths} embedded on $\surf$ connecting these corners. These paths do not intersect each other and partition $\surf$ into \emph{patches}, forming quadrilateral subsurfaces. The corners and paths do not necessarily consist of the elements (vertices and edges) of $\surf$, but must lie on the surface defined by $\surf$. In general, a corner can be positioned anywhere on a face of $\surf$ and a path can traverse the interior of faces of $\surf$. Each patch has an assigned \emph{label} corresponding to one of the six (signed) principal axes: $\pm X/\pm Y/\pm Z$. A segmentation $S$ qualifies as a polycube segmentation if there exists a polycube $Q$ whose vertices, edges, and faces correspond one-to-one with the corners, paths, and patches of $S$. Furthermore, the label of each patch must correspond to the normal vector of its associated polycube face.

Our algorithm does not compute a polycube segmentation directly, but instead focuses on computing a valid (polycube) loop structure $\mathcal{L}$. The loops of $\mathcal{L}$ are directly embedded on the surface mesh $\surf$. By ensuring that $\mathcal{L}$ satisfies the properties of Definition~\ref{def:dual}, we can guarantee that we can compute a polycube segmentation from $\mathcal{L}$ at any step. DualCube now proceeds as follows. First, we initialize a loop structure of minimum complexity on $\surf$. For a model of genus zero, this is simply a loop structure consisting of three loops (one for each axis label), which corresponds to a single cube as polycube. For models of higher genus, this initialization is more involved (see Section~\ref{sec:optimize} for more details). Next, we alter the loop structure by adding or removing loops. In Section~\ref{sec:addloop} we explain in detail how to find good loops that improve the quality of the polycube segmentation, and at the same time preserve the characterization of a polycube loop structure stated in Definition~\ref{def:dual}. Once we have obtained a good loop structure, we compute a corresponding polycube segmentation by placing a corner in every loop region, and connecting the corners of adjacent loop regions via non-intersecting paths on $\surf$. We refer to this process as \emph{primalization} and discuss this step in more detail in Section~\ref{sec:primalize}. To obtain the best loop structure $\mathcal{L}$, we actually maintain a collection of loop structures, and we utilize a type of evolutionary algorithm to optimize the quality of the corresponding polycube segmentations (see Section~\ref{sec:optimize} for details). 

\bigskip
Note that DualCube does not need to compute the polycube that corresponds to the polycube segmentation. However, for illustration purposes we frequently show a possible polycube next to the polycube segmentation that our algorithm computes. We construct these polycube examples using a simple heuristic which does not attempt to optimize the exact geometric representation; we omit further details which can, however, be found in our publicly available code, see Section~\ref{sec:results}.


To find the best loop structure via our evolutionary algorithm, we need to define a measure of \emph{quality} for the resulting polycube segmentation. There is no universal standard for measuring the quality of a polycube segmentation, as different downstream applications have different requirements. Polycube maps are commonly used in hexahedral meshing, spline fitting, and structured multi-block grid generation, each prioritizing different aspects such as minimal distortion or fewer patches in the segmentation. In general, prior work aims to to balance complexity, often tied to the number of vertices in the corresponding polycube, and distortion, which captures how well the polycube map transforms the original surface into the polycube. Following the terminology introduced by Livesu et al. ~\cite{livesu2013polycut}, we consider a segmentation to be of high quality if it defines a \emph{valid} polycube topology and achieves a favorable trade-off between \emph{fidelity} and \emph{compactness}. 
\begin{description}
    \item[Fidelity] captures the alignment between the segmentation and the input geometry. We measure this via the angular deviation between the normal of each triangle and the axis associated with its assigned label, consistent with the measure used by Dumery et al.~\cite{dumery2022evocube}.
    \item[Compactness] reflects the structural simplicity of the segmentation. We measure the compactness of a polycube segmentation via the number of loops in its loop structure.
\end{description}
While Livesu et al.~\cite{livesu2013polycut} also identify path monotonicity as a desirable property, we do not explicitly optimize for it. Our primalization step results in favorable paths by design, making the addition of a third objective term unnecessary. We opt to retain the simplicity of a two-term formulation focused on fidelity and compactness. In Section~\ref{sec:optimize} we discuss which exact metrics we use and how we handle the resulting multi-objective optimization problem.

\subsection{Adding loops}
\label{sec:addloop}

In this section we describe how we can compute a suitable loop $\ell$ to add to an existing loop structure $\mathcal{L}$. Without loss of generality we can assume that $\ell$ has axis label $X$. To add a suitable loop, we need to consider two main aspects: (1) the alignment of the loop, to ensure that the resulting patches in the polycube segmentation will also align well; and (2) the validity of the loop, to ensure that the resulting loop structure $\mathcal{L} \cup \{\ell\}$ is also a polycube loop structure. Before we discuss these aspects in detail, we first consider the representation of loops in $\mathcal{L}$ embedded on $\surf$.

\begin{figure}[t]
    \centering
    \subcaptionbox{}{%
        \includegraphics[scale=0.6]{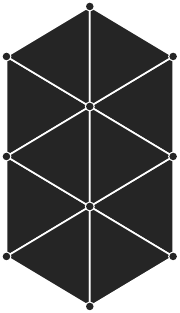}
    }
    \hspace{0.2in}
    \subcaptionbox{}{%
        \includegraphics[scale=0.6]{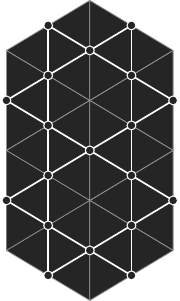}
    }
    \hspace{0.2in}
    \subcaptionbox{}{%
        \includegraphics[scale=0.6]{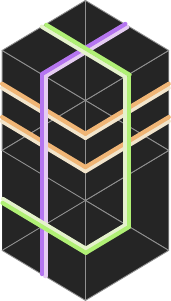}
    }
    \caption{A triangular mesh $\surf$ (a) and its edge graph $G$ (b). We compute loops on the edge graph, which results in loops crossing the edges of $\surf$ transversally (c).}
    \label{fig:representation}
\end{figure}

\mpar{Representation}
Technically speaking, it is possible to represent any possible loop $\ell$ on $\surf$ as a function $f\colon S^1 \rightarrow \surf$. However, the continuous search space makes it computationally difficult and expensive to find suitable loops. Instead, we would like to use a discretization of $\surf$ to restrict the set of possible loops. A natural choice would be to restrict loops to the vertices and edges of $\surf$. However, using this approach, two different loops cannot use the same edge of $\surf$. If we would allow two loops to share the same edge, then this could result in ambiguous overlaps between loops and degenerate loop regions. As a result, we may be unable to find any valid loop due to this restriction, making it overly restrictive. 

Instead, we want to use a representation of loops that allows us to always add another loop to the loop structure. To that end, we represent a loop $\ell$ by the sequence of edges of $\surf$ that are crossed transversely by $\ell$. Now, if two different loops cross the same edge $e$ of $\surf$, we can easily separate them along $e$ when needed (assuming that we know the order of the loops along $e$). Specifically, a loop is then represented by a sequence $[e_0, \ldots, e_n]$ of edges of $\surf$, where $e_0 = e_n$ and where $e_i$ and $e_{i+1}$ share a triangle on $\surf$ for $0 \leq i < n$. Additionally, for each edge $e$ of $\surf$, we keep track of the set of loops that cross $e$ and in which order. Using this combinatorial representation of the loops, we can easily realize the loops geometrically without introducing overlaps or degenerate loop regions, and it is always possible to add another loop (see Figure~\ref{fig:representation}).

To make it easy to work with this representation of loops, we construct the \emph{edge graph} $G$ of $\surf$. In the edge graph $G$, every node represents an edge of $\surf$, and there is a directed edge (both directions) between two nodes if the corresponding edges share a triangle in $\surf$ (see Figure~\ref{fig:representation}). We also assign a position to the nodes in $G$: if $e = (u, v)$ is an edge of $\surf$ (and hence a node in $G$), then we define $p(e) = (p(u) + p(v))/2$. Now, every loop $\ell$ using the representation chosen above simply corresponds to a path in $G$. 

\mpar{Alignment}
A good alignment of the final polycube segmentation can be achieved by creating a polycube map with small distortion. If we consider a loop $\ell$ with axis label $X$, then this corresponds to an $X$-loop on a polycube $Q$. Recall that all points on an $X$-loop have the same $x$-coordinate. Equivalently, all points on an $X$-loop lie in a single plane perpendicular to the $X$-axis. Thus, to achieve good alignment, the loop $\ell$ should have similar properties. 

Now consider a point $p$ on loop $\ell$. Let $d(p)$ be the tangent vector at $p$ in the direction of traversal of $\ell$, and let $n(p)$ be the normal vector at $p$ with respect to $\surf$. If $\ell$ were an $X$-loop on a polycube, then both $d(p)$ and $n(p)$ would be perpendicular to the $X$-axis. The cross product $d(p) \times n(p)$ would align with the $X$-axis. Therefore, for a good loop $\ell$ with axis label $X$, we would like the angle between $d(p) \times n(p)$ and the $X$-axis to be small for all points $p$ on $\ell$.

We now formulate the problem of finding a good loop $\ell$ on $\surf$ as a shortest path problem on the edge graph $G$. To that end, we assign a weight function $w_X$ to the edges of $G$. For an edge $(e, e')$ of $G$, let $d(e, e') = p(e') - p(e)$ and let $n(e, e')$ be the normal vector of the shared triangle of $e$ and $e'$ on $\surf$. We define $w_X$ as follows.
\[
w_X(e, e') = \text{angle}(d(e,e') \times n(e, e'), \overrightarrow{X})^\alpha
\]
Here, the function $\text{angle}(\cdot, \cdot)$ measures the angle between two vectors in radians, and $\overrightarrow{X}$ is the unit vector in positive $X$ direction. Furthermore, $\alpha > 1$ is the \emph{strictness factor}, on which we elaborate further below. Observe that this weight function is not symmetric, that is, $w_X(e, e') \neq w_X(e', e)$. This is necessary to enforce the loop to follow a specific orientation (clockwise), and to traverse fully around the surface mesh $\surf$. We can define weight functions $w_Y$ and $w_Z$ for the other axis labels analogously. 

We can now compute a loop $\ell$ by choosing a starting node $e$ in $G$, and then computing the shortest non-empty path from $e$ to itself in $G$ using the weight function $w_X$. To enforce the loop to traverse fully around $\surf$, we need to make sure that misaligned edges are sufficiently penalized, otherwise the loop would return instantly. This can be achieved by setting $\alpha$ sufficiently high.  

\mpar{Validity}
Not all loops are valid, that is, adding an arbitrary loop $\ell$ to $\mathcal{L}$ may not result in a valid polycube loop structure. The validity of a loop depends on its \emph{topological structure}. The topological structure of a loop $\ell$ is defined by the sequence of loop regions of $\mathcal{L}$ visited by $\ell$. In Section~\ref{sec:prelim} we discussed our approach from~\cite{snoep2025polycubes} to enumerate all valid loops.

We compute a valid loop $\ell$ in $G$ as follows. First, we use the graphs $G_V^X$, $G_V^Y$, and $G_V^Z$ from~\cite{snoep2025polycubes} to find a topological structure that corresponds to a valid loop. This topological structure then determines the sequence of loop regions in $\mathcal{L}$ that must be visited by $\ell$. However, there may be an exponential number of valid topological structures. We therefore restrict our search to the following set of topological structures. Specifically, given a starting region $R_e$ (which contains the starting edge $e$), we consider the following simple cycles. For all other regions $R'\not=R_e$, find the simple cycles that travel from $R_e$ to $R'$ and then return to $R_e$, while visiting as few loop regions as possible. This strategy results in at most as many valid topological structures as there are loop regions in $\mathcal{L}$.

We can then compute an actual embedding of the loop for a given topological structure as follows. For each pair of consecutive regions $R_i$ and $R_{i+1}$ in the sequence, we remove all directed edges in $G$ that connect $R_i$ to any region other than $R_{i+1}$. This restricted version of $G$ ensures that any computed path conforms to the desired topological structure. Finally, within this constrained graph, we compute the most aligned loop $\ell$ that matches the chosen structure. By construction, this loop is guaranteed to be valid.

\subsection{Primalization}\label{sec:primalize}
In this section we describe how we perform the primalization step, that is, how we compute the polycube segmentation $S(\surf) = (C, \mathcal{P})$ from the polycube loop structure $\mathcal{L}$. The primalization step consists of two main aspects: (1) placing a corner in each loop region $R$ of $\mathcal{L}$, and (2) computing (non-intersecting) paths between corners of adjacent loop regions.

\begin{figure}[t]
    \vspace{10pt}
    \centering
    \includegraphics[scale=0.7]{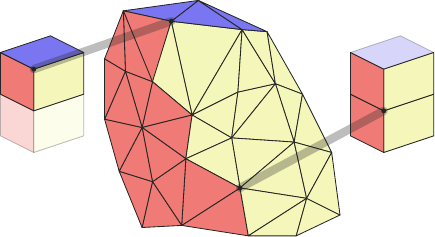}
    \caption{Polycube corners should align with vertices having similar face normals.}
    \label{fig:vertex-corner-scores}
\end{figure}

\begin{figure}[b]
    \centering
    \includegraphics[scale=0.7]{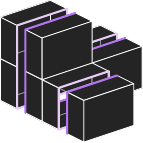}
    \hspace{0.2in}
    \includegraphics[scale=0.7]{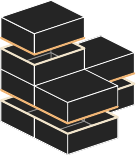}
    \hspace{0.2in}
    \includegraphics[scale=0.7]{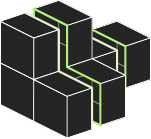}
    \caption{Polycube corners in the same zone share a coordinate.}
    \label{fig:corners-zones}
\end{figure}

\mpar{Corner placement}
Before placing corners in loop regions, we first ensure that each loop region contains at least one vertex. To this end, we refine the mesh $\surf$ into $\surf'$ using constrained Delaunay triangulation, inserting a vertex into every loop region that does not already contain one. To determine the best vertex $v \in \surf'$ for placing a corner, we apply two criteria: (1) The local neighborhood of $v$ in $\surf'$ should be as similar as possible to the local neighborhood of the corresponding corner in the polycube $Q$ (see Figure~\ref{fig:vertex-corner-scores}); (2) Corners that belong to the same zone should be placed on vertices with similar coordinates. Specifically, if two corners belong to the same $X$-zone, then they should be placed on vertices in $\surf'$ with a similar $x$-coordinate (see Figure~\ref{fig:corners-zones}).

\begin{figure*}[b]
    \addtocounter{figure}{+1}
    \vspace{10pt}
    \centering
    \includegraphics[width=0.9\linewidth]{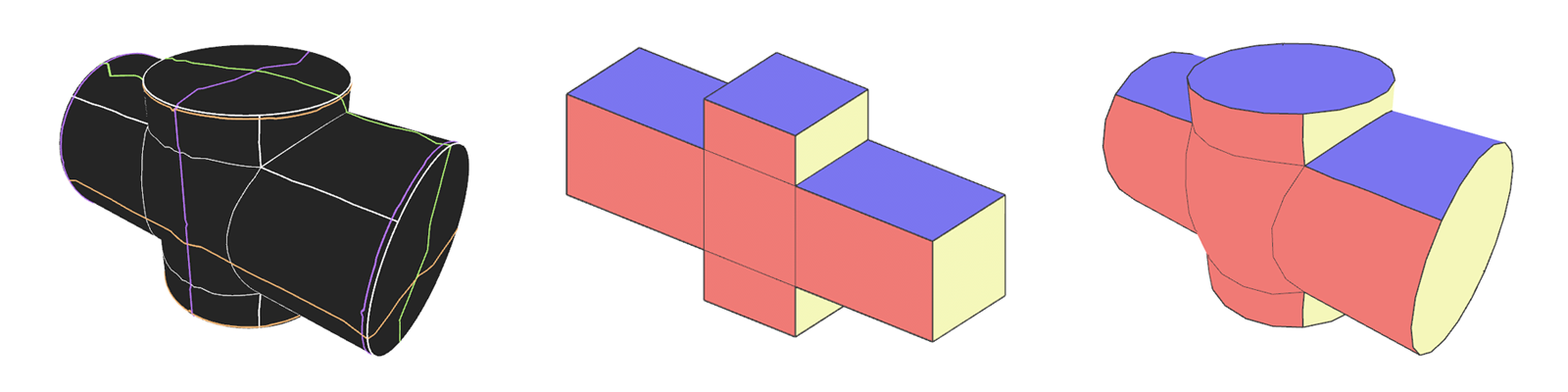}
    \caption{Primalization of the polycube loop structure to a polycube segmentation. Notice the placement of the corners and paths.}
    \label{fig:primalization}
\end{figure*}

We satisfy these two criteria using a two-step process. First, for every loop region $R$, we compute a candidate set $\mathcal{C}(R)$ of vertices based on the first criterion. Next, we choose the best vertices from each candidate set that optimize the second criterion. 

For the first step, consider a loop region $R$ and let $c$ be the corresponding corner of the polycube $Q$. The corner $c$ is incident to a number of faces in $Q$ that are perpendicular to one of the (signed) principal axes: $-X, +X, -Y, +Y, -Z, +Z$. Let $L(c)$ be the set of face types of all faces incident to $c$ in $Q$. Now consider a triangle $t$ of $\surf'$ in $R$. We can also assign a label among $\{-X, +X, -Y, +Y, -Z, +Z\}$ to $t$ based on which principal direction is closest to the normal $n(t)$. Then, for each vertex $v$, we can similarly define $L(v)$ as the set of all labels of triangles incident to $v$ on $\surf'$. We now define the similarity $s_c(v)$ of $v$ with respect to $c$:
\[
s_c(v) = |L(v) \cap L(c)| - |L(v)\setminus L(c)|
\]
The candidate set $\mathcal{C}(R)$ simply consists of all vertices $v$ in $R$ that achieve the highest similarity score $s_c(v)$ over all vertices in $R$.

For the second step, we need to consider the zones to which the loop regions belong. Let $\zeta_x$ be an $X$-zone and let $R_1, \ldots, R_k$ be the loop regions that belong to $\zeta_x$. We first compute the smallest interval $I(\zeta_x) = [x^-, x^+]$ such that there exists a vertex $v \in \mathcal{C}(R_i)$ with $x(v) \in I$ for all $i \in [1, k]$. We can compute this interval efficiently by first sorting the candidate vertices of the respective loop regions on $x$-coordinate, and then maintaining a valid interval using two pointers, as we move the left endpoint of the interval over all possible $x$-coordinates in increasing order. Let $m(\zeta_x)$ be the midpoint of the interval $I(\zeta_x)$. 

Now again consider a single loop region $R$. This loop region must be part of exactly one $X$-zone $\zeta_x$, one $Y$-zone $\zeta_y$, and one $Z$-zone $\zeta_z$. To choose the vertex $v$ on which to place the corner of $R$, we simply pick the vertex $v \in \mathcal{C}(R)$ that minimizes the Euclidean distance between $p(v)$ and $p^* = (m(\zeta_x), m(\zeta_y), m(\zeta_z))$. 

\mpar{Paths}
The next step is to compute non-intersecting paths connecting the corners of adjacent loop regions. We compute these paths incrementally (one after the other), which is both efficient and results in high-quality paths. To give these paths enough freedom to route between corners, we allow them to route on both the edge graph of $\surf'$ and on the vertices and edges of $\surf'$ itself. Specifically, we construct a graph $G^+$ from $\surf'$, where every node of $G^+$ corresponds to either a vertex or edge of $\surf'$, and where two nodes of $G^+$ are connected by an edge if one of the following holds: (1) the two corresponding vertices are connected by an edge in $\surf'$, (2) the two corresponding edges share a triangle in $\surf'$, or (3) the corresponding vertex/edge pair shares a triangle in $\surf'$.

We compute the paths (roughly) in order from shortest to longest, where we use the Euclidean distance between corners as a proxy for the length of the path between them. Consider computing the path $P$ between the two corners at $v$ and $v'$ or $R$ and $R'$, respectively. Similar to the corner placement, we have two main quality criteria: (1) the local neighborhood around $P$ should be as similar as possible to the local neighborhood of the corresponding edge in polycube $Q$, and (2) the length of $P$ should be as short as possible. The second criterion can easily be achieved by turning $G^+$ into a weighted graph, where the weight of an edge is determined by the Euclidean distance between its endpoints (note that every node in $G^+$ has a position in $\mathbb{R}^3$). To ensure that $P$ crosses only the loop segment between $R$ and $R'$, we exclude all edges from $G^+$ that cross another loop segment. Similarly, to avoid intersecting with already computed paths, we exclude the edges of other paths as well. We can then compute $P$ via a shortest-path algorithm.

\begin{figure}[t]
    \addtocounter{figure}{-2}
    \vspace{10pt}
    \centering
    \includegraphics[scale=0.7]{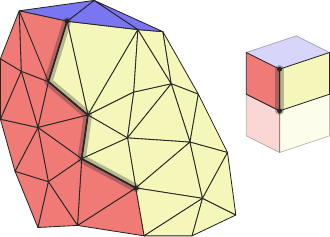}
    \caption{Patch boundaries should follow mesh edges where adjacent face normals resemble those along the polycube edge.}
    \label{fig:vertex-path-scores}
\end{figure}

\begin{figure*}[t]
    \addtocounter{figure}{+1}
    \vspace{20pt}
    \centering
    \subcaptionbox{}{%
        \includegraphics[scale=0.7]{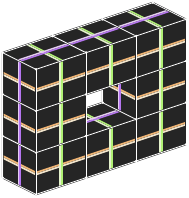}
    }
    \hspace{0.2in}
    \subcaptionbox{}{%
        \includegraphics[scale=0.7]{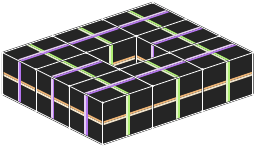}
    }
    \hspace{0.2in}
    \subcaptionbox{}{%
        \includegraphics[scale=0.7]{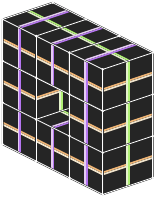}
    }
    \hspace{0.5in}
    \subcaptionbox{}{%
        \includegraphics[scale=0.7]{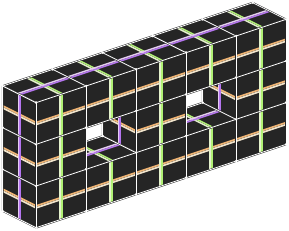}
    }
    \hspace{0.2in}
    \subcaptionbox{}{%
        \includegraphics[scale=0.7]{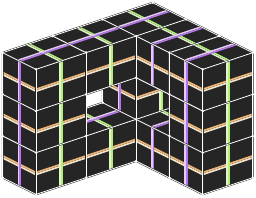}
    }
    \caption{Minimal-complexity polycube configurations for higher genus shapes.}
    \label{fig:initialization}
\end{figure*}

To also incorporate the first quality criterion, we adapt the weights of the edges of $G^+$ based on the orientation of the incident triangles on $\surf'$. If we consider the edge of the polycube $Q$ that corresponds to $P$, then the normals of the incident faces left and right from the edge are constant in $Q$ and align with one of the principal axes (see Figure~\ref{fig:vertex-path-scores}). Let these normals be $n_l$ and $n_r$, respectively. Similarly, for a directed edge $e$ of $G^+$, we consider the triangles $t_l$ and $t_r$ to the left and right of $e$, respectively. Note that, if $e$ is not an edge of $\surf'$, then $e$ lies on a single triangle $t$ of $\surf'$, and we define $t_l = t_r = t$. We define alignment penalties $a_l(e)$ and $a_r(e)$ for the left and right side of an edge $e$, respectively, where $a_l(e)$ is defined as:
\begin{align*}
a_l(e) = \begin{cases}
1 & \text{if } \theta \leq 0.955\\
(1 + \theta - 0.955)^2 & \text{otherwise}\\
\end{cases}
\\
\text{with } \theta = \text{angle}(n_l, n(t_l))
\end{align*}

Note that the alignment penalty is $1$ (no penalty) if the angle between $n_l$ and $n(t_l)$ is smaller than $\cos^{-1}(1/\sqrt{3}) \approx 0.955$ radians. This is the largest angle between a normal vector and its most similar principal direction. The penalty $a_r(e)$ is defined analogously. We now adapt the weight of $e$ by multiplying its current weight (Euclidean distance) by $(a_l(e) + a_r(e))/2$. 

We can now compute a path $P$ between two adjacent corners by finding the shortest path in $G^+$ with the given weights. Afterwards, we add the edges of $P$ to the underlying mesh $\surf'$, and update $\surf'$ and $G^+$ accordingly. We repeat this process until all paths have been computed, after which we obtain a valid polycube segmentation. Note that it is straightforward to assign labels ($\pm X/\pm Y/\pm Z$) to the triangles/patches of $\surf'$ given the placement of the corners, the paths, and the loop structure, see Figure~\ref{fig:primalization}.

\subsection{Loop structure optimization}\label{sec:optimize}
In this section we describe our evolutionary algorithm for finding a good polycube loop structure. We first describe how to initialize a valid polycube loop structure, then we establish the optimization function for polycube loop structure, and finally we discuss the details of our evolutionary algorithm. 

\mpar{Initialization} 
For genus-zero models, the initial polycube loop structure should correspond to a single cube. This requires exactly one $X$-loop, one $Y$-loop, and one $Z$-loop. Using the approach from Section~\ref{sec:addloop}, we can add these loops in any order. Although their topological structures are fixed, we can still vary the starting loop region for each. To obtain a good starting configuration, we generate 100 candidate solutions as follows: randomly choose an order for the $X$, $Y$, and $Z$ loops. For each loop, sample a random edge as the starting point. The first loop can be any valid loop. The second must intersect the first loop twice. The third must intersect both the first and second loops twice, in alternating fashion. This guarantees a valid initial loop structure. Among the 100 samples, we initialize with the one that yields the best polycube segmentation according to the criteria in Section~\ref{sec:algorithm}.

For models with higher genus, initialization becomes significantly more challenging. Unlike the genus-zero case, which always starts with a single cube, higher-genus shapes admit many distinct minimal-complexity polycube configurations. This is because each handle introduces degrees of freedom in its orientation. For example, a polycube torus can be oriented in three different ways, see Figure~\ref{fig:initialization}(a,b,c). A genus-2 surface has two independent holes, each of which can be oriented separately, further increasing the number of valid configurations, see Figure~\ref{fig:initialization}(d,e). There is currently no robust method to automatically determine an optimal initial polycube or to embed the corresponding loops on the surface. To address this, we initialize the loop structures manually. The user selects starting edges on the mesh, and for each selected edge, the method from Section~\ref{sec:addloop} computes a valid loop. By iteratively adding such loops, the user builds a minimal-complexity polycube loop structure. This process is relatively straightforward for users familiar with the appearance of higher-genus polycube structures on the surface. For genus up to 5, this manual initialization typically takes less than 10 minutes. For significantly higher genus, however, manual initialization becomes more difficult. Automating this process remains an open problem for future work.

\begin{figure}[b]
    \centering
    \subcaptionbox{$\beta = 0$}[0.48\linewidth]{%
        \includegraphics[scale=0.19]{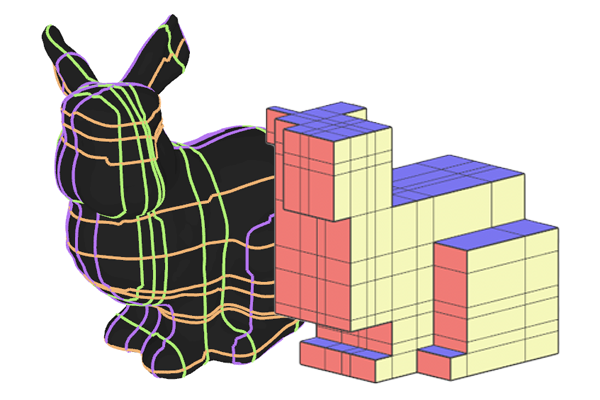}
    }
    \subcaptionbox{$\beta = 0.001$}[0.48\linewidth]{%
        \includegraphics[scale=0.19]{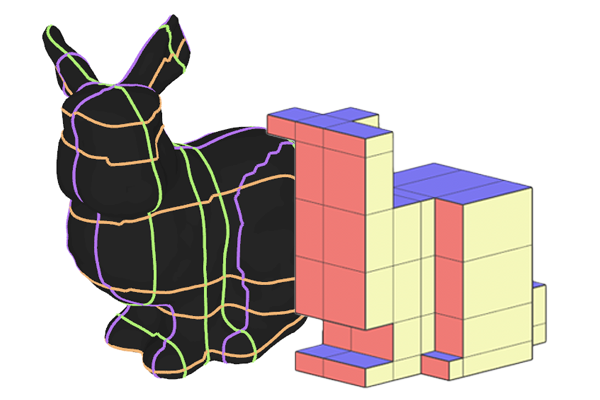}
    }

    \vspace{20pt}
    
    \subcaptionbox{$\beta = 0.01$}[0.48\linewidth]{%
        \includegraphics[scale=0.19]{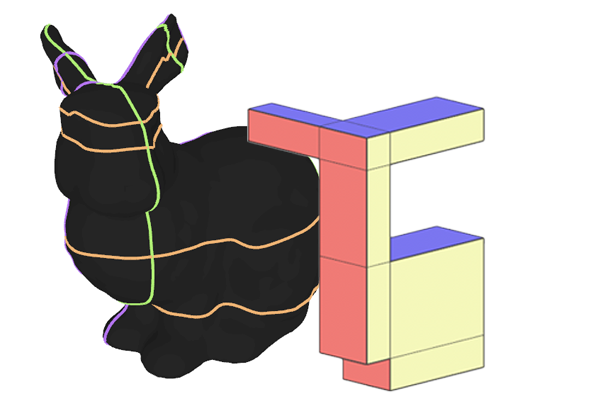}
    }
    \subcaptionbox{$\beta = 0.1$}[0.48\linewidth]{%
        \includegraphics[scale=0.19]{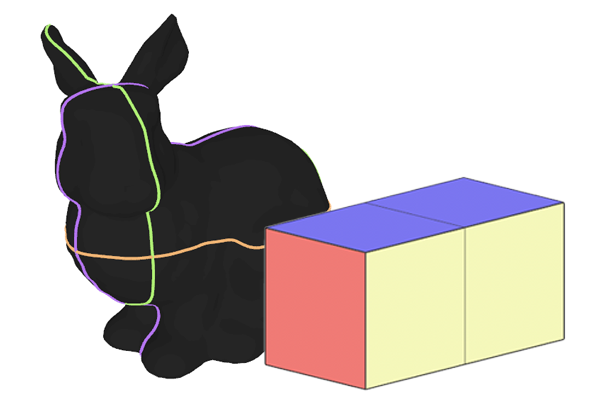}
    }
    \caption{Influence of $\beta$ on compactness of resulting polycube.}
    \label{fig:beta}
\end{figure}

\mpar{Optimization function} As already discussed at the start of Section~\ref{sec:algorithm}, we consider two optimization criteria: fidelity and compactness. The compactness $C(\mathcal{L})$ of a loop structure $\mathcal{L}$ is simply measured by the number of loops. To compute the fidelity, we first need to primalize the loop structure $\mathcal{L}$. Then, for each triangle $t$ in the resulting mesh $\surf'$, we consider its normal $n(t)$ and the vector $l(t)$ that corresponds to the label given to $t$ by the polycube segmentation. Let $f(t)$ be the alignment between $n(t)$ and $l(t)$ computed using the dot product:
\[
f(t) = n(t) \cdot l(t) 
\]
Observe that the dot product measures the cosine of the angle between $n(t)$ and $l(t)$. As such, $f(t)$ is a value between $-1$ and $1$, where $1$ indicates perfect alignment. 

To compute the total fidelity $F(\mathcal{S}(\surf'))$ of a polycube segmentation $\mathcal{S}(\surf')$, we simply use an area-weighted sum:
\[
F(\mathcal{S}(\surf')) = \frac{1}{\text{area}(\surf')}\sum_{t \in \surf'} f(t) * \text{area}(t)
\]
Observe that the value for fidelity is always between $-1$ and $1$.

Since we have two optimization criteria, we essentially obtain a multi-objective optimization  problem. To solve this problem, we use linear scalarization to turn this problem into a single-objective optimization (maximization) problem. Specifically, we define the quality of a solution as
\[
\text{quality}(\mathcal{L}, \mathcal{S}(\surf')) = F(\mathcal{S}(\surf')) - \beta * C(\mathcal{L}),
\]
where $\beta$ is a parameter that controls the trade-off between compactness and fidelity. This optimization function can be interpreted as follows. For every additional loop in the loop structure, the total fidelity score should increase by at least $\beta$. The parameter $\beta$ can then also be used to control the compactness of the resulting polycube, see Figure~\ref{fig:beta}.

\mpar{Evolutionary algorithm}  
We optimize the polycube loop structure using an evolutionary algorithm with mutation and selection only (no crossover). The algorithm maintains a population of $\mu$ loop structures, initialized as described before. In each generation, $\lambda$ offspring are produced by mutating randomly selected parents.

For each offspring, we draw a random value in $[0, 1]$; if it exceeds $\tau$, we apply an \emph{addition mutation}, otherwise a \emph{removal mutation}. In the addition mutation, we sample $n_x, n_y, n_z \in [\delta_{\min}, \delta_{\max}]$ and insert the corresponding number of $X$-, $Y$-, and $Z$-loops at random positions (Section~\ref{sec:addloop}). Note that for this we must compute the $G_V^X$, $G_V^Y$, $G_V^Z$ graphs first. Then for each loop to be added, we first compute a set of valid topological structures, and grab a random valid topological structure to be embedded on $\surf$. In the removal mutation, we sample $n_r \in [\delta'_{\min}, \delta'_{\max}]$ and delete that many valid loops (Section~\ref{sec:prelim}). All mutations preserve validity.

The next generation consists of the top $\lceil \mu/2 \rceil$ offspring and $\lfloor \mu/2 \rfloor$ parents, selected by the optimization criteria explained above. The process terminates after $\chi$ generations without improvement, returning the best solution found.

We use $\mu=10$, $\lambda=30$, $\tau=0.5$, $\delta_{\min}=0$, $\delta_{\max}=2$, $\delta'_{\min}=1$, $\delta'_{\max}=1$, and $\chi=10$, which we found to yield satisfactory results across a range of input meshes.

\begin{algorithm}
\fontsize{8}{10}\selectfont
\DontPrintSemicolon
\caption{DualCube}
\label{pseudocode}
\KwIn{Input triangle mesh $\surf = (V, T)$}
\KwOut{Polycube segmentation $S(\surf)$}

$\mathcal{P} \gets$ initialize population of $\mu$ loop structures on $\surf$\;

$\text{best} \gets \max(\mathcal{P})$\;
$\text{no\_improvement} \gets 0$\;

\While{$\text{no\_improvement} < \chi$}{
    $\mathcal{O} \gets \emptyset$\;

    \For{$i \gets 1$ \KwTo $\lambda$}{
        $L \gets$ random element from $\mathcal{P}$\;

        \eIf{random value $\in [0, 1]$ $> \tau$}{
            $n_X, n_Y, n_Z \gets$ random integers in $[\delta_{\min}, \delta_{\max}]$\;
            Add $n_X$ $X$-loops, $n_Y$ $Y$-loops, and $n_Z$ $Z$-loops to $L$\;
        }{
            $n_r \gets$ random integer in $[\delta'_\text{min}, \delta'_\text{max}]$\;
            Remove $n_r$ loops from $L$\;
        }

        Add $L$ to $\mathcal{O}$\;
    }
    
    $\mathcal{P} \gets$ $\{$ top $\lfloor \mu/2 \rfloor$ of $\mathcal{P}$ $\}$ $\cup$ $\{$ top $\lceil \mu/2 \rceil$ of $\mathcal{O}$ $\}$\;

    $\text{no\_improvement} \gets \text{no\_improvement} + 1$\;
    $\text{best'} \gets \max(\mathcal{P})$\;
    \If{$\text{best'} > \text{best}$}{
        $\text{best} \gets \text{best'}$\;
        $\text{no\_improvement} \gets 0$\;
    }
}
\end{algorithm}

\subsection{Analysis}

We analyze the computational complexity of our method using $n'$ to denote the number of vertices in the refined mesh $\surf'$. Since $\surf'$ is a refinement of a triangular mesh, the number of faces and edges is $O(n')$. The edge graph $G$ also has $O(n')$ elements, as it contains a constant number of elements per element in $\surf'$. Let $l$, $s$, $r$, and $z$ denote the number of loops, loop segments, loop regions, and zones, respectively. Note that, if we have $r$ loop regions, then we can have at most $6r$ loop segments. Furthermore, if we have $6r$ loop segments, then we can have at most $6r$ loops. Lastly, if we have $r$ loop regions, then we have at most $3r$ zones.

\mpar{Loop addition}
To add a loop, we construct graph $G_L$ with $s$ vertices and at most $6s$ edges. We filter $G_L$ to obtain the valid graphs $G_V^X$, $G_V^Y$, and $G_V^Z$ in $O(s)$ time. We run $r$ modified Dijkstra runs on $G_V^X$, $G_V^Y$, or $G_V^Z$ to find at most $r$ simple cycles in $O(r s \log s)$. We embed one of the loops on $\surf'$ with one Dijkstra run on edge graph $G$, which takes $O(n' \log n')$ time. Thus, each addition costs $O(rs \log s + n' \log n')$ time, or $O(r^2 \log r + n' \log n')$ time.

\mpar{Loop removal}  
Loop removal checks Conditions 1--4 in $O(s)$ time. Condition 5 is checked via its level graph with $z$ nodes and $l$ edges. We detect cycles in these graphs using DFS in $O(z + l)$ time. Thus, each removal costs $O(s + z + l)$ time, or $O(r)$ time.

\mpar{Primalization}
Constrained Delaunay refinement on $\surf'$ takes $O(n' \log n')$ time. Assigning naive labels to faces and computing local label sets for vertices takes $O(n')$. We compute label similarity in $O(n')$ time and sort the vertices by similarity in $O(n' \log n')$ time. For each zone, we sort candidates three times by $x$, $y$, and $z$ coordinates in $O(n' \log n')$ time. For each loop region, we choose the best candidate in $O(n')$ time. For each loop segment, we compute a path via Dijkstra on $\surf'$ in $O(s  n' \log n')$ time. Thus, primalization costs $O(s  n'\log n' )$ time, or $O(r n' \log n')$ time. Fidelity and compactness can be calculated in $O(n'+l)$ time, or $O(n'+r)$ time.

Automatic initialization involves constructing $\mu$ initial loop structures using 3 loop additions in $O(r^2 \log r + n' \log n')$ time. In each iteration, each offspring requires at most $3\delta_\text{max}$ loop additions, $\delta'_\text{max}$ loop removals, and a primalization step. Thus, the cost per individual per iteration is $O(\delta_\text{max}(r^2 \log r + n' \log n') + \delta'_\text{max}r)$.

Expressing this in terms of the input mesh vertex count $n$, we observe that $n' \leq n + r$ (each loop region introduces at most one new vertex). Furthermore, the number of loop regions $r$ corresponds to the number of polycube corners in the final solution. It is fair to assume that the resulting polycube contains significantly fewer corners than the number of vertices in the input mesh $\surf$, i.e., $r \ll n$. Under these assumptions, we get that the runtimes simplify to:
\[
\begin{aligned}
\text{initialization per individual} \quad & O(n^2 \log n) \\
\text{per iteration per individual} \quad & O(\delta_\text{max}(n^2 \log n) + \delta'_\text{max}n)
\end{aligned}
\]

\section{Results}
\label{sec:results}

We implemented DualCube as described in Section~\ref{sec:algorithm}. The code is available at~\href{https://github.com/tue-alga/DualCube}{github.com/tue-alga/DualCube}. We evaluate DualCube across three aspects:  direct quality of polycube segmentations, quality of the resulting hexahedral meshes, and a stress test on challenging inputs.

Our dataset (\href{https://github.com/tue-alga/meshes/tree/v1.0.1}{github.com/tue-alga/meshes/tree/v1.0.1}) consists of a diverse set of 3D models collected from online sources categorized by both geometric style and topological complexity. Geometrically, we distinguish between smooth and CAD models. Smooth models typically represent organic shapes, such as animals or humans, and are widely used in computer graphics applications like animation and game development. In contrast, CAD models depict mechanical components characterized by sharp edges and flat surfaces, commonly found in engineering contexts such as aerospace or automotive design. We also separate models based on genus, as our method currently requires manual initialization for inputs with genus $g > 0$. For a fair overview, results for higher-genus inputs are reported separately.

\begin{enumerate}
\item \textbf{Smooth ($g=0$, $n=22$)}: \texttt{airplane1}, \texttt{airplane2}, \texttt{amogus}, \texttt{armadillo}, \texttt{bimba}, \texttt{blub}, \texttt{bone}, \texttt{bumpysphere}, \texttt{bunny}, \texttt{buste}, \texttt{cat}, \texttt{chineselion}, \texttt{dino2}, \texttt{ghost}, \texttt{goathead}, \texttt{homer}, \texttt{igea}, \texttt{koala}, \texttt{moai}, \texttt{sphinx}, \texttt{spot}, \texttt{venus}
\item \textbf{Smooth} ($g>0$, $n=6$): \texttt{bottle1}, \texttt{bottle2}, \texttt{bumpytorus}, \texttt{cup1}, \texttt{dtorus}, \texttt{teapot}
\item \textbf{CAD} ($g=0$, $n=42$): \texttt{B0}, \texttt{B11}, \texttt{B12}, \texttt{B14}, \texttt{B15}, \texttt{B16}, \texttt{B17}, \texttt{B18}, \texttt{B19}, \texttt{B2}, \texttt{B20}, \texttt{B21}, \texttt{B23}, \texttt{B25}, \texttt{B27}, \texttt{B28}, \texttt{B30}, \texttt{B34}, \texttt{B35}, \texttt{B36}, \texttt{B38}, \texttt{B39}, \texttt{B40}, \texttt{B41}, \texttt{B43}, \texttt{B46}, \texttt{B48}, \texttt{B49}, \texttt{B5}, \texttt{B50}, \texttt{B57}, \texttt{B59}, \texttt{B60}, \texttt{B61}, \texttt{B68}, \texttt{B7}, \texttt{B70}, \texttt{B71}, \texttt{B75}, \texttt{B8}, \texttt{B9}, \texttt{fandisk}
\item \textbf{CAD ($g>0$, $n=14$)}: \texttt{B1}, \texttt{B3}, \texttt{B10}, \texttt{B13}, \texttt{B32}, \texttt{B33},  \texttt{B51}, \texttt{B62}, \texttt{B65}, \texttt{B66}, \texttt{B73}, \texttt{block}, \texttt{rocker}, \texttt{rod}
\end{enumerate}

We compare DualCube against two publicly available polycube segmentation algorithms: \emph{PolyCut}~\cite{livesu2013polycut} and \emph{EvoCube}~\cite{dumery2022evocube}. Table~\ref{table:results} reports segmentation quality in terms of fidelity and compactness, two commonly used metrics in polycube evaluation. However, as these metrics do not always predict downstream performance, we additionally measure the quality of hexahedral meshes generated by a polycube-based hexahedral meshing pipeline (RobustPolycube~\cite{protais2022robust}) using as input the polycube segmentations generated by the three methods. These results are presented in Table~\ref{table:results2}. Figure~\ref{fig:showcase} shows an overview of our results, and Figure~\ref{fig:comparison} shows side-by-side visual comparisons of polycube segmentations and resulting hexahedral meshes generated via the three different methods. To further test robustness, we evaluate all methods on the \emph{Nightmare of Polycubes} dataset (\href{https://github.com/LIHPC-Computational-Geometry/nightmare_of_polycubes}{github.com/LIHPC-Computational-Geometry/nightmare\_of\_polycubes}), which is designed to expose failure cases in polycube methods. Results on these nightmares are shown in Figure~\ref{fig:nightmare}. 

For PolyCut, EvoCube, and RobustPolycube we used default parameters; for our method we used $\alpha = 10$, $\beta=0.001$, $\mu=10$, $\lambda=30$, $\tau=0.5$, $\delta_{\min}=0$, $\delta_{\max}=2$, $\delta'_{\min}=1$, $\delta'_{\max}=1$, and $\chi=10$. It is important to note that EvoCube is a randomized algorithm, as is our own. Therefore, different runs may yield different results.


\subsection{Polycube segmentation quality}

\begin{table}[t]
\centering
    \begin{tabular}{l r r}
    \multicolumn{1}{c}{} &
    \multicolumn{1}{c}{fidelity} &
    \multicolumn{1}{c}{compactness}
    \\ \\

    PolyCut & \textbf{0,911} & 29,1 \\ 
    \midrule
    \quad Smooth ($g=0$) & \textbf{0,843} & \textbf{37,8} \\ 
    \quad Smooth ($g>0$) & 0,840 & 67,0 \\ 
    \quad CAD ($g=0$) & \textbf{0,943} & 21,6 \\ 
    \quad CAD ($g>0$) & \textbf{0,943} & \textbf{27,0} \\ \\

    EvoCube & 0,884 & 98,7 \\ 
    \midrule
    \quad Smooth ($g=0$) & 0,796 & 296,5 \\ 
    \quad Smooth ($g>0$) & 0,847 & 67,6 \\ 
    \quad CAD ($g=0$) & 0,916 & \textbf{20,1} \\ 
    \quad CAD ($g>0$) & \textbf{0,943} & 29,1 \\ \\

    \textbf{DualCube} & 0,907 & \textbf{28,6} \\ 
    \midrule
    \quad Smooth ($g=0$) & 0,836 & 38,0 \\ 
    \quad Smooth ($g>0$) & \textbf{0,848} & \textbf{55,3} \\ 
    \quad CAD ($g=0$) & 0,941 & 20,3 \\ 
    \quad CAD ($g>0$) & 0,942 & 27,1 \\ \\

    \end{tabular}
    \caption{Evaluation and comparison of polycube segmentations.}
    \label{table:results}
\end{table}

We evaluate and compare the polycube segmentations of the three methods in terms of fidelity and compactness, as reported in Table~\ref{table:results}. Fidelity is measured by the dot product between each triangle’s normal and the normal of its assigned axis-aligned polycube face, averaged over all triangle faces of the resulting polycube segmentation~\cite{dumery2022evocube}. Compactness is quantified by counting the number of corners in the segmentation~\cite{dumery2022evocube}, defined as vertices incident to more than two distinct labels.

While DualCube explicitly optimizes for both fidelity and compactness, potentially biasing the comparison, both PolyCut and EvoCube also target similar objectives, either directly or through related proxy terms. We are unable to verify validity of results produced by PolyCut or EvoCube. Our method guarantees valid polycube segmentations by construction.

\subsection{Hexahedral meshing}
\label{subsec:hex}

We compare hexahedral meshes generated by the polycube-based hexahedral meshing pipeline RobustPolycube~\cite{protais2022robust}, using as input the polycube segmentations generated by the three methods: PolyCut, EvoCube, and our DualCube, see Table~\ref{table:results2}. We evaluate the meshes by measuring cell quality via scaled Jacobian~\cite{pietroni2022hex}, regularity via vertex irregularity~\cite{pietroni2022hex}, and similarity via the Hausdorff distance~\cite{guo2020cut,fang2016all,huang2014l1}. Furthermore, we report on the number of failure cases, instances where the pipeline cannot process the given segmentation (typically due to an invalid polycube segmentation). These failure cases are not penalized in the average quality metrics; instead, all evaluations are computed over the subset of successful outputs. For reproducibility, we briefly explain how these metrics are computed. 

\bigskip
The Scaled Jacobian (SJ) of a hexahedral cell $h$ that is part of a hexahedral mesh $\mathcal{H}$ is the minimum of the Jacobian determinants at each vertex of $h$, normalized by the lengths of the three adjacent edges ($e_1$, $e_2$, and $e_3$):
\[
\text{SJ}(h) = \min_{v \in h}\left(\frac{\text{det J}(v)}{|e_1| \cdot |e_2| \cdot |e_3|}\right)
\]
We report on the minimum and average SJ across all cells in $\mathcal{H}$. Hexahedral meshes with cells with $\text{SJ}(h)<0$ (inverted or degenerate cells) are typically unusable for simulation.

We measure mesh irregularity (\%irr) via the percentage of irregular vertices. A vertex of a hexahedral mesh is \emph{regular} if it is adjacent to $2$ or $4$ hexahedral cells (on the boundary), or $8$ hexahedral cells (in the interior), and \emph{irregular} otherwise. Similarity between the input model $\mathcal{M}$ and the surface of hexahedral mesh $\mathcal{H}$ is measured using the symmetric Hausdorff distance $\overline{\text{HD}}$. We normalize the Hausdorff distance by dividing it by the length of the diagonal of the axis-aligned bounding box of $\mathcal{M}$. We multiply the Hausdorff distance by $100$ for a concise presentation of the results.  

\begin{table}[t]
\centering
    \begin{tabular}{l c r r r r}
    \multicolumn{1}{c}{} &
    \multicolumn{1}{c}{fail} &
    \multicolumn{1}{c}{$\text{SJ}_\text{min}$} &
    \multicolumn{1}{c}{$\text{SJ}_\text{avg}$} & 
    \multicolumn{1}{c}{\%irr} & 
    \multicolumn{1}{c}{$\overline{\text{HD}}$}
    \\ \\

    PolyCut & 7 & 0,042 & 0,913 & 0,032 & 2,212 \\ 
    \midrule
    \quad Smooth ($g=0$) & 1 & -0,041 & \textbf{0,911} & \textbf{0,026} & 2,009 \\ 
    \quad Smooth ($g>0$) & 4 & 0,017 & 0,883 & 0,027 & \textbf{1,249} \\
    \quad CAD ($g=0$) & 2 & 0,063 & 0,908 & 0,036 & \textbf{2,421} \\ 
    \quad CAD ($g>0$) & \textbf{0} & 0,112 & 0,937 & 0,030 & 2,058 \\ \\

    EvoCube & 3 & 0,025 & 0,908 & 0,035 & 2,534 \\ 
    \midrule
    \quad Smooth ($g=0$) & \textbf{0} & -0,081 & 0,909 & 0,036 & 3,037 \\ 
    \quad Smooth ($g>0$) & 1 & 0,046 & 0,904 & 0,023 & 1,450 \\
    \quad CAD ($g=0$) & 2 & 0,048 & 0,899 & 0,038 & 2,511 \\ 
    \quad CAD ($g>0$) & \textbf{0} & \textbf{0,117} & 0,933 & 0,030 & 2,197 \\ \\

    \textbf{DualCube} & \textbf{0} & \textbf{0,076} & \textbf{0,919} & \textbf{0,031} & \textbf{2,200} \\
    \midrule
    \quad Smooth ($g=0$) & \textbf{0} & \textbf{0,059} & \textbf{0,911} & 0,030 & \textbf{1,921} \\ 
    \quad Smooth ($g>0$) & \textbf{0} & \textbf{0,079} & \textbf{0,913} & \textbf{0,019} & 1,485 \\
    \quad CAD ($g=0$) & \textbf{0} & \textbf{0,073} & \textbf{0,917} & \textbf{0,034} & 2,522 \\
    \quad CAD ($g>0$) & \textbf{0} & 0,111 & \textbf{0,941} & \textbf{0,027} & \textbf{1,977} \\ \\

    \end{tabular}
    \caption{Evaluation and comparison of resulting hex-meshes.}
    \label{table:results2}
\end{table}

\subsection{Nightmare of polycubes}

\begin{figure*}
\centering
    \subcaptionsetup{skip=3pt}

    \subcaptionbox*{}{%
        \hspace{0.13\linewidth}
    }
    \hfill
    \subcaptionbox*{
    \textbf{PolyCut}
    }{%
        \hspace{0.13\linewidth}
        \hspace{0.13\linewidth}
    }
    \hfill
    \subcaptionbox*{
    \textbf{EvoCube}
    }{%
        \hspace{0.13\linewidth}
        \hspace{0.13\linewidth}
    }
    \hfill
    \subcaptionbox*{
    \textbf{DualCube}
    }{%
        \hspace{0.13\linewidth}
        \hspace{0.13\linewidth}
    }

    \vspace{5pt}
    
    \subcaptionbox{\texttt{screw}}{%
        \includegraphics[width=0.13\linewidth]{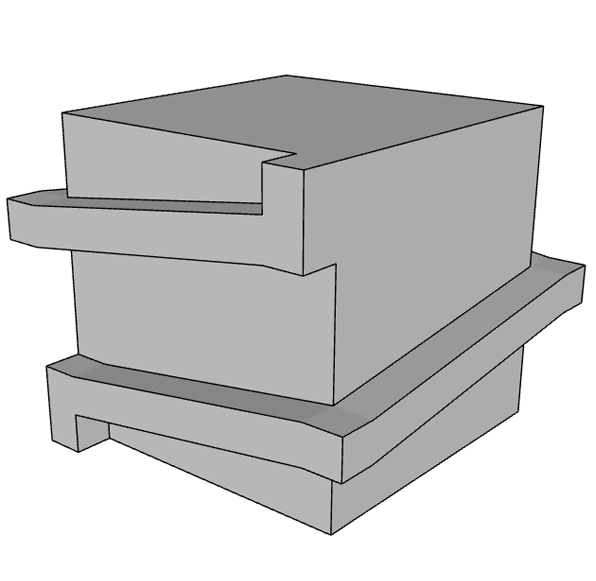}
    }
    \hfill
    \subcaptionbox*{
    $\text{SJ}_\text{min}=-1,00$\enskip$\text{SJ}_\text{avg}=0,91$\enskip$\overline{\text{HD}}=51,1$
    }{%
        \includegraphics[width=0.13\linewidth]{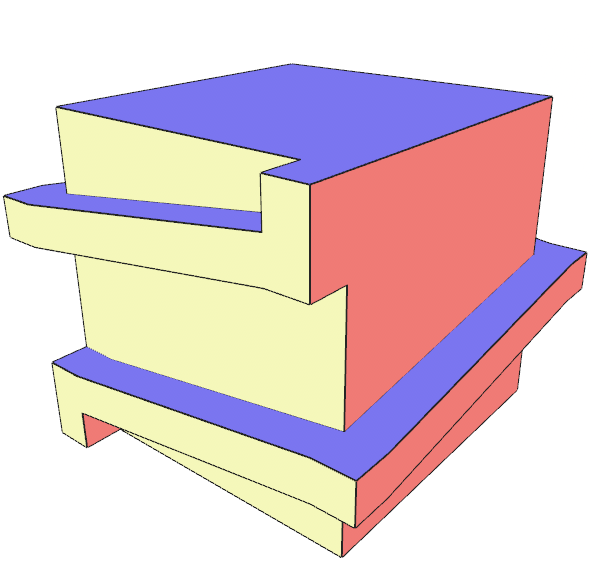}
        \includegraphics[width=0.13\linewidth]{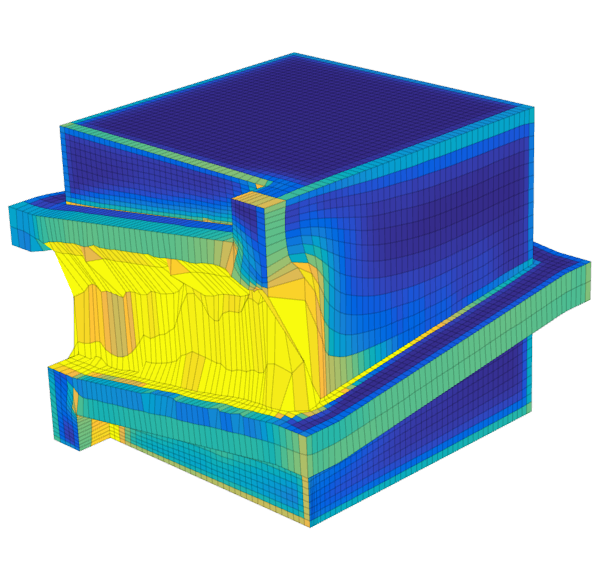}
    }
    \hfill
    \subcaptionbox*{
    $\text{SJ}_\text{min}=-0,99$\enskip$\text{SJ}_\text{avg}=0,92$\enskip$\overline{\text{HD}}=55,2$
    }{%
        \includegraphics[width=0.13\linewidth]{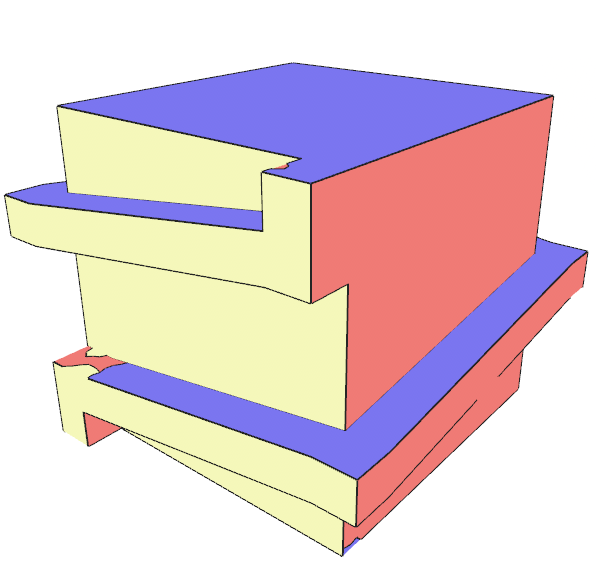}
        \includegraphics[width=0.13\linewidth]{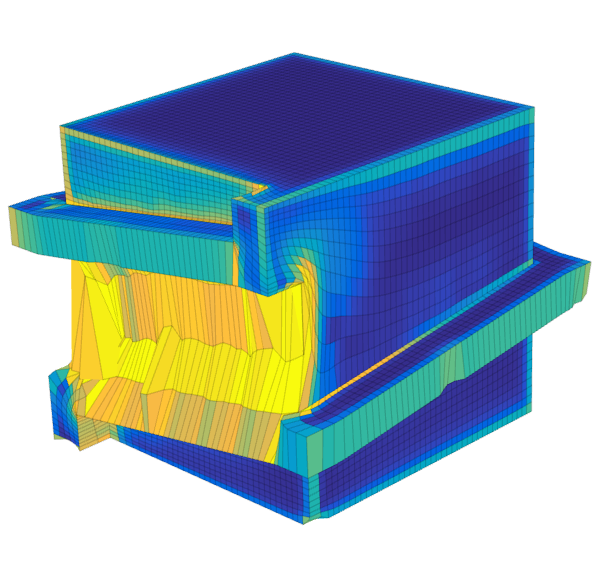}
    }
    \hfill
    \subcaptionbox*{
    $\text{SJ}_\text{min}=\mathbf{0,01}$\enskip$\text{SJ}_\text{avg}=\mathbf{0,96}$\enskip$\overline{\text{HD}}=\mathbf{20,0}$
    }{%
        \includegraphics[width=0.13\linewidth]{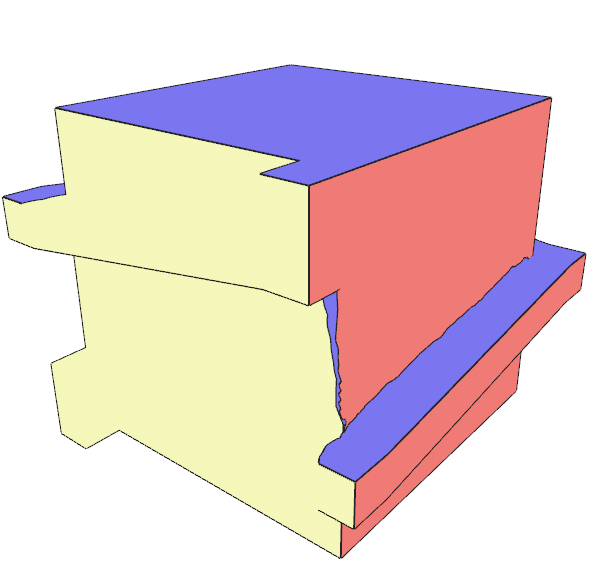}
        \includegraphics[width=0.13\linewidth]{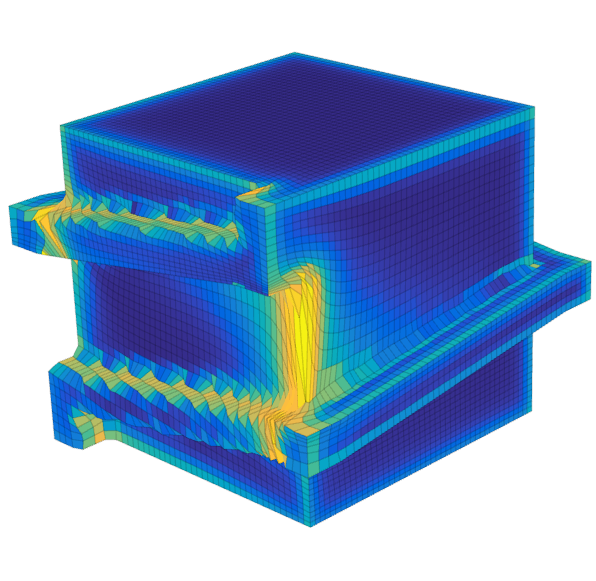}
    }
    
    \vspace{15pt}
    
    \subcaptionbox{\texttt{torus\_step}}{%
        \includegraphics[width=0.13\linewidth]{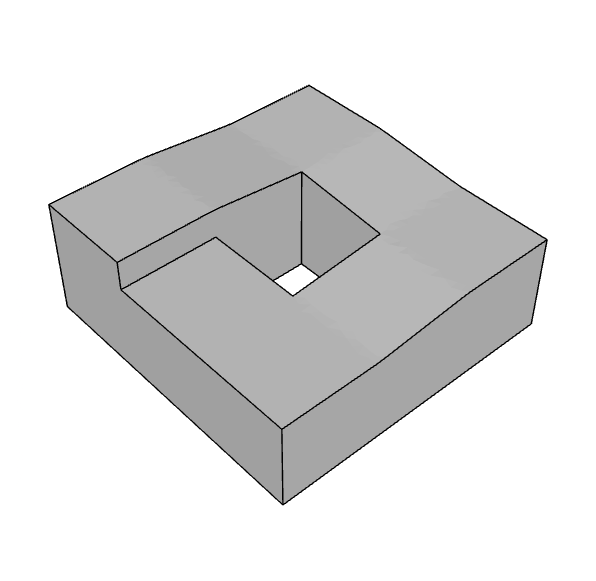}
    }
    \hfill
    \subcaptionbox*{
    $\text{SJ}_\text{min}=0,06$\enskip$\text{SJ}_\text{avg}=\mathbf{0,96}$\enskip$\overline{\text{HD}}=38,7$
    }{%
        \includegraphics[width=0.13\linewidth]{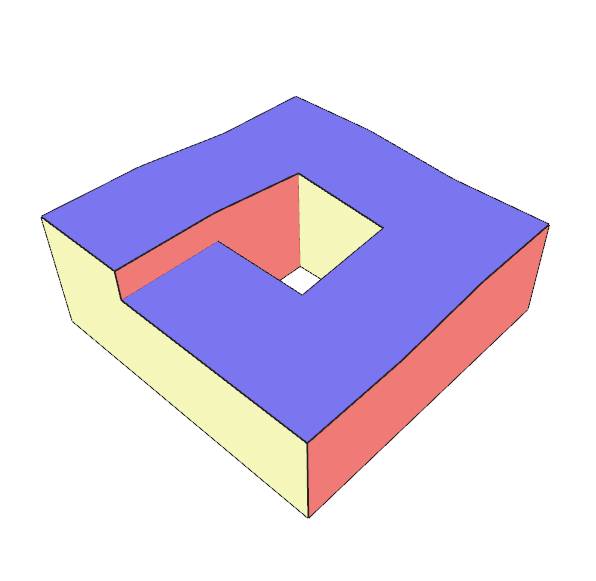}
        \includegraphics[width=0.13\linewidth]{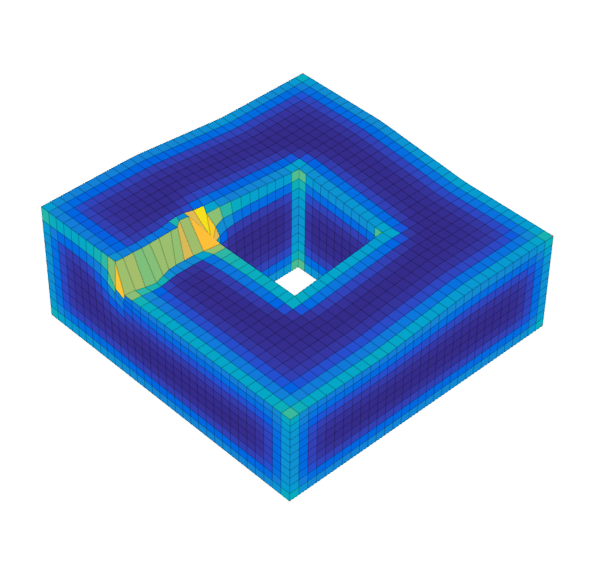}
    }
    \hfill
    \subcaptionbox*{
    $\text{SJ}_\text{min}=0,06$\enskip$\text{SJ}_\text{avg}=0,95$\enskip$\overline{\text{HD}}=62,1$
    }{%
        \includegraphics[width=0.13\linewidth]{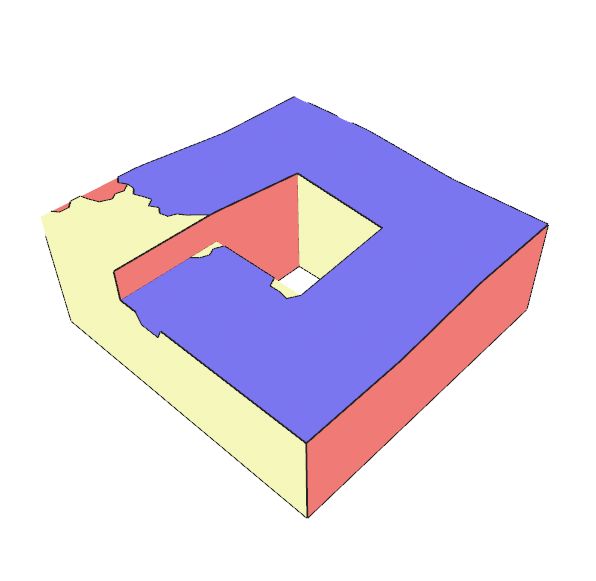}
        \includegraphics[width=0.13\linewidth]{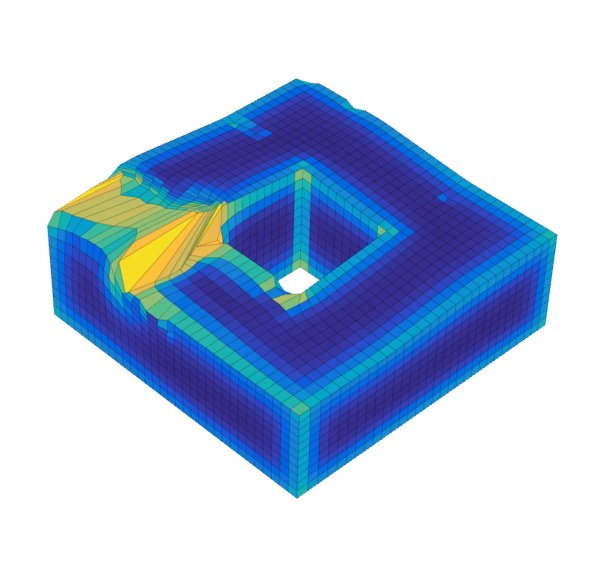}
    }
    \hfill
    \subcaptionbox*{
    $\text{SJ}_\text{min}=\mathbf{0,33}$\enskip$\text{SJ}_\text{avg}=\mathbf{0,96}$\enskip$\overline{\text{HD}}=\mathbf{25,1}$
    }{%
        \includegraphics[width=0.13\linewidth]{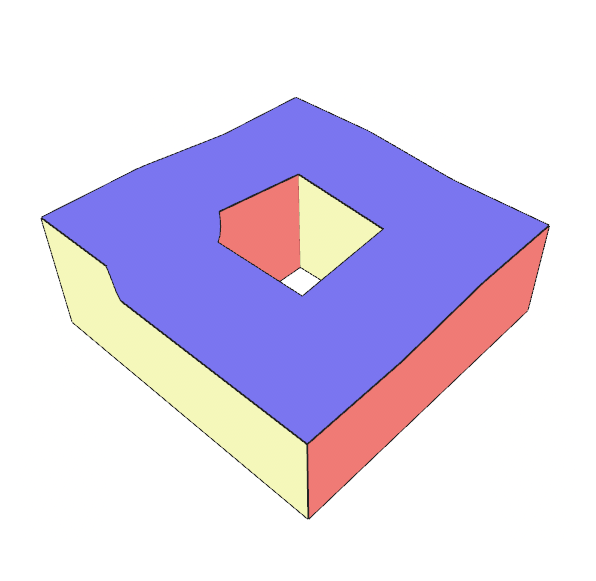}
        \includegraphics[width=0.13\linewidth]{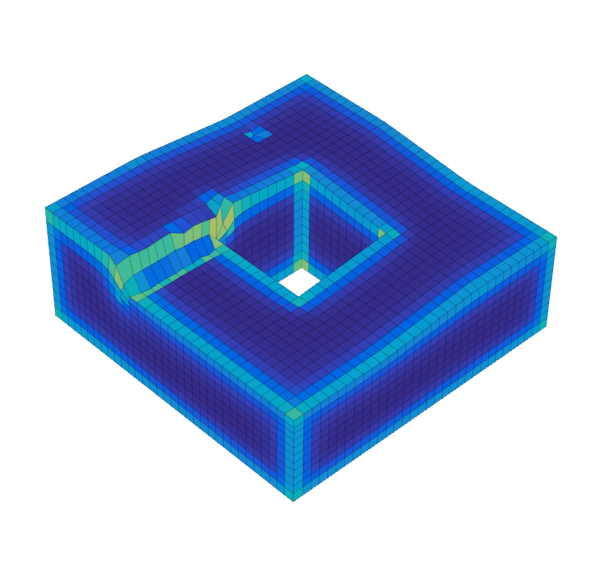}
    }
    
    \vspace{15pt}
    
    \subcaptionbox{\texttt{helix}}{%
        \includegraphics[width=0.13\linewidth]{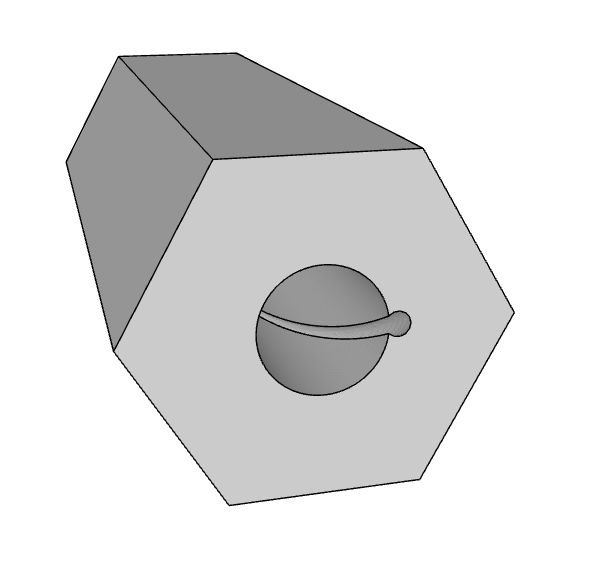}
    }
    \hfill
    \subcaptionbox*{
    no solution
    }{%
        \includegraphics[width=0.13\linewidth]{results/armadillo_poly_polycut.png}
        \includegraphics[width=0.13\linewidth]{results/armadillo_polycut.png}
    }
    \hfill
    \subcaptionbox*{
    $\text{SJ}_\text{min}=-0,97$\enskip$\text{SJ}_\text{avg}=0,93$\enskip$\overline{\text{HD}}=39,1$
    }{%
        \includegraphics[width=0.13\linewidth]{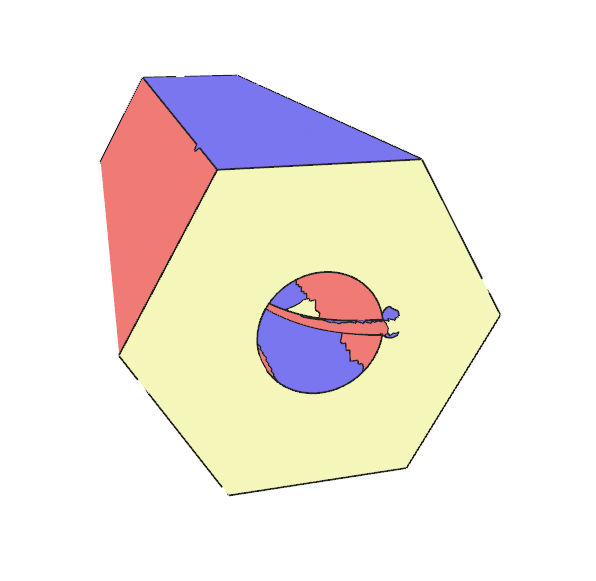}
        \includegraphics[width=0.13\linewidth]{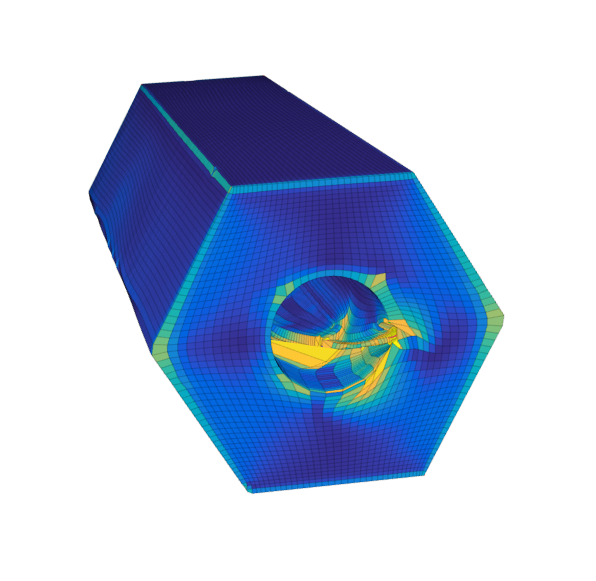}
    }
    \hfill
    \subcaptionbox*{
    $\text{SJ}_\text{min}=\mathbf{0,03}$\enskip$\text{SJ}_\text{avg}=\mathbf{0,94}$\enskip$\overline{\text{HD}}=\mathbf{8,99}$
    }{%
        \includegraphics[width=0.13\linewidth]{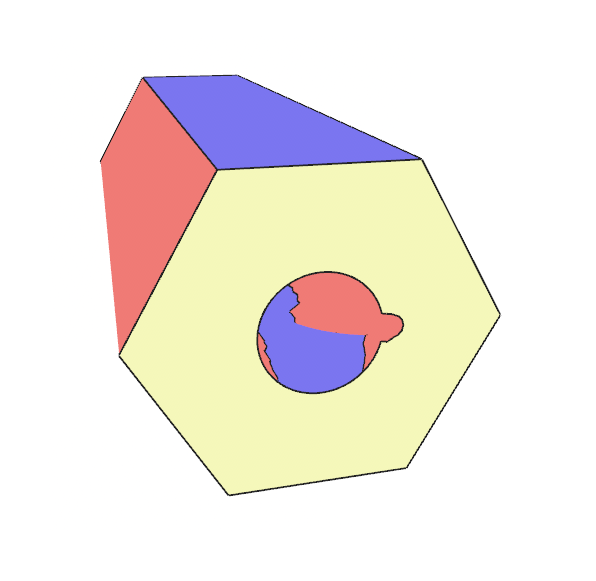}
        \includegraphics[width=0.13\linewidth]{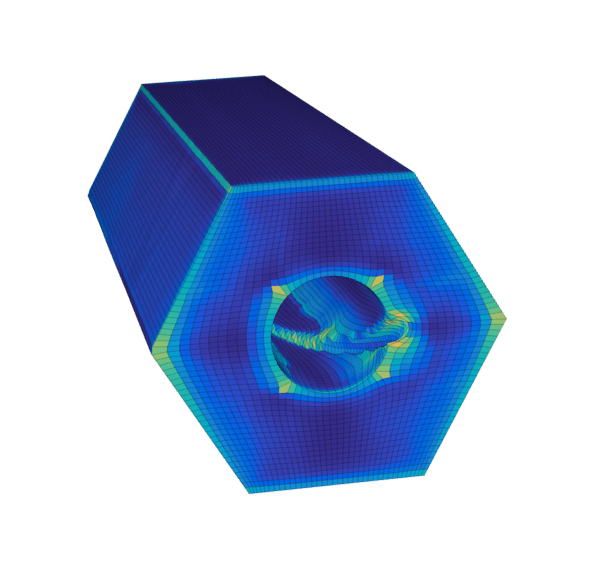}
    }

    \vspace{15pt}

        \subcaptionbox{\texttt{encrusted}}{%
        \includegraphics[width=0.13\linewidth]{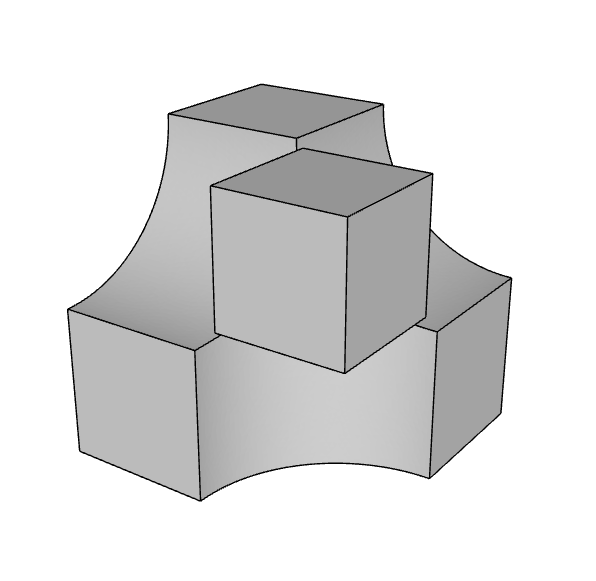}
    }
    \hfill
    \subcaptionbox*{
    $\text{SJ}_\text{min}=0,02$\enskip$\text{SJ}_\text{avg}=\mathbf{0,98}$\enskip$\overline{\text{HD}}=17,0$
    }{%
        \includegraphics[width=0.13\linewidth]{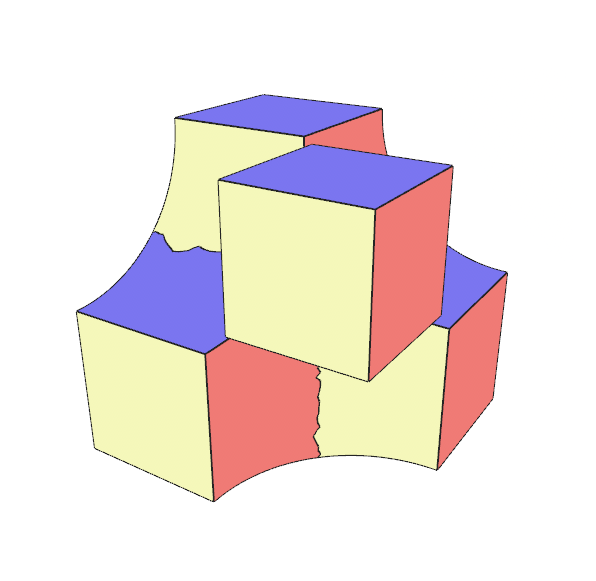}
        \includegraphics[width=0.13\linewidth]{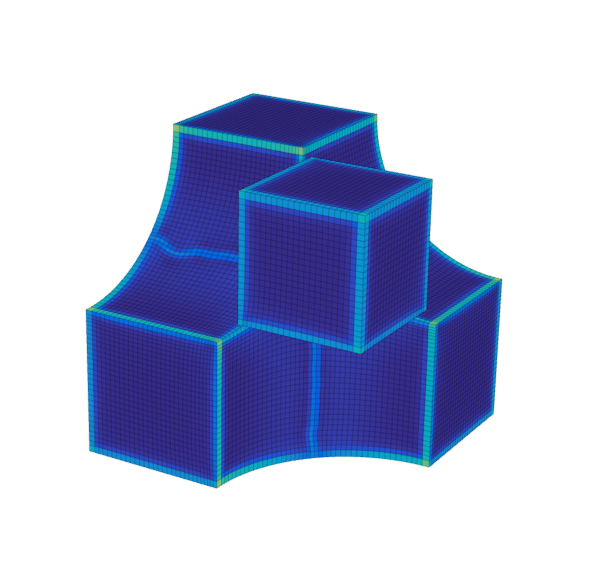}
    }
    \hfill
    \subcaptionbox*{
    $\text{SJ}_\text{min}=-0,68$\enskip$\text{SJ}_\text{avg}=0,97$\enskip$\overline{\text{HD}}=32,1$
    }{%
        \includegraphics[width=0.13\linewidth]{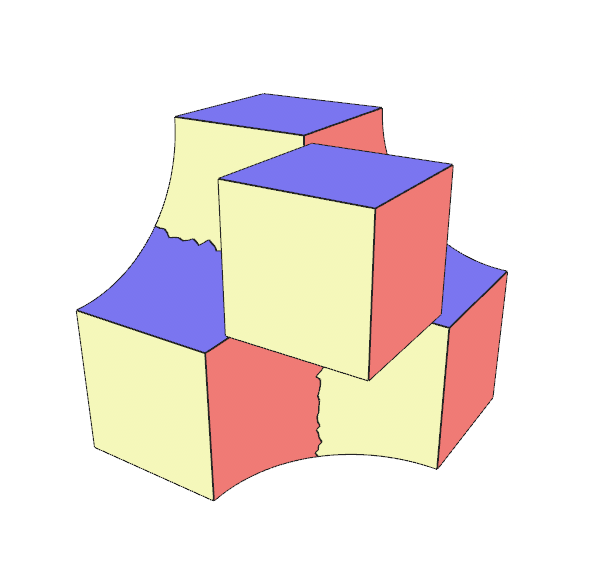}
        \includegraphics[width=0.13\linewidth]{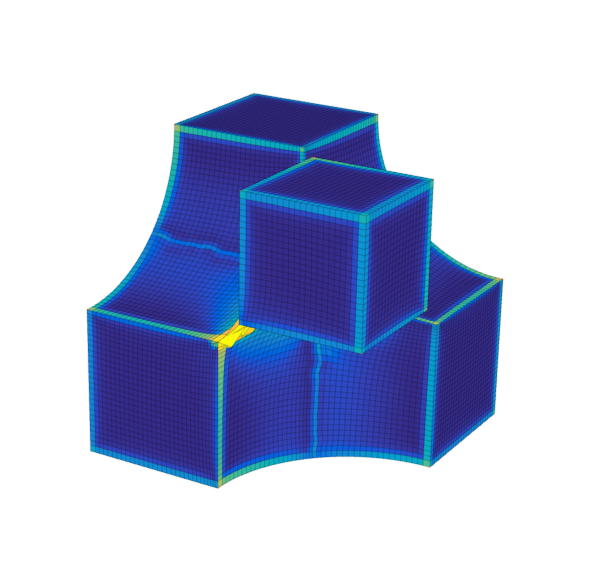}
    }
    \hfill
    \subcaptionbox*{
    $\text{SJ}_\text{min}=\mathbf{0,10}$\enskip$\text{SJ}_\text{avg}=\mathbf{0,98}$\enskip$\overline{\text{HD}}=\mathbf{16,8}$
    }{%
        \includegraphics[width=0.13\linewidth]{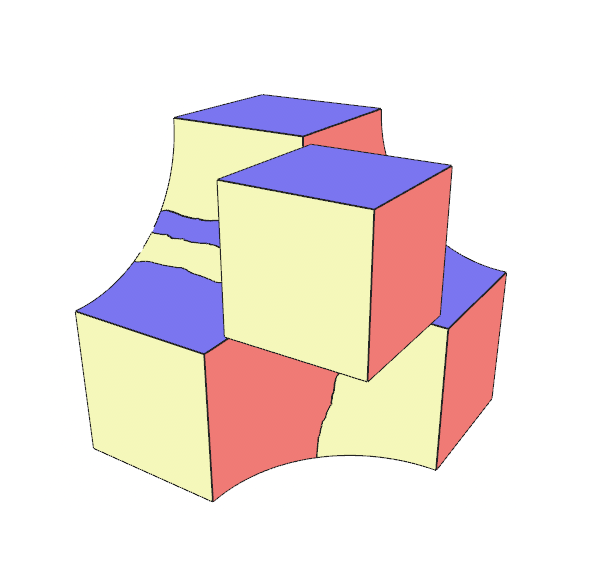}
        \includegraphics[width=0.13\linewidth]{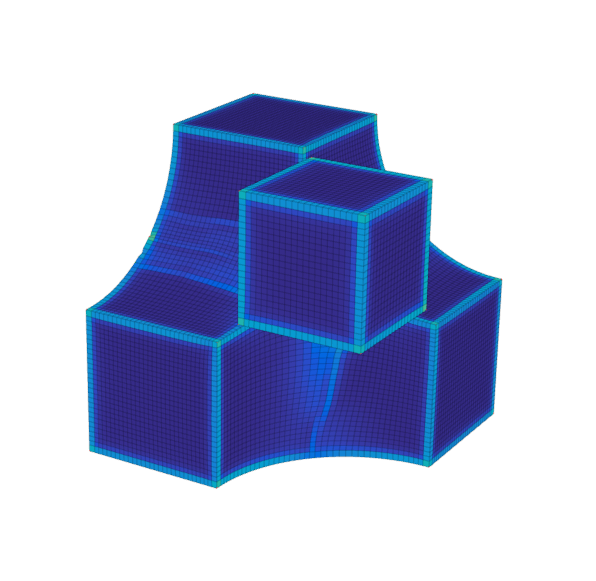}
    }

    \vspace{15pt}
    
    \subcaptionbox{\texttt{8connected}}{%
        \includegraphics[width=0.13\linewidth]{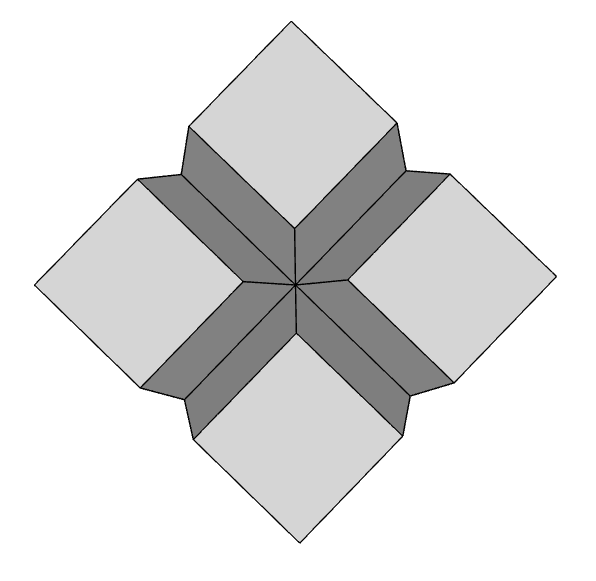}
    }
    \hfill
    \subcaptionbox*{
    $\text{SJ}_\text{min}=\mathbf{0,04}$\enskip$\text{SJ}_\text{avg}=\mathbf{0,98}$\enskip$\overline{\text{HD}}=\mathbf{14,7}$
    }{%
        \includegraphics[width=0.13\linewidth]{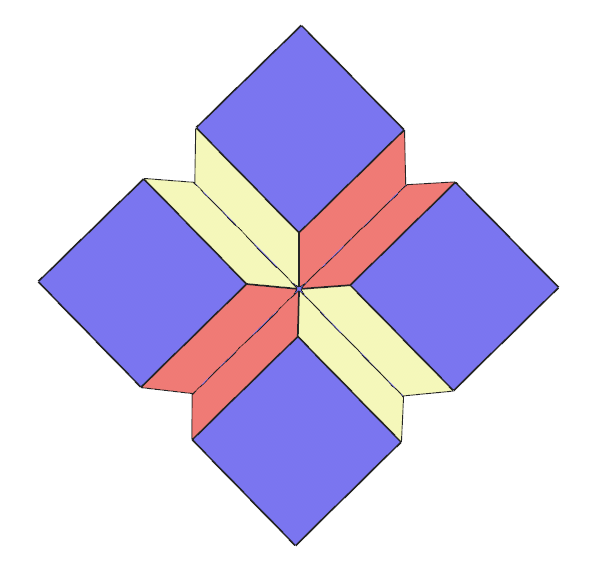}
        \includegraphics[width=0.13\linewidth]{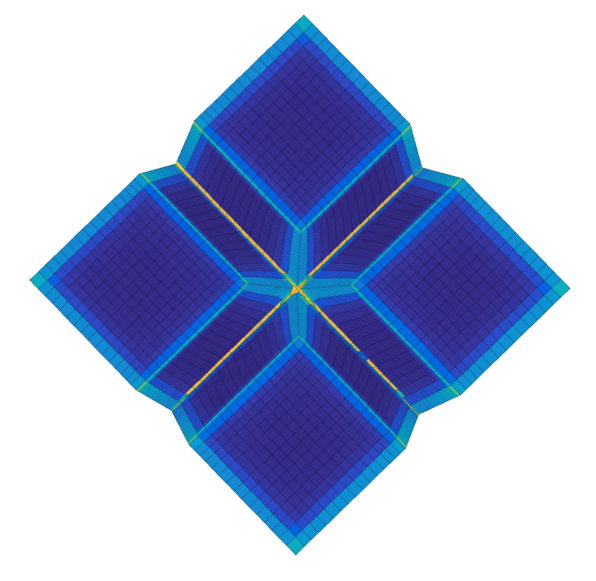}
    }
    \hfill
    \subcaptionbox*{
    $\text{SJ}_\text{min}=-0,92$\enskip$\text{SJ}_\text{avg}=0,97$\enskip$\overline{\text{HD}}=53,5$
    }{%
        \includegraphics[width=0.13\linewidth]{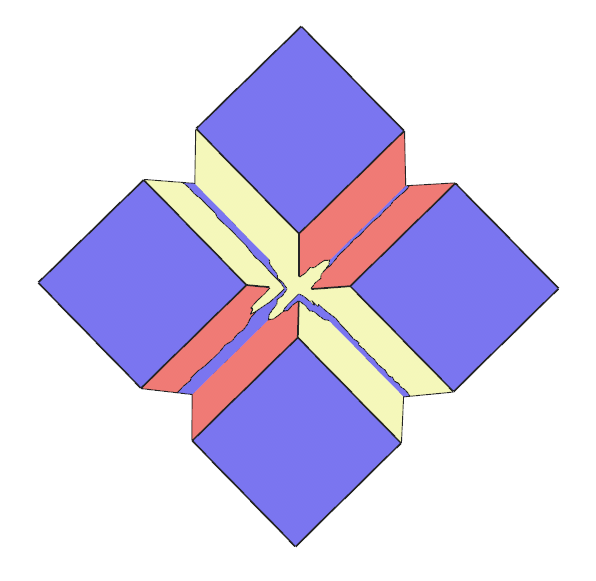}
        \includegraphics[width=0.13\linewidth]{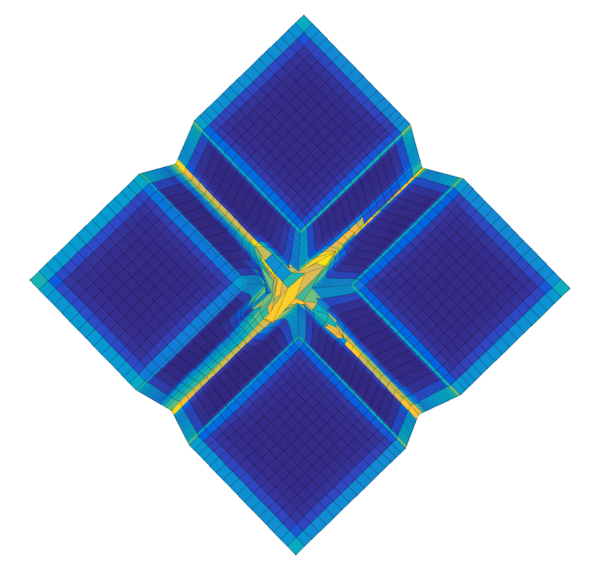}
    }
    \hfill
    \subcaptionbox*{
    $\text{SJ}_\text{min}=\mathbf{0,04}$\enskip$\text{SJ}_\text{avg}=0,90$\enskip$\overline{\text{HD}}=44,0$
    }{%
        \includegraphics[width=0.13\linewidth]{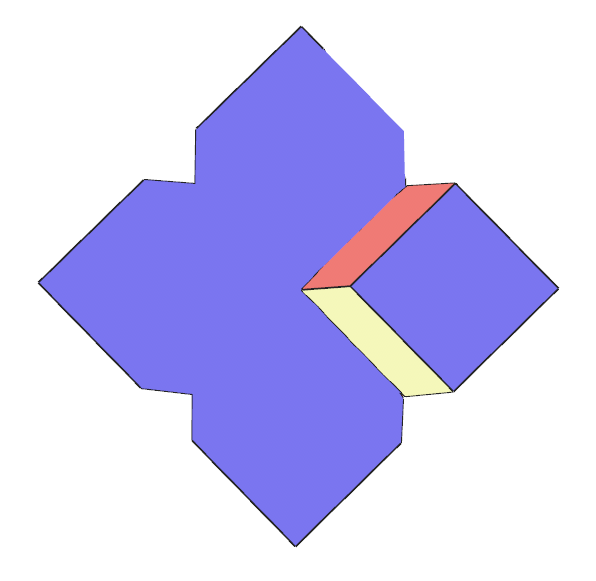}
        \includegraphics[width=0.13\linewidth]{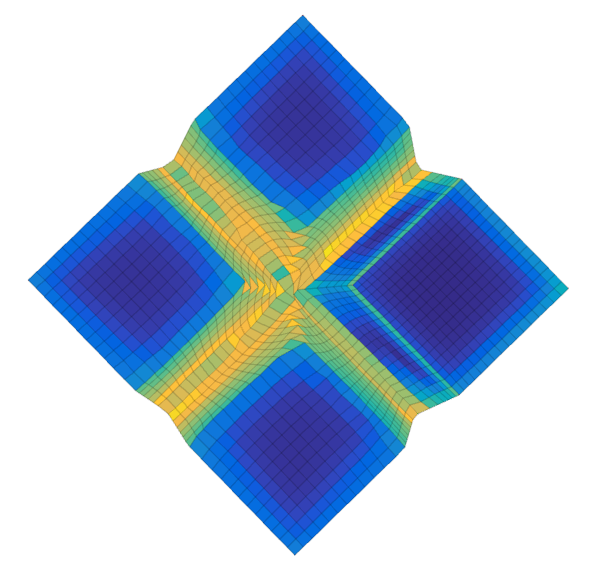}
    }

    \vspace{15pt}

    \subcaptionbox{Not handling:}{%
        \includegraphics[width=0.13\linewidth]{results/armadillo_polycut.png}
    }
    \hfill
    \subcaptionbox*{
    \texttt{24connected}
    }{%
        \includegraphics[width=0.13\linewidth]{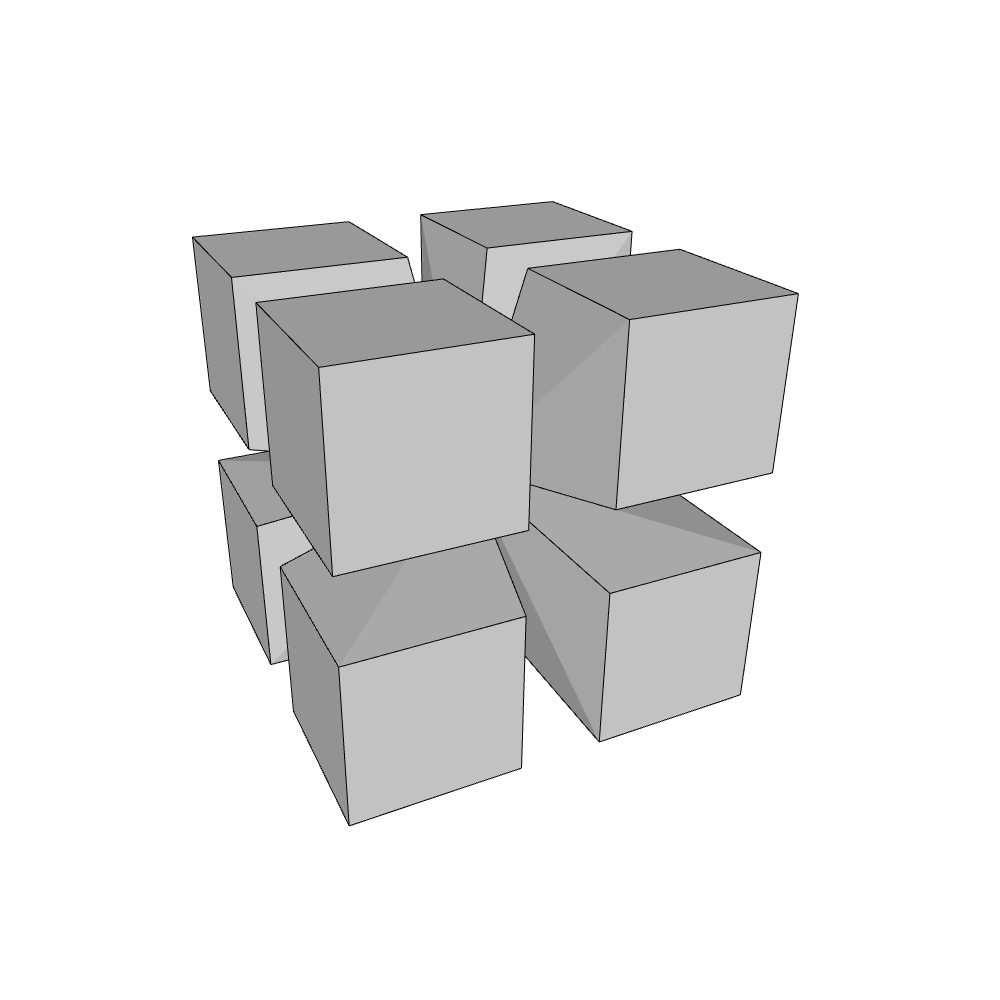}
        \includegraphics[width=0.13\linewidth]{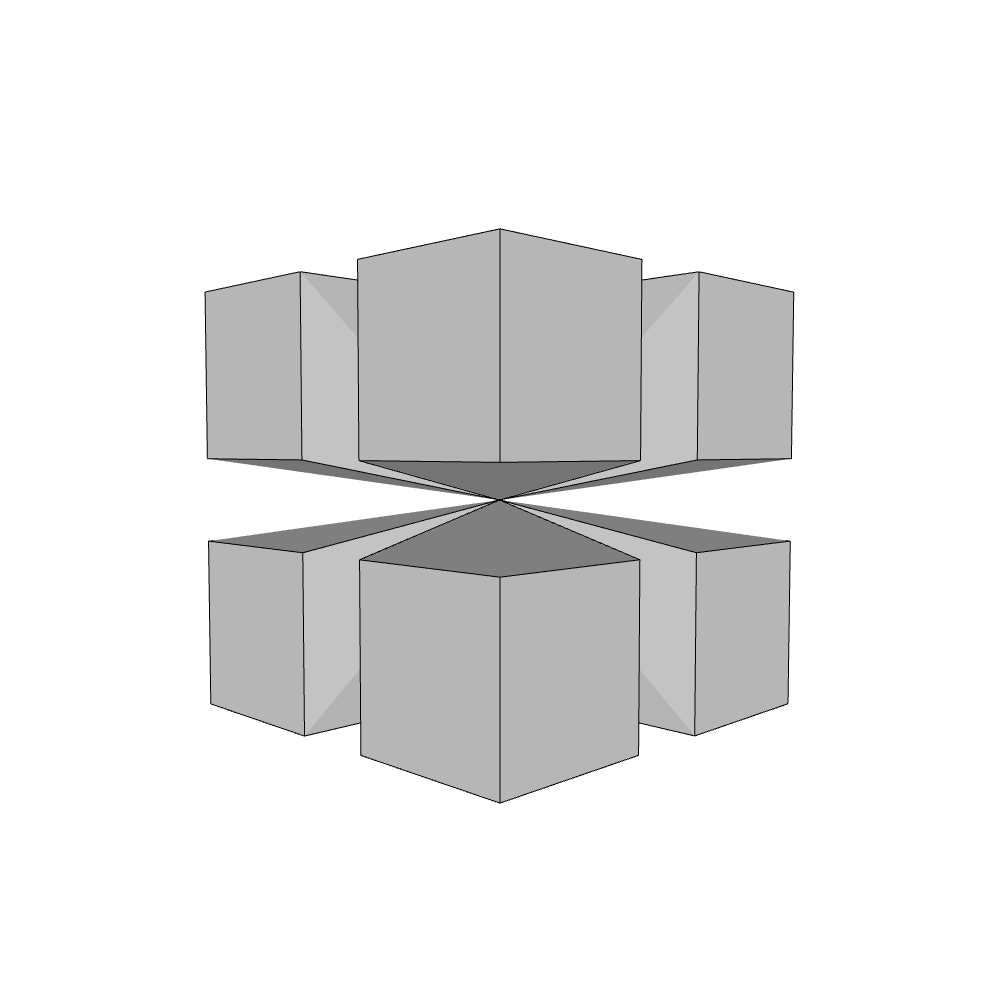}
    }
    \hfill
    \subcaptionbox*{
    \texttt{twist}
    }{%
        \includegraphics[width=0.13\linewidth]{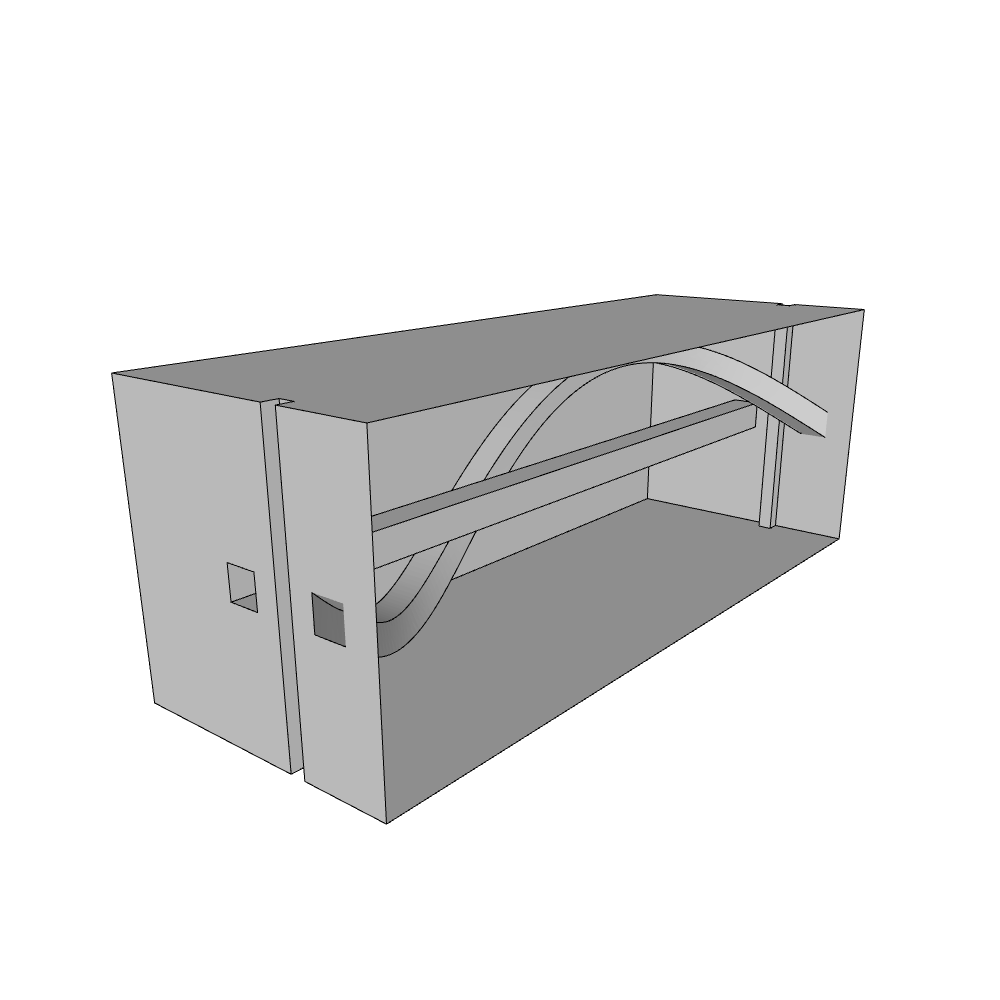}
        \includegraphics[width=0.13\linewidth]{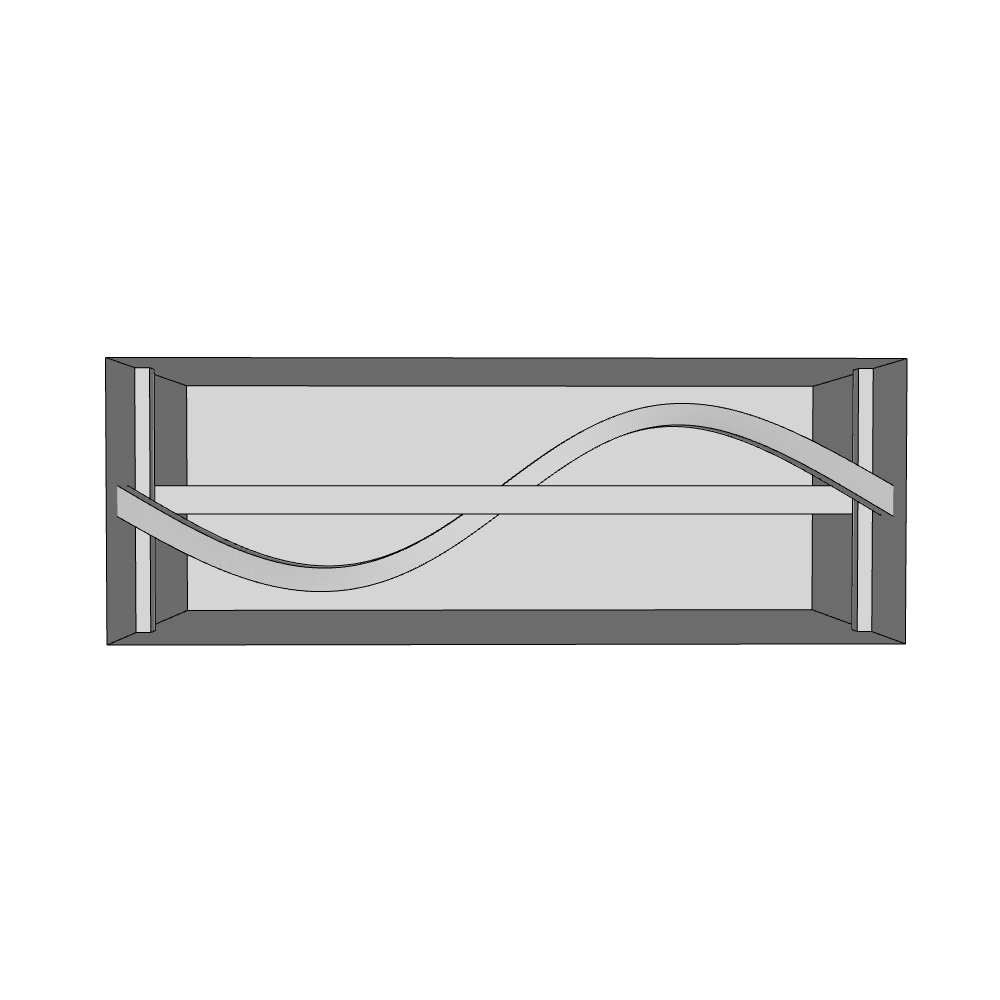}
    }
    \hfill
    \subcaptionbox*{
    \texttt{knot}
    }{%
        \includegraphics[width=0.13\linewidth]{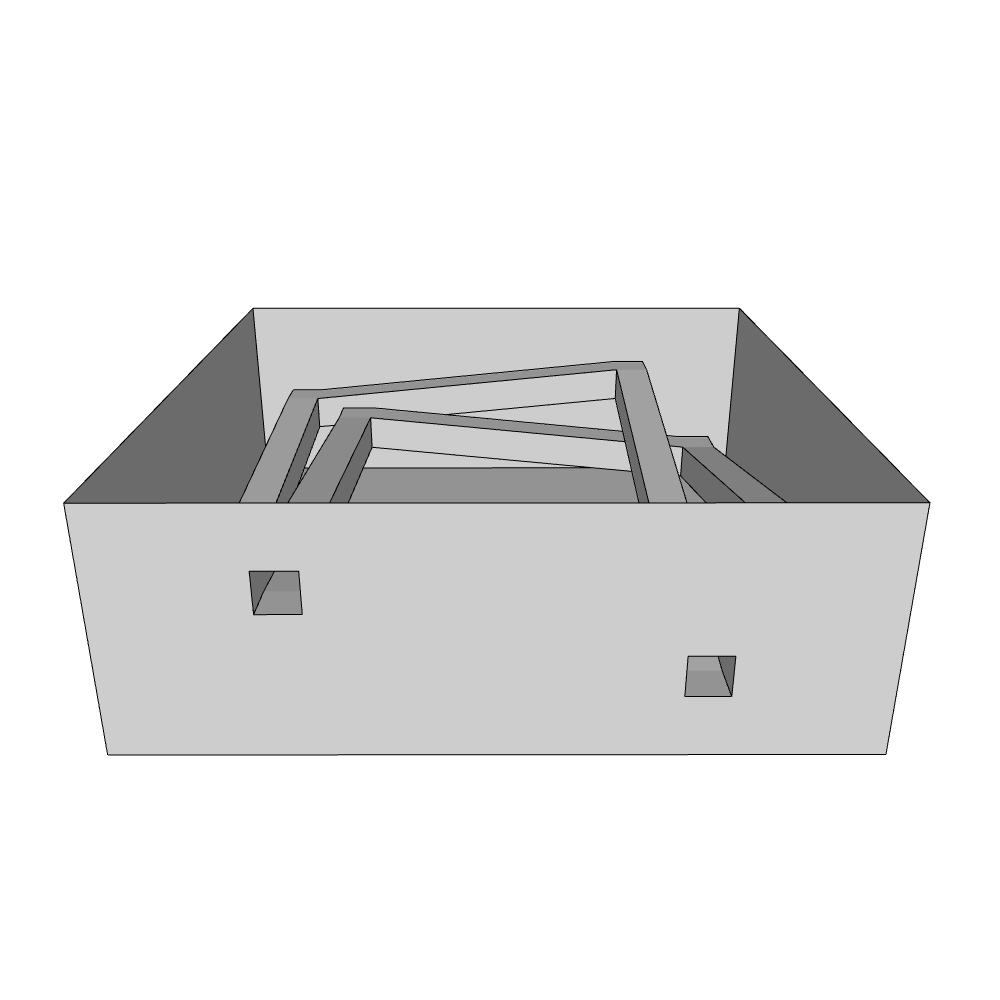}
        \includegraphics[width=0.13\linewidth]{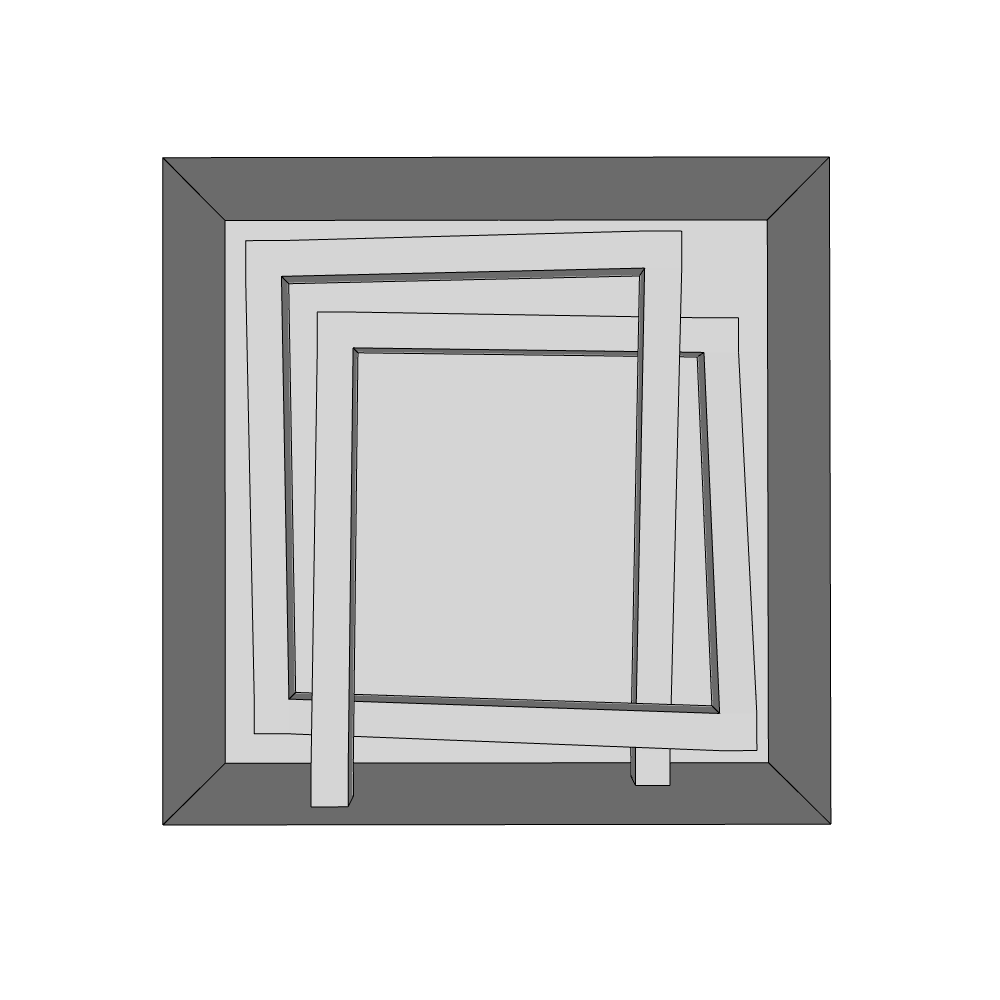}
    }
    \hfill\quad

    \caption{
    Comparison of the three methods PolyCut~\cite{livesu2013polycut}, EvoCube~\cite{dumery2022evocube}, and our method DualCube on the nightmares. We show the input mesh, and polycube segmentation and resulting hexahedral meshes per method. We include the values for $\text{SJ}_\text{min}$, $\text{SJ}_\text{avg}$, and $\overline{\text{HD}}$ per hexahedral mesh.
    }
    \label{fig:nightmare}
\end{figure*}

We evaluated all three methods also on the notoriously difficult input models from the Nightmare of Polycubes dataset; see Figure~\ref{fig:nightmare}. 
The nightmares contain input models for which PolyCut and EvoCube either do not return a solution or an invalid polycube segmentation. The hexahedral meshing pipeline~\cite{protais2022robust} appears to handle some of these invalid polycube segmentations well. Specifically, we observe that collapsing patches (for example \texttt{screw} and \texttt{torus\_step} by PolyCut and EvoCube) result in low-quality hex-meshes. However, polycubes with locally overlapping edges (for example \texttt{8connected} by PolyCut and EvoCube) result in high-quality hex-meshes.

Our method expects a watertight model bounding a single volume, and thus, we cannot handle \texttt{24connected}. Furthermore, for higher genus inputs we require manual initialization. For this reason, we do not handle the complex higher genus inputs: \texttt{twist}, \texttt{knot}, and \texttt{helix7} (not shown). For the other models, we return strictly valid polycube segmentations. For \texttt{screw}, \texttt{torus\_step}, \texttt{tray\_step} (not shown), and \texttt{helix}, the quality of the resulting hexahedral meshes is higher than those based on PolyCut and EvoCube. For \texttt{encrusted} the quality of the hex-mesh is similar, and in addition, our polycube has no locally overlapping edges. For \texttt{7connected} (not shown), \texttt{8connected}, and \texttt{twins} (not shown), our method returns a valid polycube segmentation, however, the quality of the resulting hex-mesh is not high. Currently our method routes loops only across faces. Hence, we cannot adequately capture edges with a dihedral angle approaching 0 degrees. To remedy this shortcoming in the future, we plan to extend our method to allow loops routed directly along edges instead of only across faces. 

\subsection{Discussion}

For most models, all three methods (PolyCut, EvoCube, and our approach, DualCube) completed within a few minutes, followed by hexahedral mesh generation via the RobustPolycube pipeline in an additional few minutes. The most time-consuming case was the \texttt{rocker} model, where PolyCut required 80 minutes, DualCube 25 minutes, and EvoCube 10 minutes. All experiments were performed on a standard laptop (Intel i7-11th Gen, 32 GB RAM).

Quantitative results can be found in Table~\ref{table:results} and~\ref{table:results2}. The resulting polycube segmentations from PolyCut and our method (DualCube) are of mostly similar quality. The polycube segmentations by EvoCube have a slightly lower fidelity, and a higher compactness. The hex-meshing pipeline failed for some results of PolyCut (7) and EvoCube (3). The hex-meshing pipeline always succeeded with our polycube segmentations, likely due to our polycube segmentations to be valid by definition. Furthermore, the hexahedral meshes resulting from our polycube segmentations often are of higher quality than those constructed via the other methods in terms of the Scaled Jacobian, irregularity, and similarity (measured through the Hausdorff distance).

\begin{figure*}
\centering

    \vspace{15pt}

    \hfill
    \subcaptionbox*{}{%
        \includegraphics[width=0.14\linewidth]{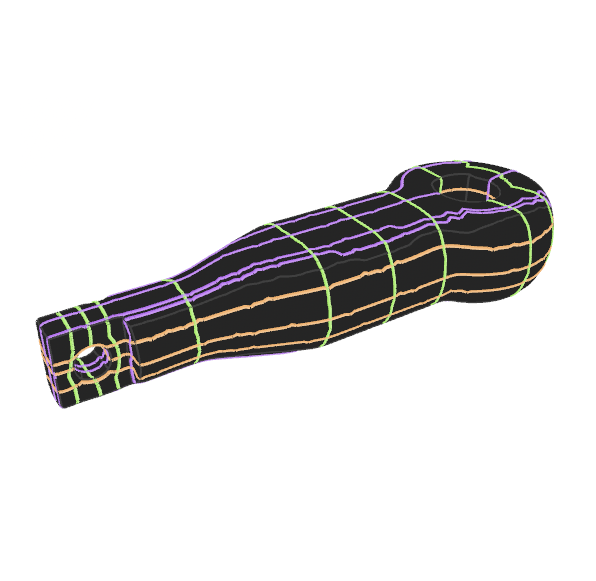}
    }
    \hfill
    \subcaptionbox{\texttt{rod}}{%
        \includegraphics[width=0.14\linewidth]{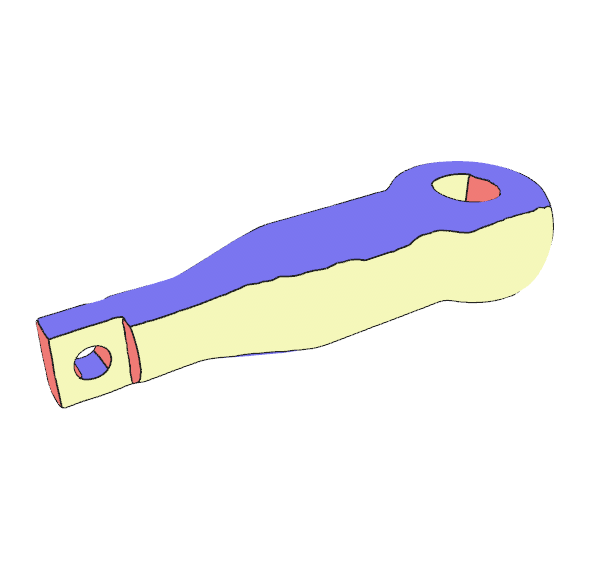}
    }
    \hfill
    \subcaptionbox*{}{%
        \includegraphics[width=0.14\linewidth]{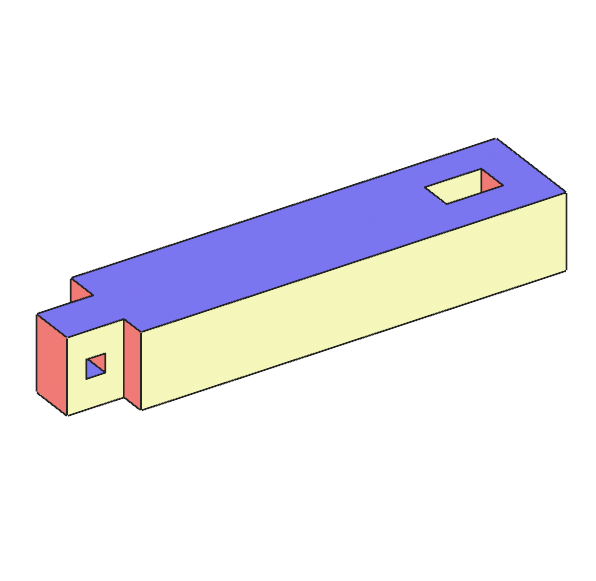}
    }
    \hspace{40pt}
    \subcaptionbox*{}{%
        \includegraphics[width=0.14\linewidth]{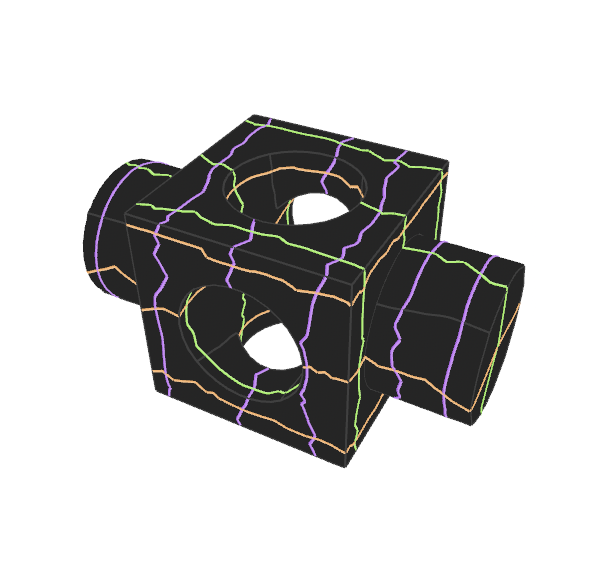}
    }
    \hfill
    \subcaptionbox{\texttt{block}}{%
        \includegraphics[width=0.14\linewidth]{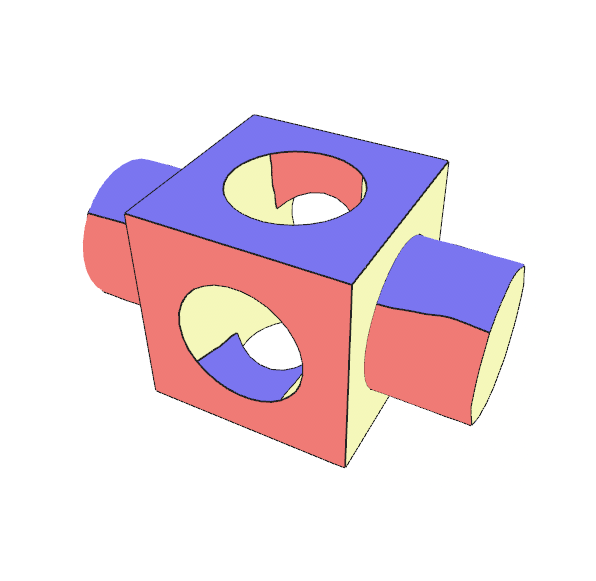}
    }
    \hfill
    \subcaptionbox*{}{%
        \includegraphics[width=0.14\linewidth]{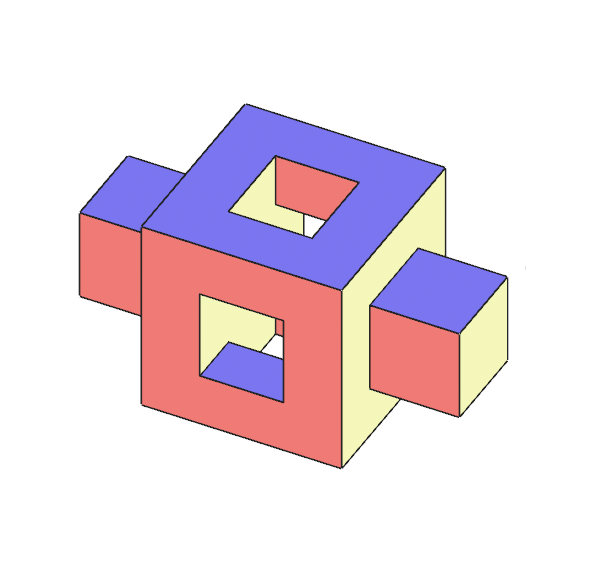}
    }
    \hfill\quad

    \vspace{15pt}

    \hfill
    \subcaptionbox*{}{%
        \includegraphics[width=0.14\linewidth]{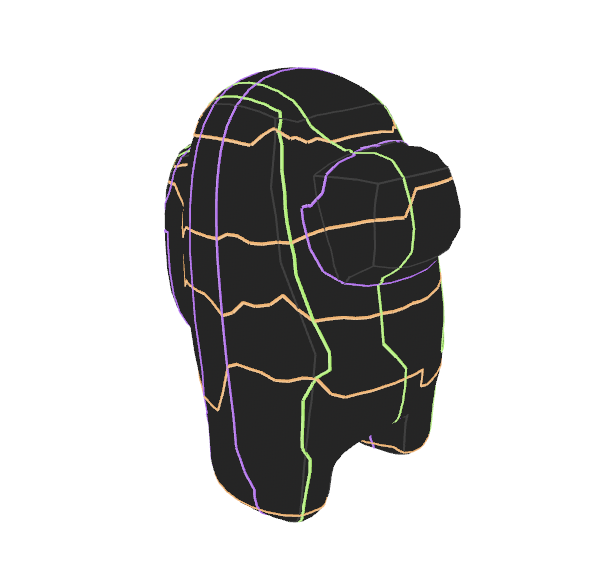}
    }
    \hfill
    \subcaptionbox{\texttt{amogus}}{%
        \includegraphics[width=0.14\linewidth]{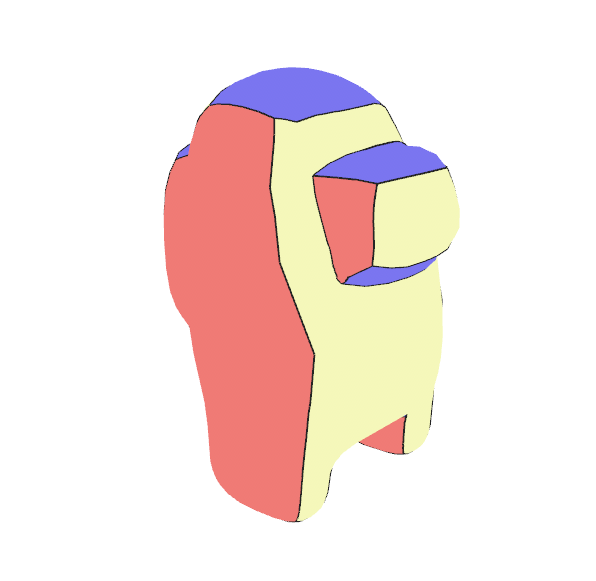}
    }
    \hfill
    \subcaptionbox*{}{%
        \includegraphics[width=0.14\linewidth]{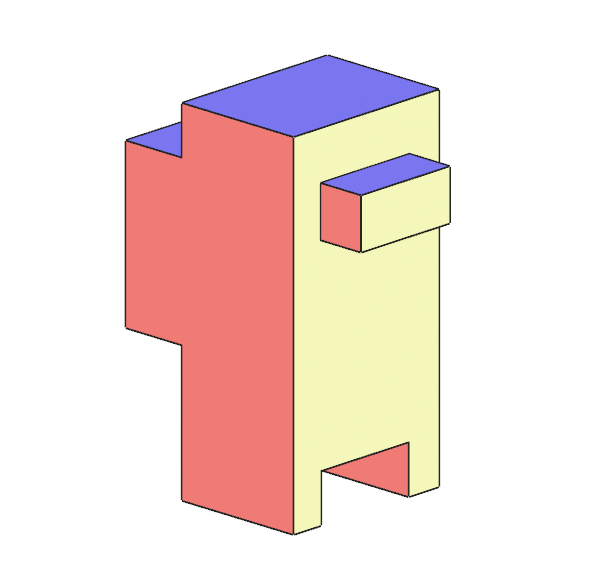}
    }
    \hspace{40pt}
    \subcaptionbox*{}{%
        \includegraphics[width=0.14\linewidth]{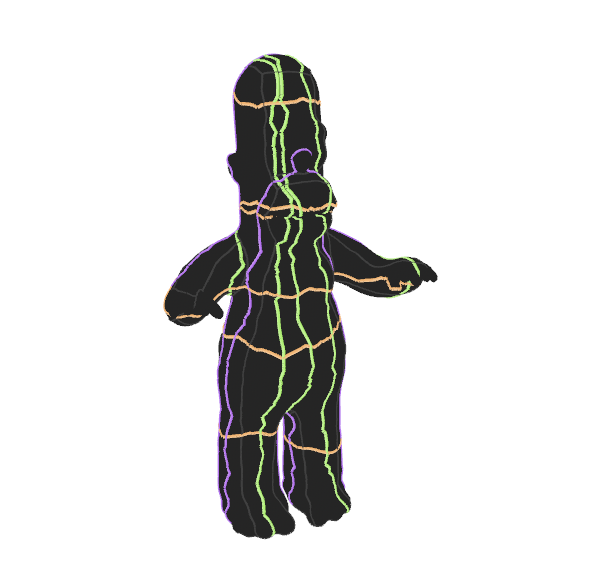}
    }
    \hfill
    \subcaptionbox{\texttt{homer}}{%
        \includegraphics[width=0.14\linewidth]{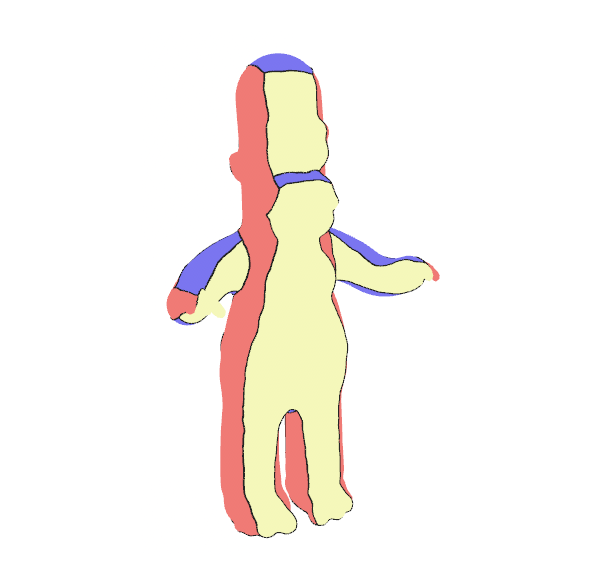}
    }
    \hfill
    \subcaptionbox*{}{%
        \includegraphics[width=0.14\linewidth]{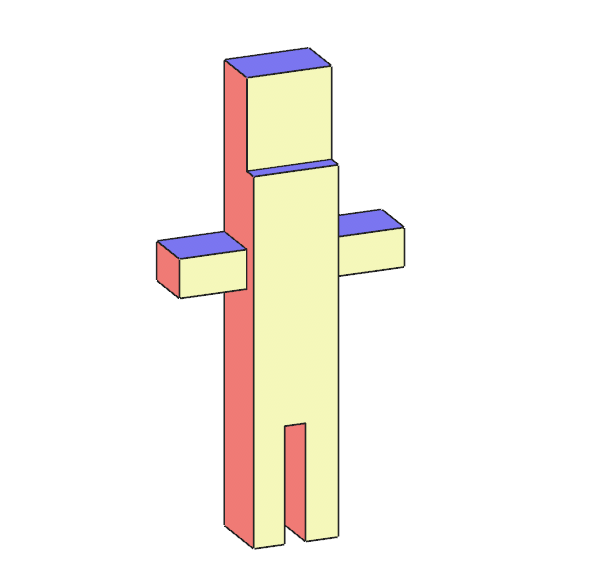}
    }
    \hfill\quad

    \vspace{15pt}

    \hfill
    \subcaptionbox*{}{%
        \includegraphics[width=0.14\linewidth]{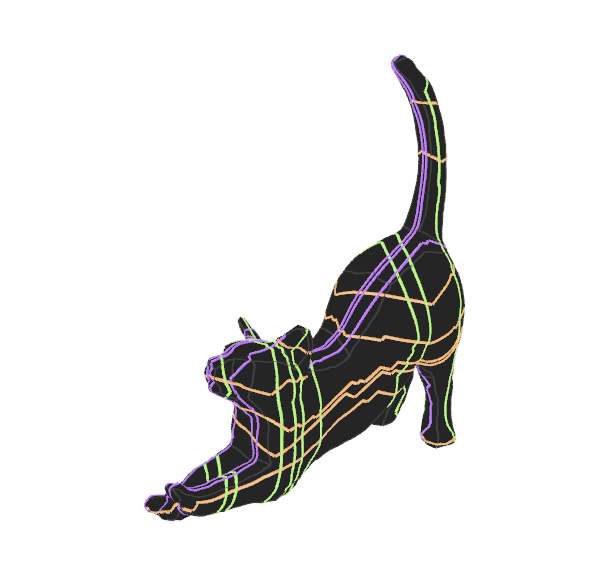}
    }
    \hfill
    \subcaptionbox{\texttt{cat}}{%
        \includegraphics[width=0.14\linewidth]{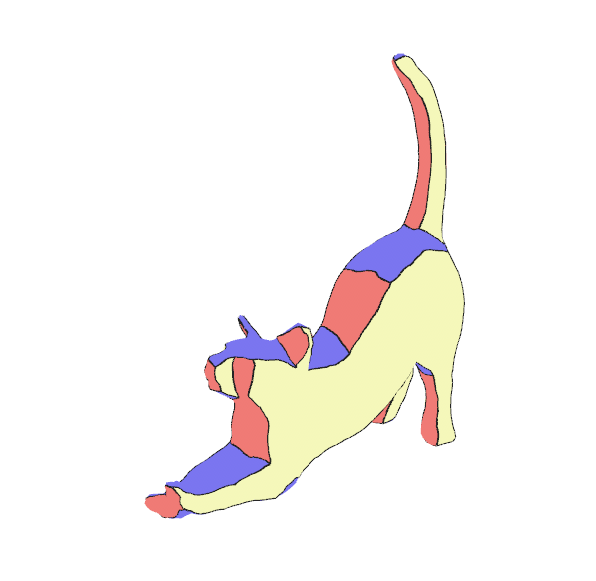} 
    }
    \hfill
    \subcaptionbox*{}{%
        \includegraphics[width=0.14\linewidth]{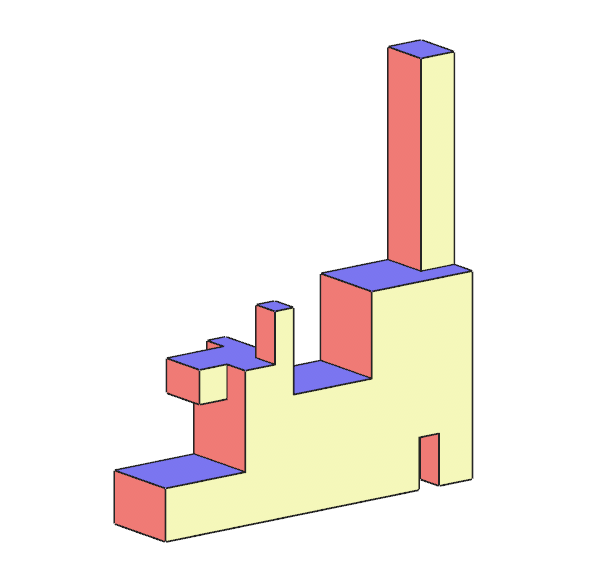}
    }
    \hspace{40pt}
    \subcaptionbox*{}{%
        \includegraphics[width=0.14\linewidth]{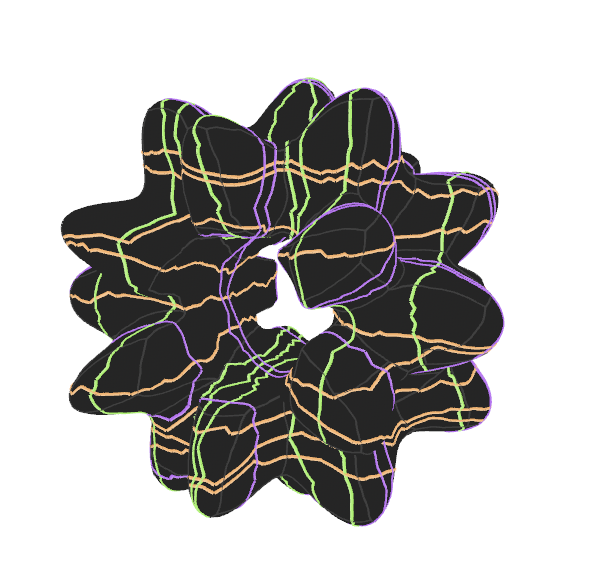}
    }
    \hfill
    \subcaptionbox{\texttt{bumpytorus}}{%
        \includegraphics[width=0.14\linewidth]{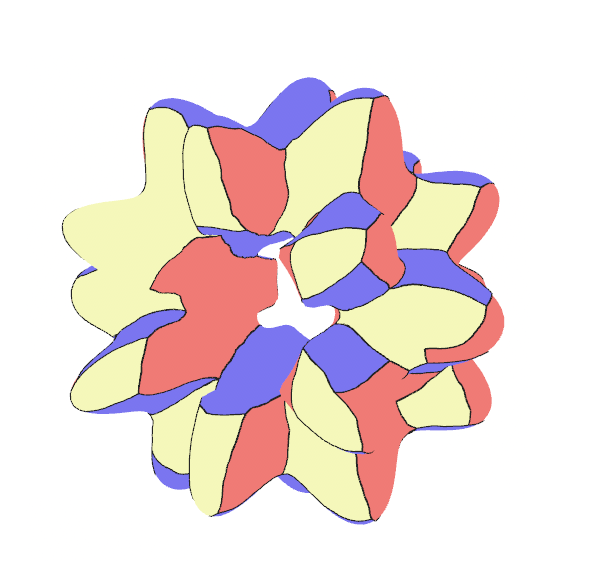}
    }
    \hfill
    \subcaptionbox*{}{%
        \includegraphics[width=0.14\linewidth]{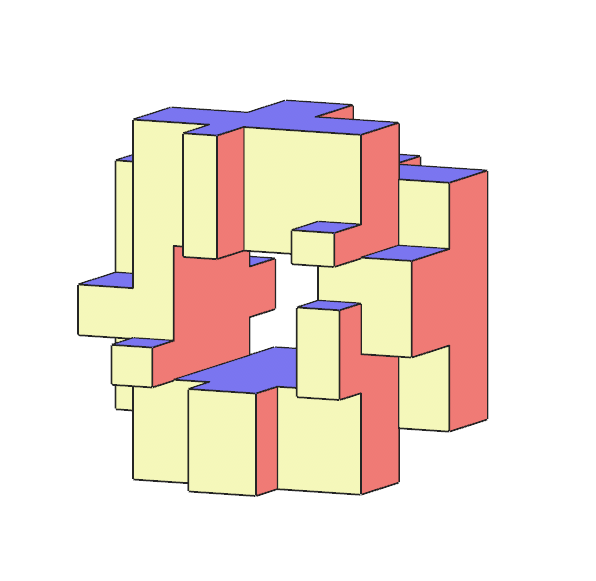}
    }
    \hfill\quad
    
    \vspace{15pt}

    \hfill
    \subcaptionbox*{}{%
        \includegraphics[width=0.14\linewidth]{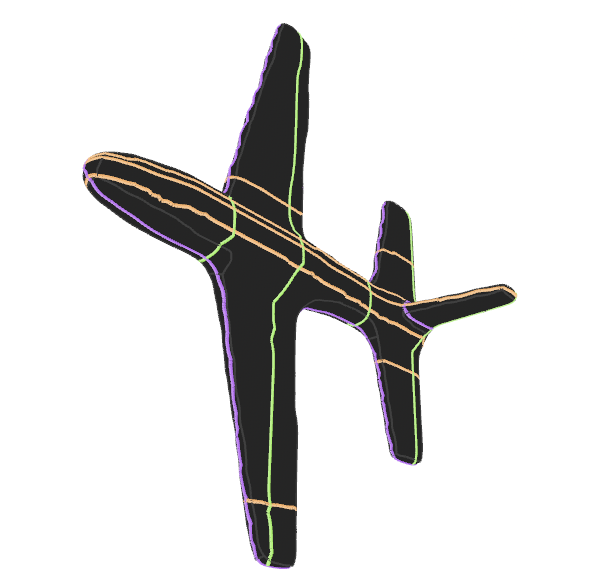}
    }
    \hfill
    \subcaptionbox{\texttt{airplane2}}{%
        \includegraphics[width=0.14\linewidth]{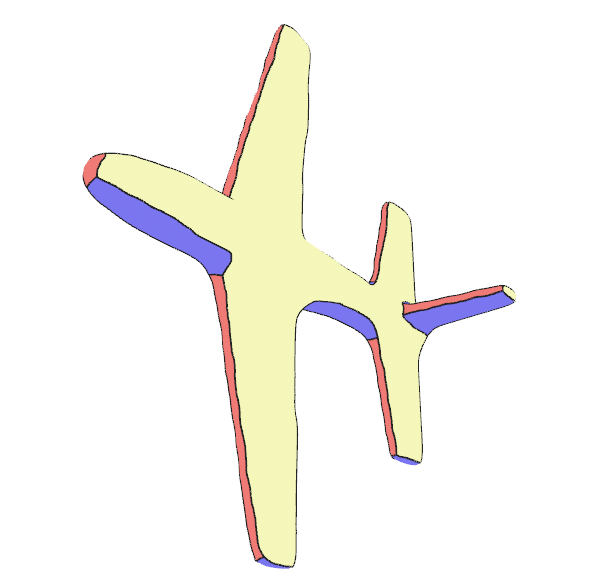}
    }
    \hfill
    \subcaptionbox*{}{%
        \includegraphics[width=0.14\linewidth]{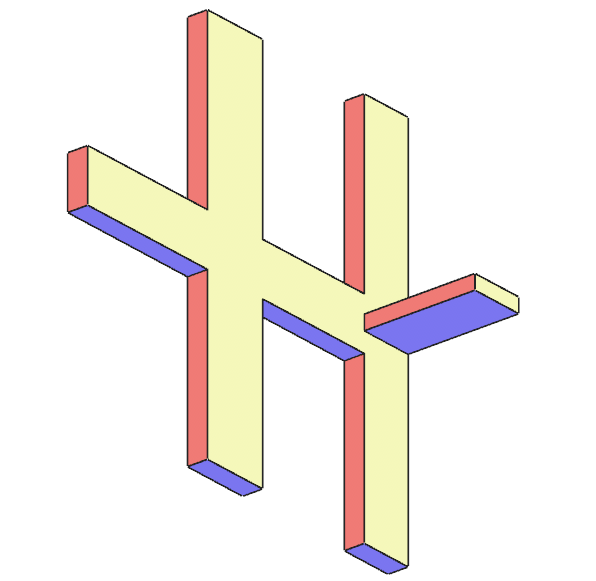}
    }
    \hspace{40pt}
    \subcaptionbox*{}{%
        \includegraphics[width=0.14\linewidth]{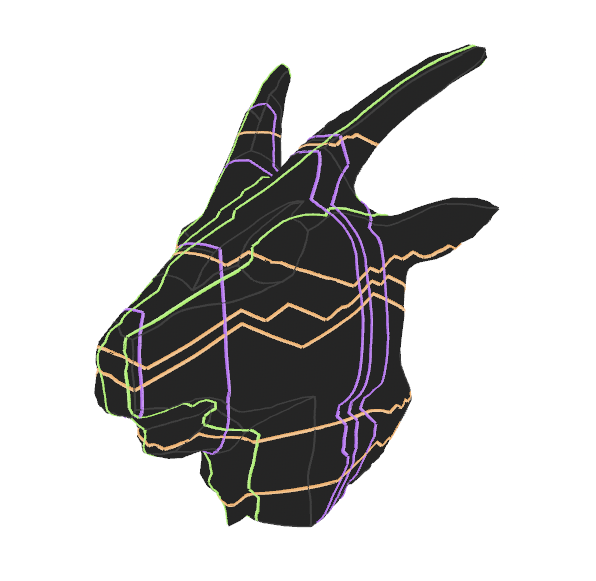}
    }
    \hfill
    \subcaptionbox{\texttt{goathead}}{%
        \includegraphics[width=0.14\linewidth]{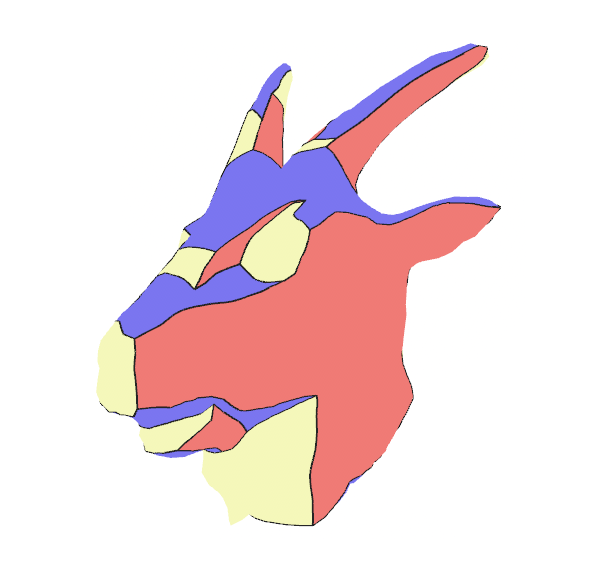}
    }
    \hfill
    \subcaptionbox*{}{%
        \includegraphics[width=0.14\linewidth]{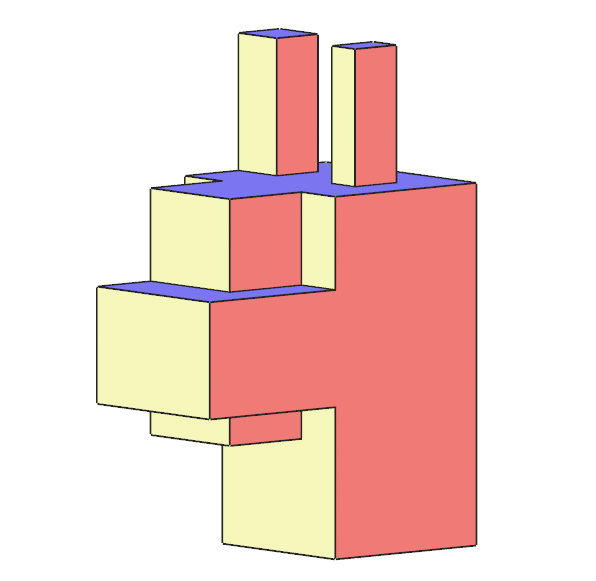}
    }
    \hfill\quad

    \vspace{15pt}

    \hfill
    \subcaptionbox*{}{%
        \includegraphics[width=0.14\linewidth]{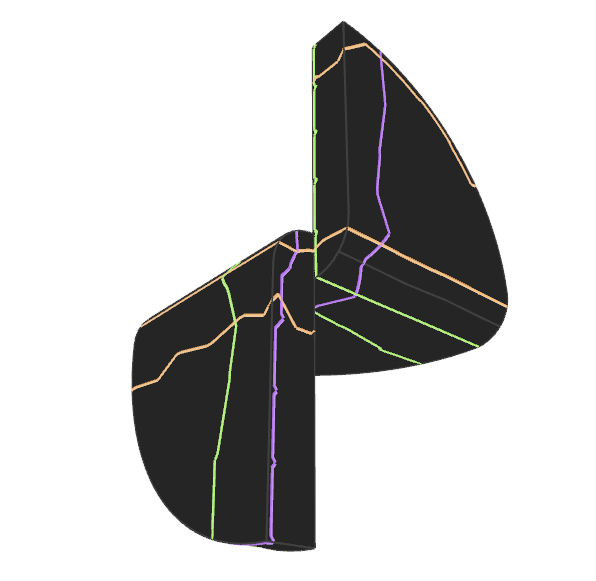}
    }
    \hfill
    \subcaptionbox{\texttt{B48}}{%
        \includegraphics[width=0.14\linewidth]{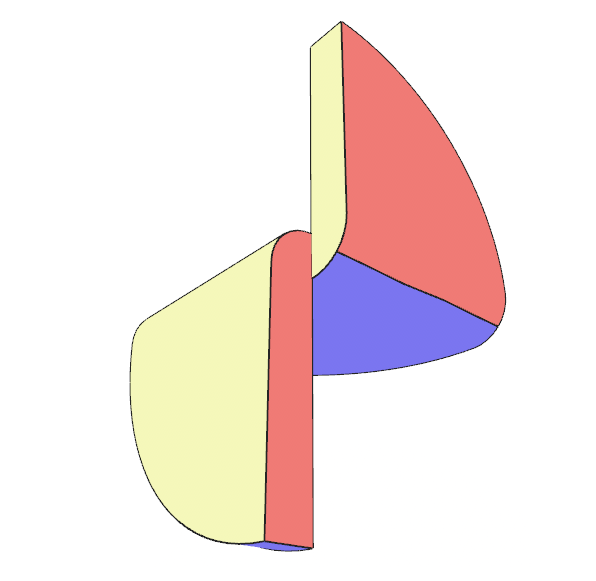}
    }
    \hfill
    \subcaptionbox*{}{%
        \includegraphics[width=0.14\linewidth]{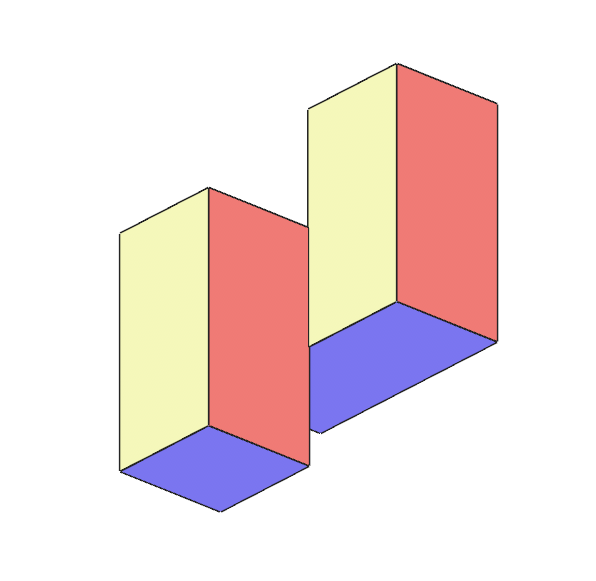}
    }
    \hspace{40pt}
    \subcaptionbox*{}{%
        \includegraphics[width=0.14\linewidth]{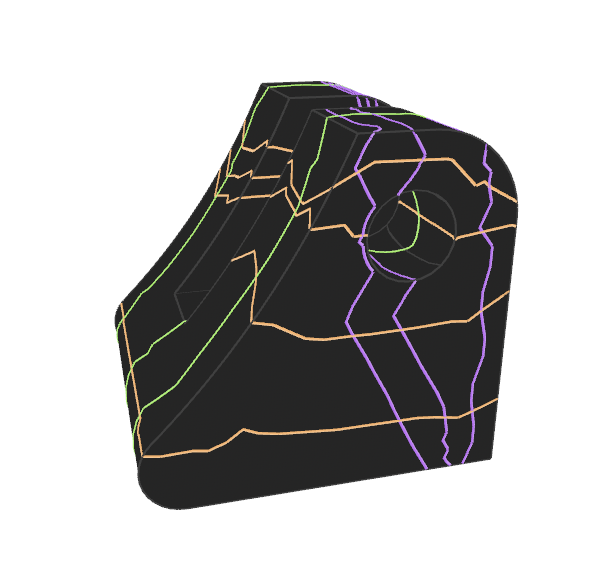}
    }
    \hfill
    \subcaptionbox{\texttt{B33}}{%
        \includegraphics[width=0.14\linewidth]{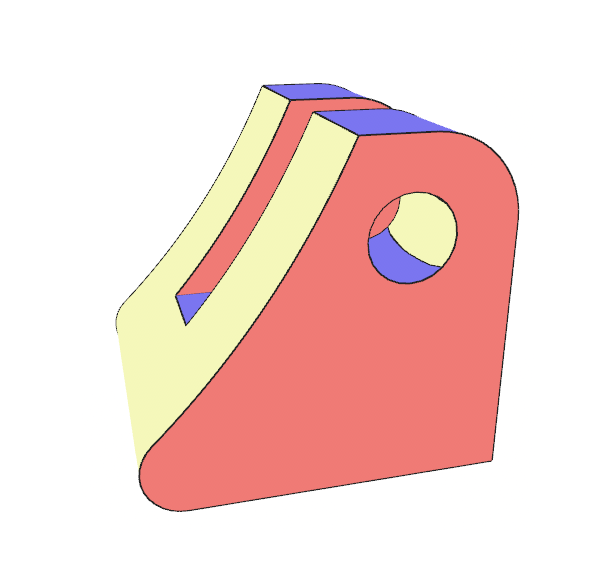}
    }
    \hfill
    \subcaptionbox*{}{%
        \includegraphics[width=0.14\linewidth]{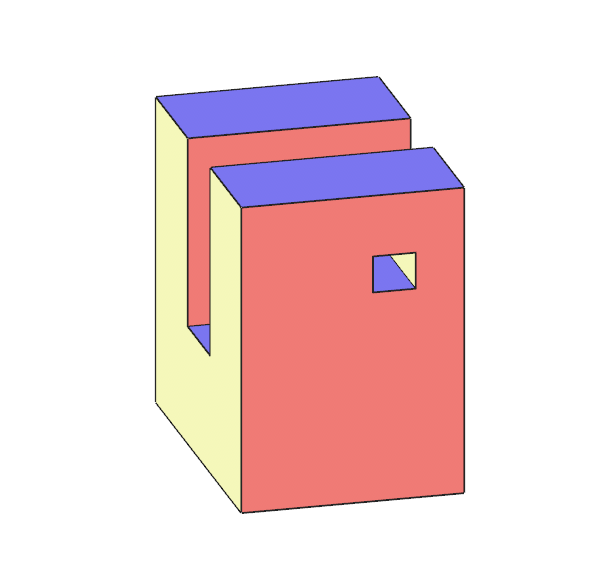}
    }
    \hfill\quad

    \vspace{15pt}

    \hfill
    \subcaptionbox*{}{%
        \includegraphics[width=0.14\linewidth]{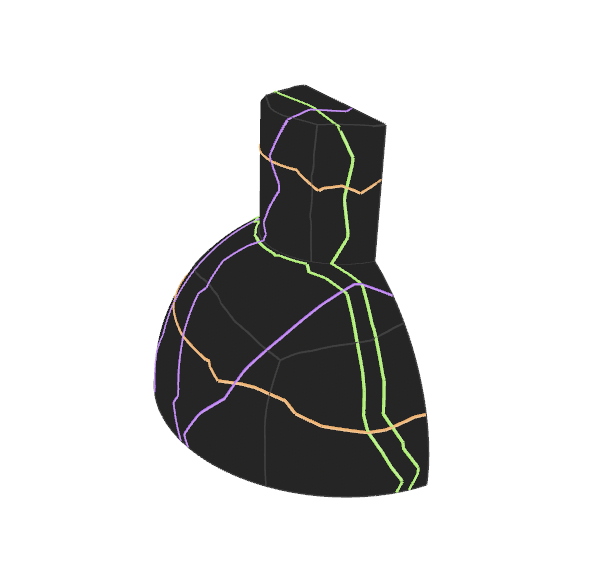}
    }
    \hfill
    \subcaptionbox{\texttt{B61}}{%
        \includegraphics[width=0.14\linewidth]{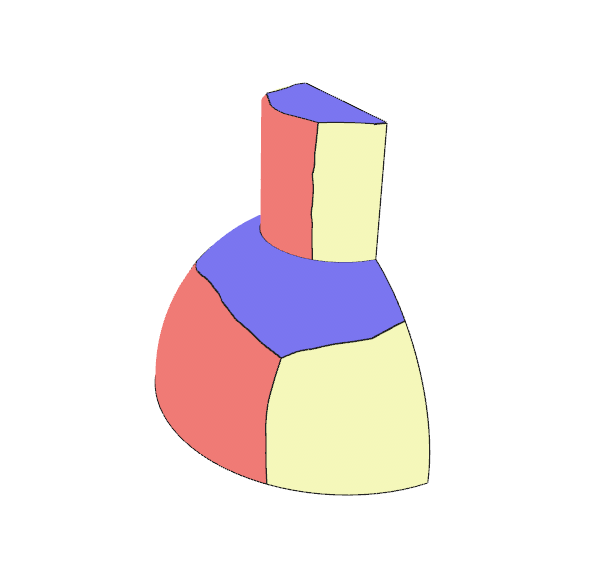}
    }
    \hfill
    \subcaptionbox*{}{%
        \includegraphics[width=0.14\linewidth]{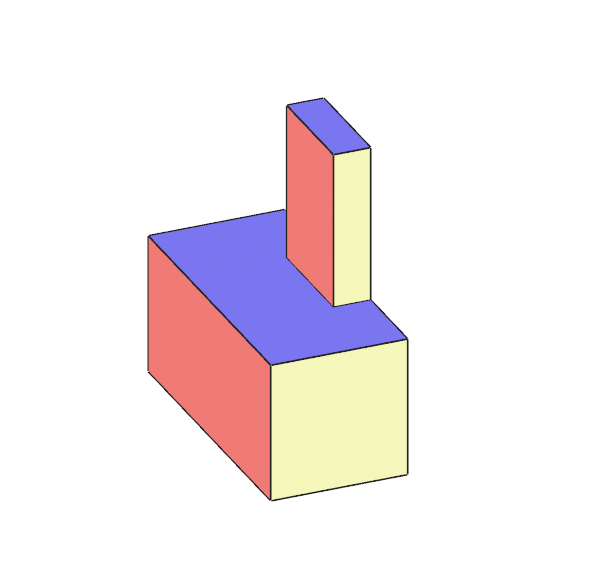}
    }
    \hspace{40pt}
    \subcaptionbox*{}{%
        \includegraphics[width=0.14\linewidth]{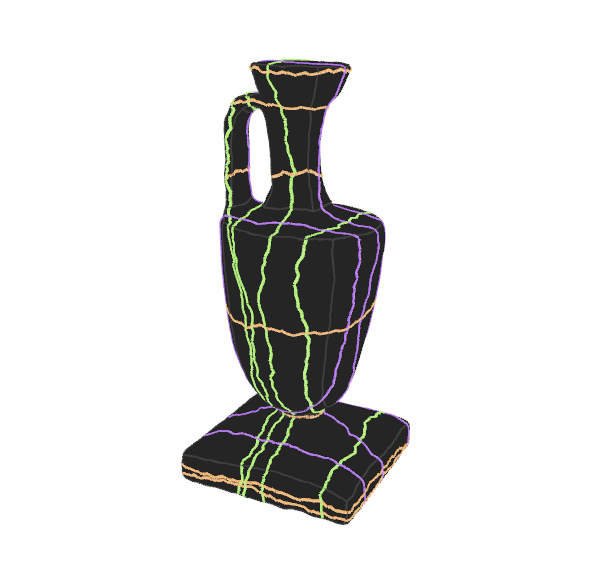}
    }
    \hfill
    \subcaptionbox{\texttt{bottle1}}{%
        \includegraphics[width=0.14\linewidth]{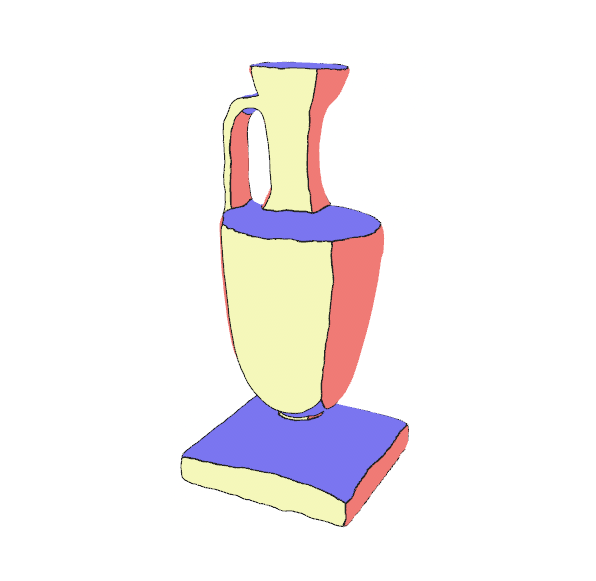}
    }
    \hfill
    \subcaptionbox*{}{%
        \includegraphics[width=0.14\linewidth]{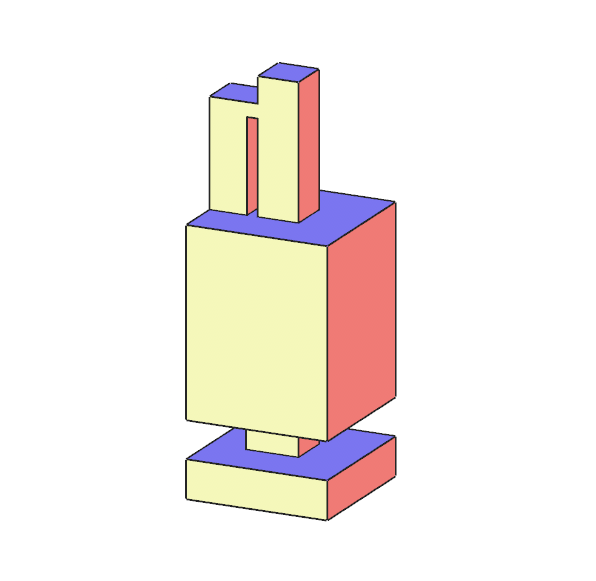}
    }
    \hfill\quad
  
    \caption{
    Showcase of results generated by DualCube, illustrating the transformation from loop structure to polycube segmentation and corresponding polycube.
    }
    \label{fig:showcase}
\end{figure*}

\begin{figure*}
\centering
    \subcaptionsetup{skip=3pt}

    \subcaptionbox*{}{%
        \hspace{0.13\linewidth}
    }
    \hfill
    \subcaptionbox*{
    \textbf{PolyCut}
    }{%
        \hspace{0.13\linewidth}
        \hspace{0.13\linewidth}
    }
    \hfill
    \subcaptionbox*{
    \textbf{EvoCube}
    }{%
        \hspace{0.13\linewidth}
        \hspace{0.13\linewidth}
    }
    \hfill
    \subcaptionbox*{
    \textbf{DualCube}
    }{%
        \hspace{0.13\linewidth}
        \hspace{0.13\linewidth}
    }

    \vspace{5pt}

    \subcaptionbox{\texttt{blub}}{%
        \includegraphics[width=0.13\linewidth]{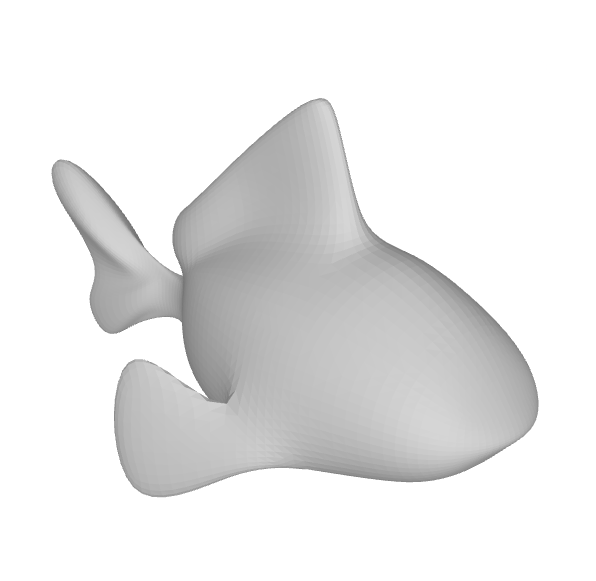}
    }
    \hfill
    \subcaptionbox*{
    $\text{SJ}_\text{min}=-0,99$\enskip$\text{SJ}_\text{avg}=0,85$\enskip$\overline{\text{HD}}=4,31$
    }{%
        \includegraphics[width=0.13\linewidth]{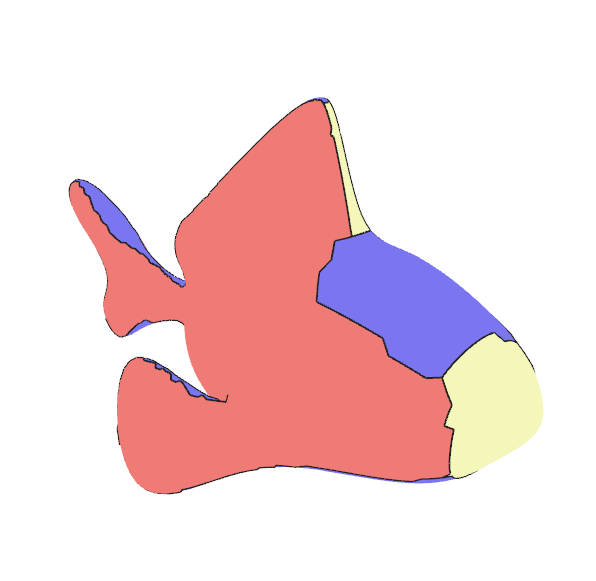}
        \includegraphics[width=0.13\linewidth]{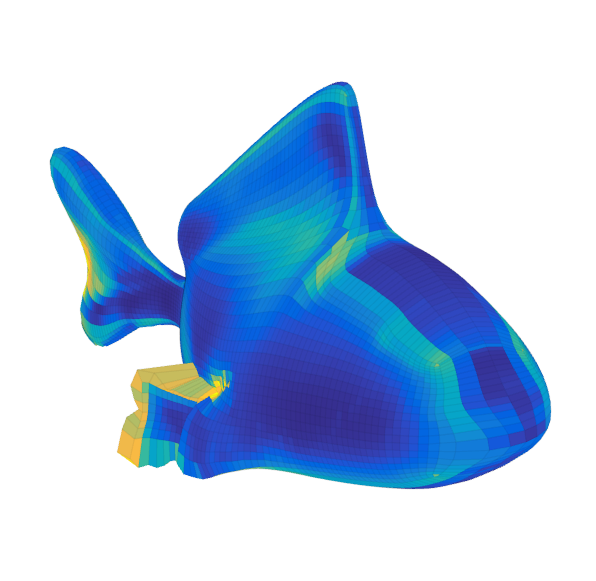}
    }
    \hfill
    \subcaptionbox*{
    $\text{SJ}_\text{min}=0,03$\enskip$\text{SJ}_\text{avg}=0,84$\enskip$\overline{\text{HD}}=9,03$
    }{%
        \includegraphics[width=0.13\linewidth]{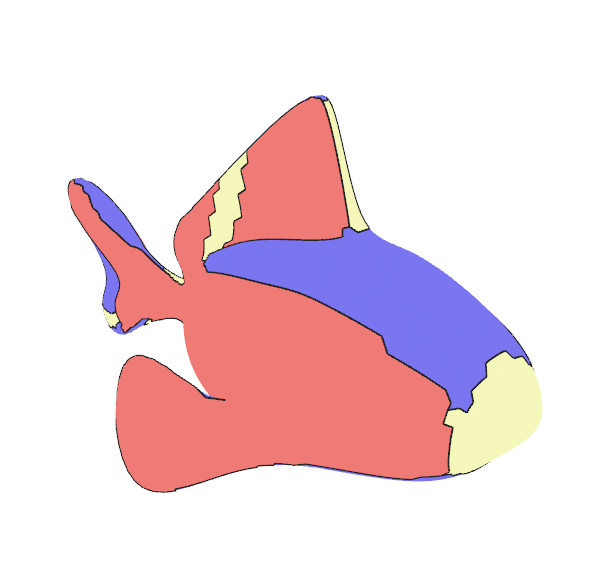}
        \includegraphics[width=0.13\linewidth]{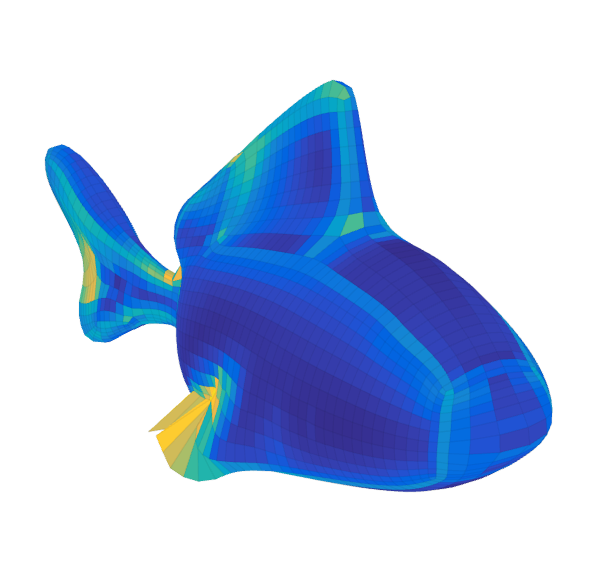}
    }
    \hfill
    \subcaptionbox*{
    $\text{SJ}_\text{min}=\mathbf{0,07}$\enskip$\text{SJ}_\text{avg}=\mathbf{0,87}$\enskip$\overline{\text{HD}}=\mathbf{2,21}$
    }{%
        \includegraphics[width=0.13\linewidth]{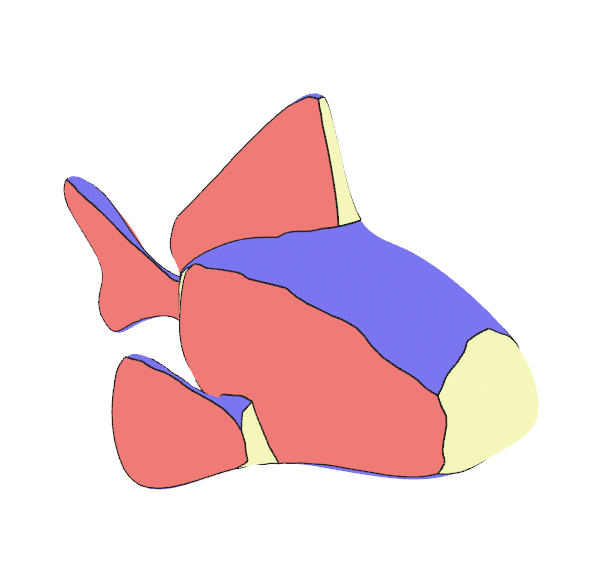}
        \includegraphics[width=0.13\linewidth]{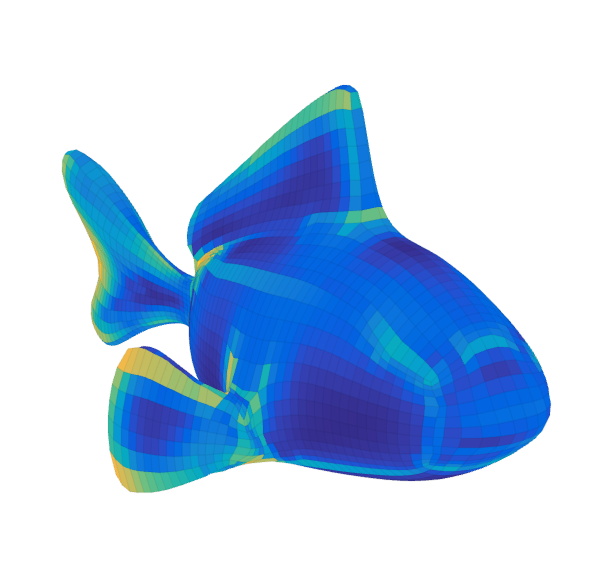}
    }

    \vspace{15pt}
    
    \subcaptionbox{\texttt{bunny}}{%
        \includegraphics[width=0.13\linewidth]{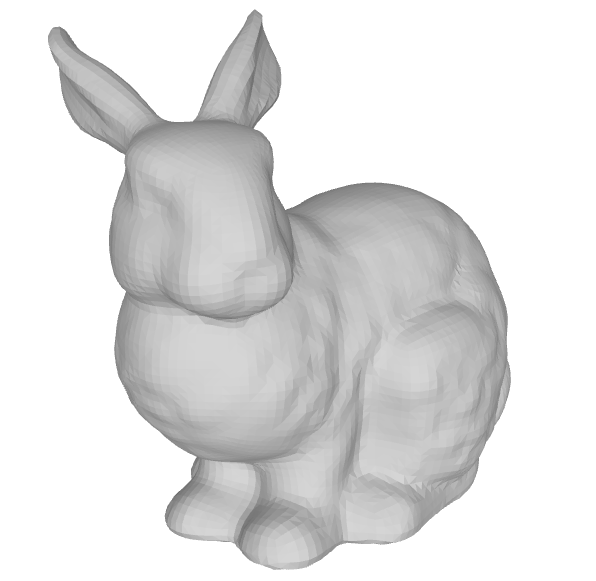}
    }
    \hfill
    \subcaptionbox*{
    $\text{SJ}_\text{min}=\mathbf{0,13}$\enskip$\text{SJ}_\text{avg}=0,92$\enskip$\overline{\text{HD}}=1,33$
    }{%
        \includegraphics[width=0.13\linewidth]{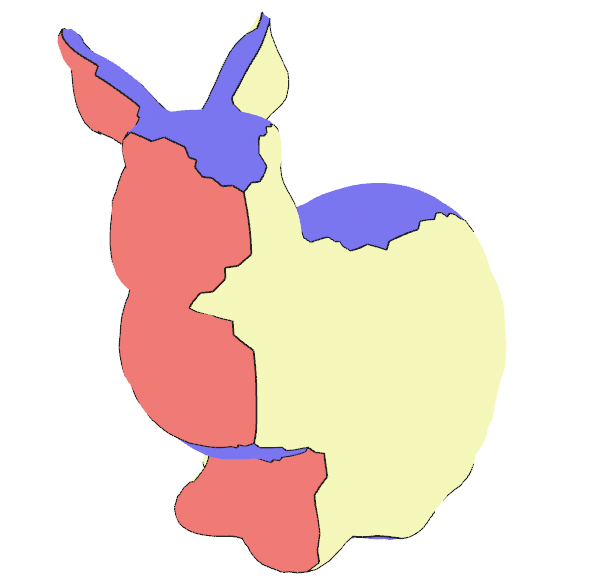}
        \includegraphics[width=0.13\linewidth]{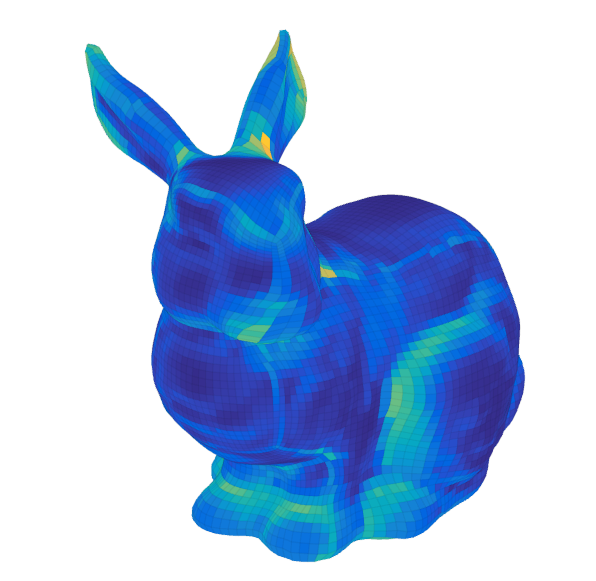}
    }
    \hfill
    \subcaptionbox*{
    $\text{SJ}_\text{min}=0,07$\enskip$\text{SJ}_\text{avg}=\mathbf{0,93}$\enskip$\overline{\text{HD}}=1,34$
    }{%
        \includegraphics[width=0.13\linewidth]{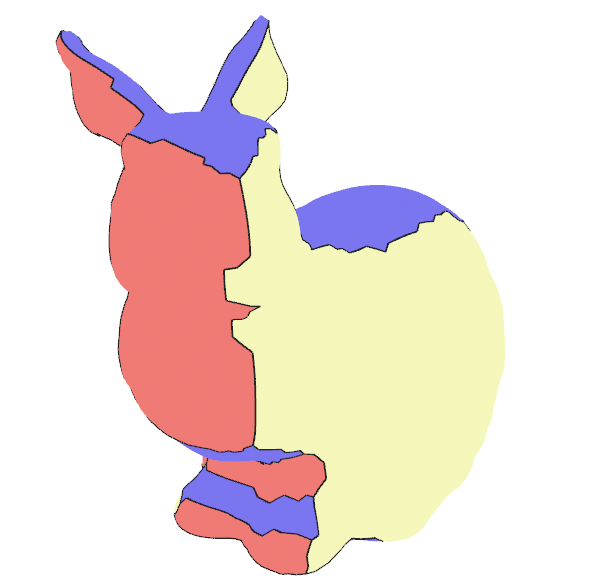}
        \includegraphics[width=0.13\linewidth]{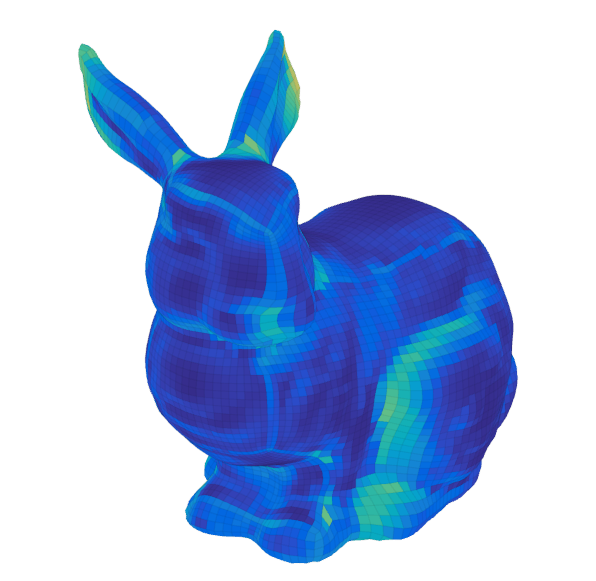}
    }
    \hfill
    \subcaptionbox*{
    $\text{SJ}_\text{min}=0,08$\enskip$\text{SJ}_\text{avg}=0,92$\enskip$\overline{\text{HD}}=\mathbf{1,16}$
    }{%
        \includegraphics[width=0.13\linewidth]{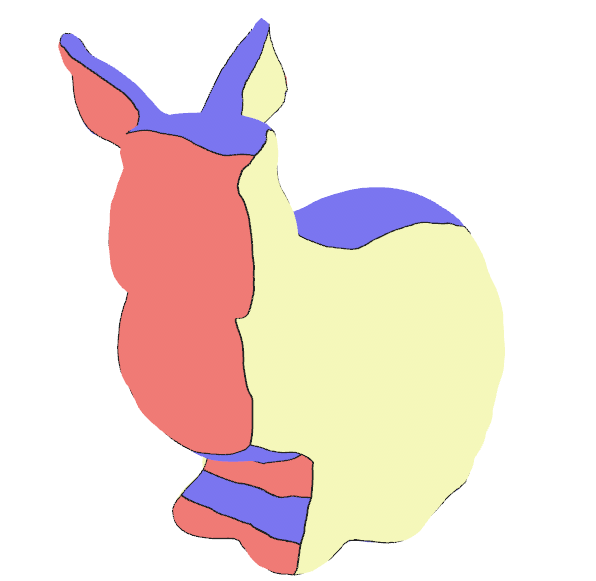}
        \includegraphics[width=0.13\linewidth]{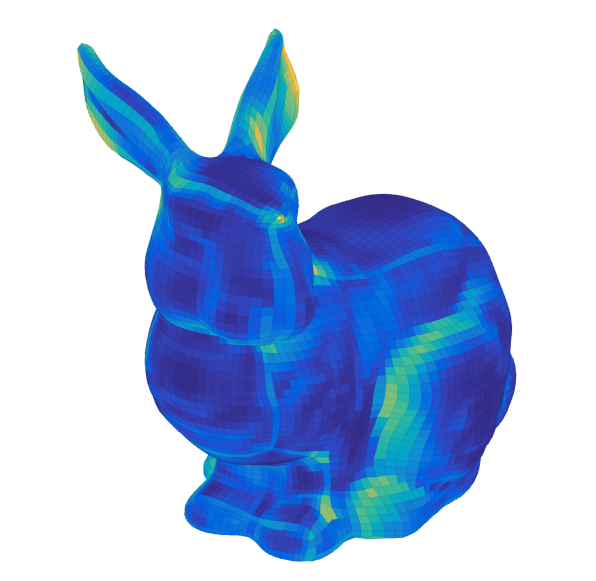}
    }
    
    \vspace{15pt}
    
    \subcaptionbox{\texttt{igea}}{%
        \includegraphics[width=0.13\linewidth]{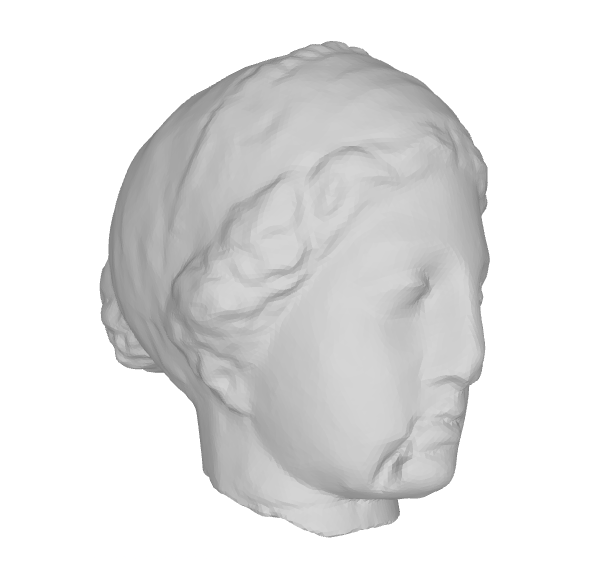}
    }
    \hfill
    \subcaptionbox*{
    $\text{SJ}_\text{min}=\mathbf{0,08}$\enskip$\text{SJ}_\text{avg}=\mathbf{0,94}$\enskip$\overline{\text{HD}}=0,80$
    }{%
        \includegraphics[width=0.13\linewidth]{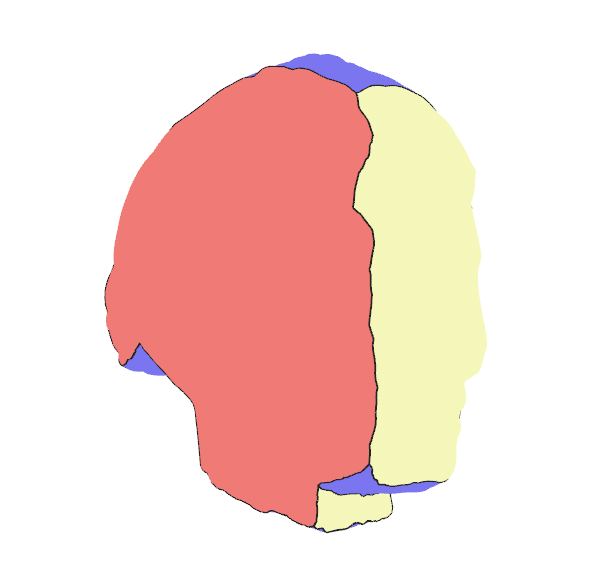}
        \includegraphics[width=0.13\linewidth]{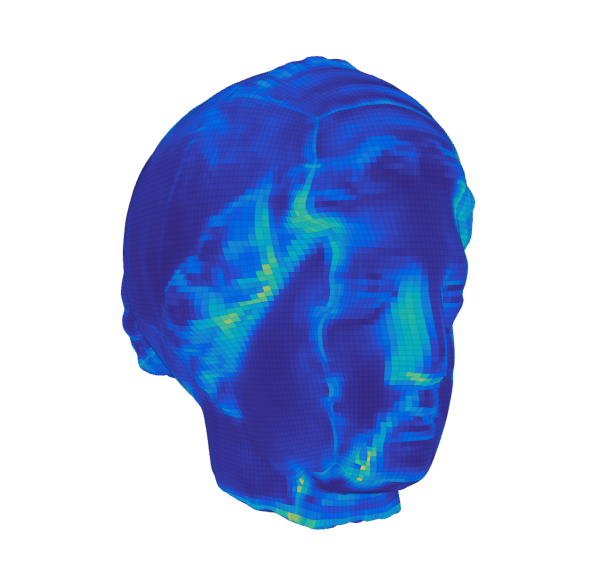}
    }
    \hfill
    \subcaptionbox*{
    $\text{SJ}_\text{min}=-0,76$\enskip$\text{SJ}_\text{avg}=0,93$\enskip$\overline{\text{HD}}=2,35$
    }{%
        \includegraphics[width=0.13\linewidth]{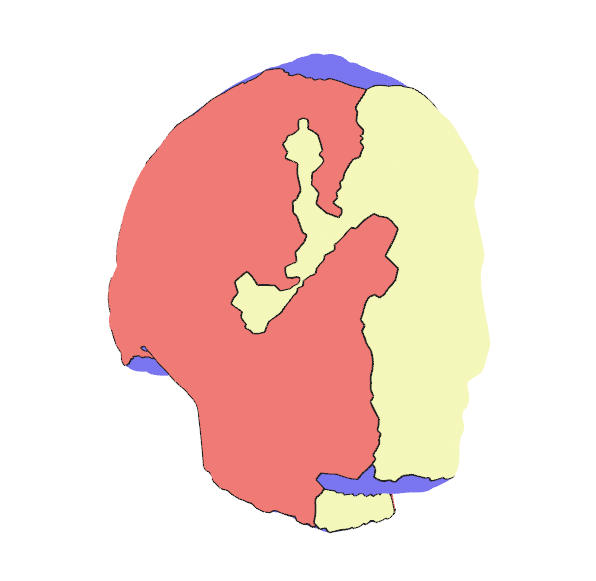}
        \includegraphics[width=0.13\linewidth]{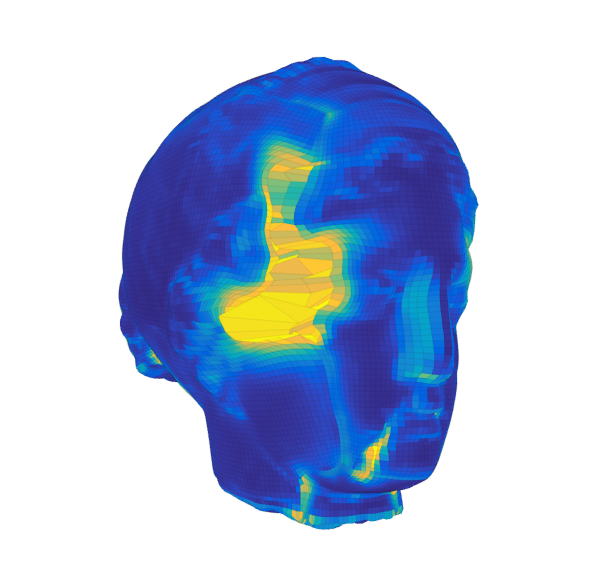}
    }
    \hfill
    \subcaptionbox*{
    $\text{SJ}_\text{min}=0,05$\enskip$\text{SJ}_\text{avg}=\mathbf{0,94}$\enskip$\overline{\text{HD}}=\mathbf{0,72}$
    }{%
        \includegraphics[width=0.13\linewidth]{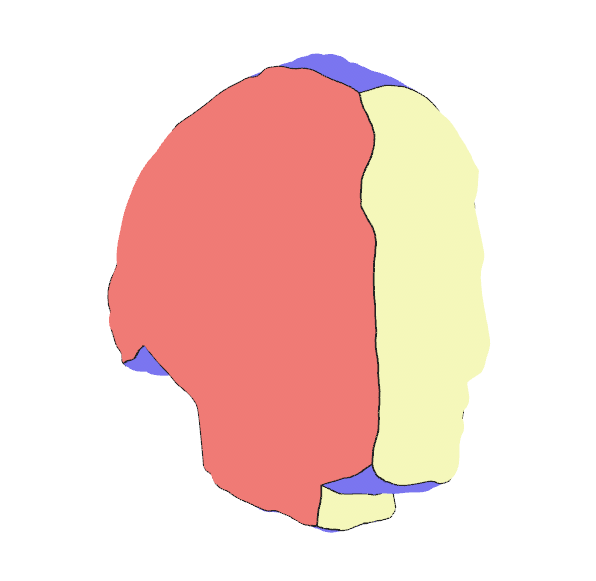}
        \includegraphics[width=0.13\linewidth]{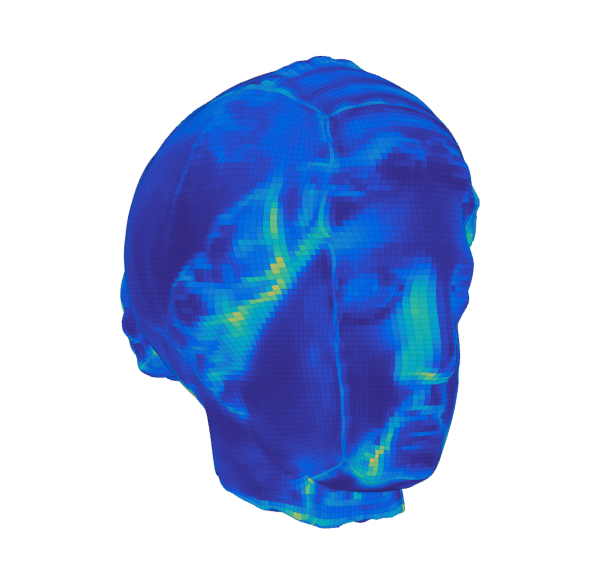}
    }
    
    \vspace{15pt}
    
    \subcaptionbox{\texttt{armadillo}}{%
        \includegraphics[width=0.13\linewidth]{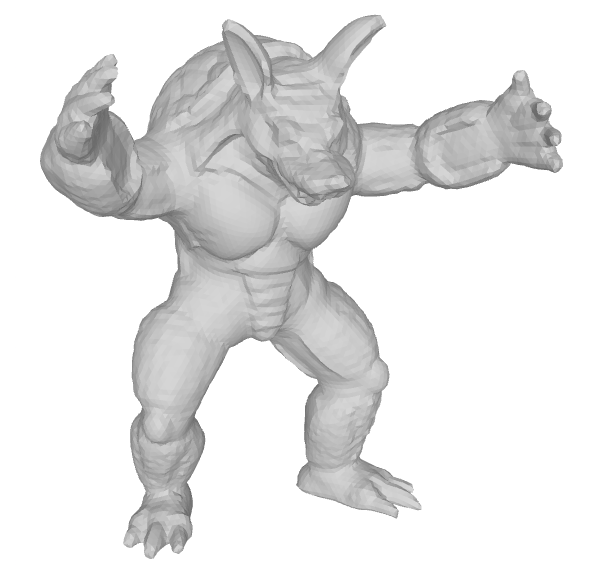}
    }
    \hfill
    \subcaptionbox*{
    no solution
    }{%
        \includegraphics[width=0.13\linewidth]{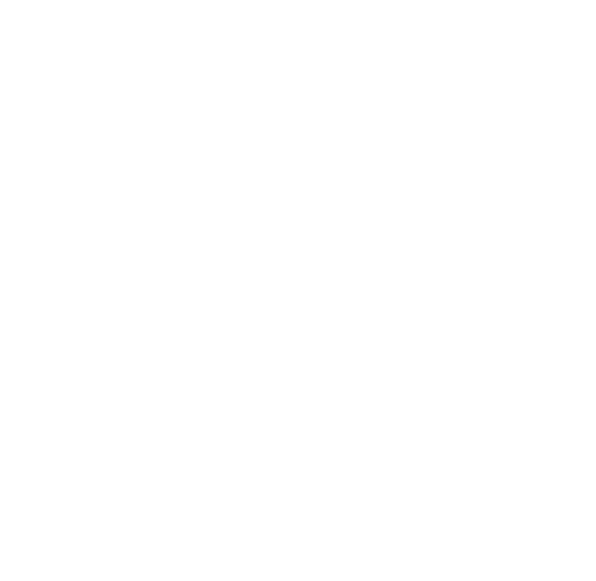}
        \includegraphics[width=0.13\linewidth]{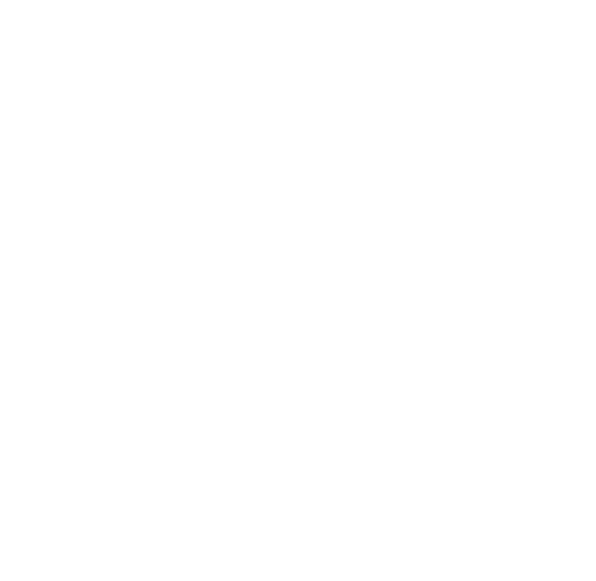}
    }
    \hfill
    \subcaptionbox*{
    $\text{SJ}_\text{min}=-0,88$\enskip$\text{SJ}_\text{avg}=\mathbf{0,87}$\enskip$\overline{\text{HD}}=\mathbf{2,25}$
    }{%
        \includegraphics[width=0.13\linewidth]{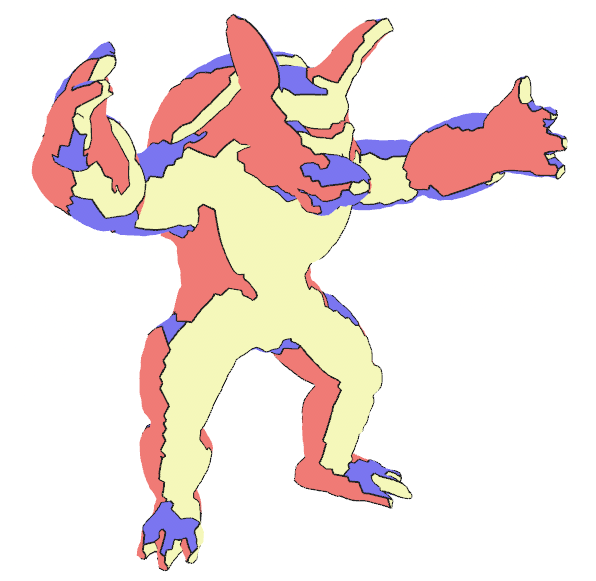}
        \includegraphics[width=0.13\linewidth]{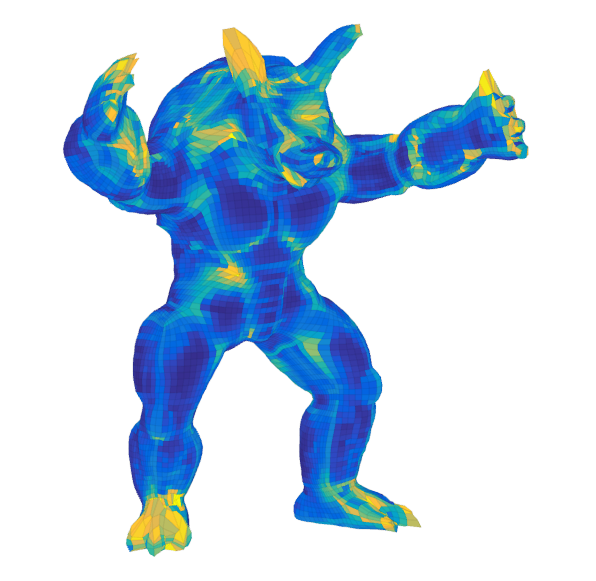}
    }
    \hfill
    \subcaptionbox*{
    $\text{SJ}_\text{min}=\mathbf{0,02}$\enskip$\text{SJ}_\text{avg}=0,84$\enskip$\overline{\text{HD}}=3,15$
    }{%
        \includegraphics[width=0.13\linewidth]{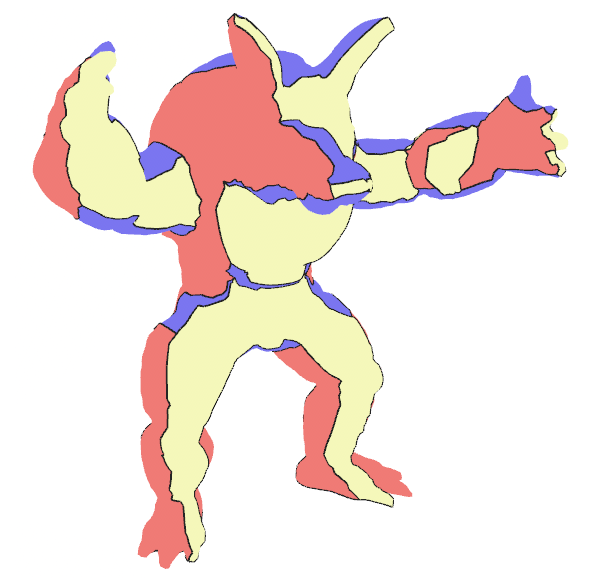}
        \includegraphics[width=0.13\linewidth]{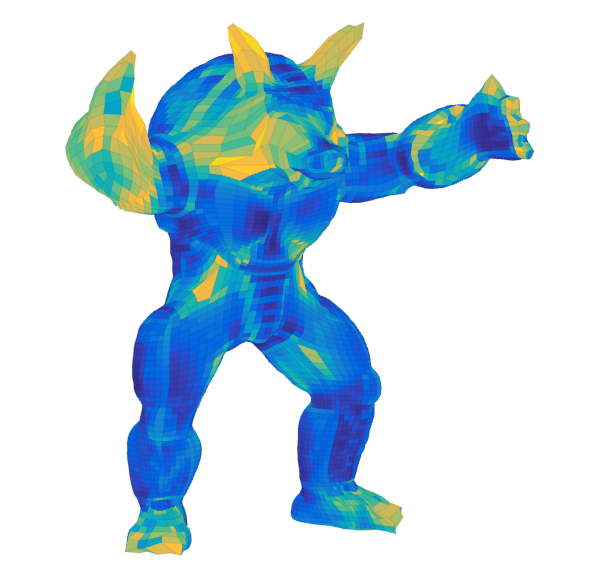}
    }

    \vspace{15pt}
    
    \subcaptionbox{\texttt{rocker}}{%
        \includegraphics[width=0.13\linewidth]{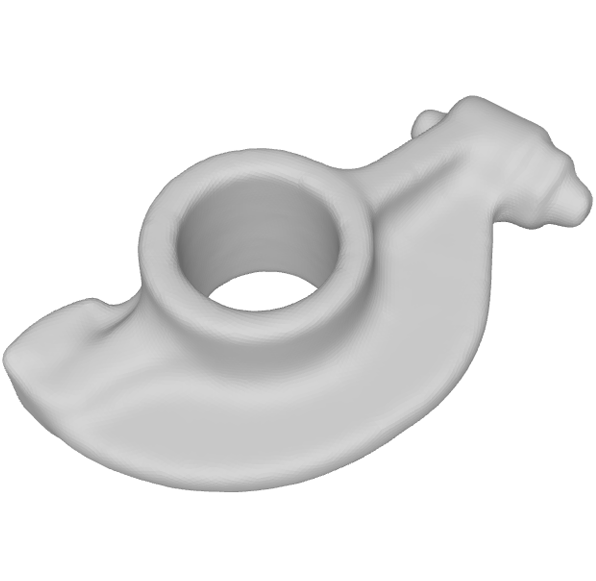}
    }
    \hfill
    \subcaptionbox*{
    $\text{SJ}_\text{min}=0,05$\enskip$\text{SJ}_\text{avg}=0,91$\enskip$\overline{\text{HD}}=2,01$
    }{%
        \includegraphics[width=0.13\linewidth]{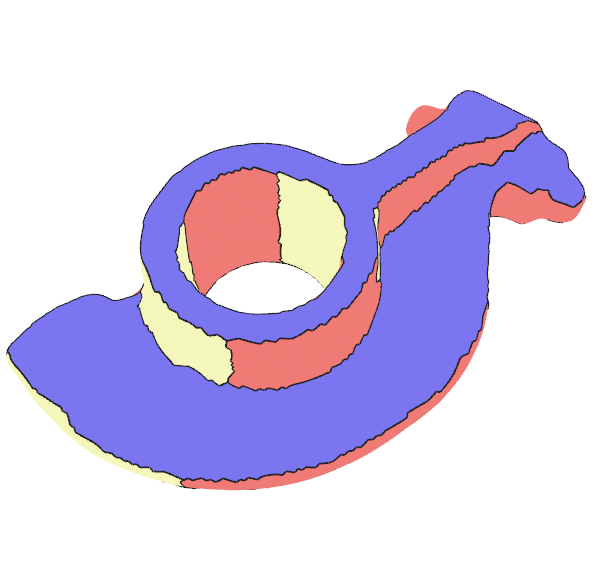}
        \includegraphics[width=0.13\linewidth]{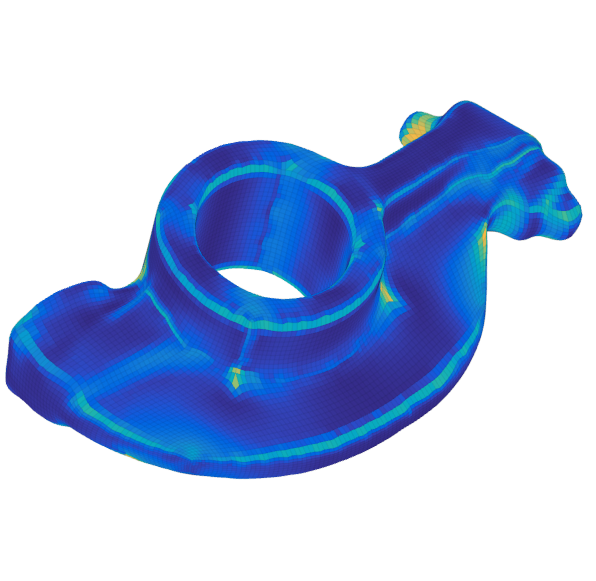}
    }
    \hfill
    \subcaptionbox*{
    $\text{SJ}_\text{min}=0,01$\enskip$\text{SJ}_\text{avg}=\mathbf{0,92}$\enskip$\overline{\text{HD}}=\mathbf{0,97}$
    }{%
        \includegraphics[width=0.13\linewidth]{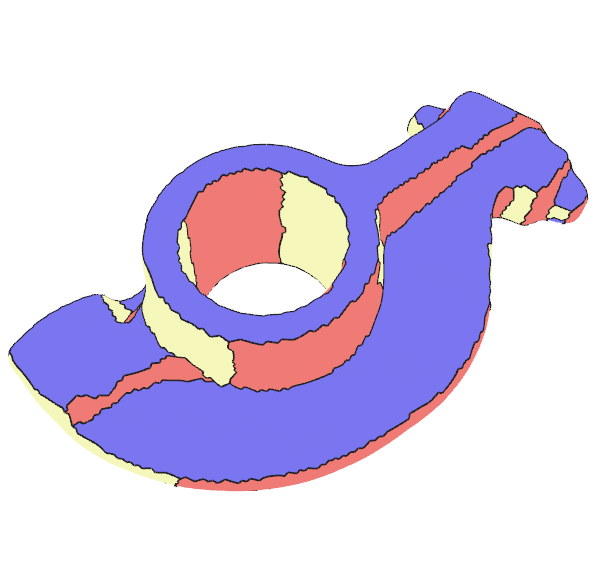}
        \includegraphics[width=0.13\linewidth]{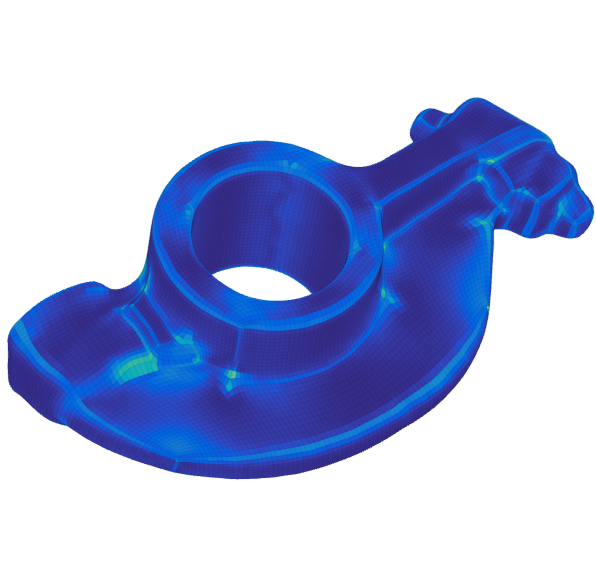}
    }
    \hfill
    \subcaptionbox*{
    $\text{SJ}_\text{min}=\mathbf{0,06}$\enskip$\text{SJ}_\text{avg}=0,91$\enskip$\overline{\text{HD}}=1,25$
    }{%
        \includegraphics[width=0.13\linewidth]{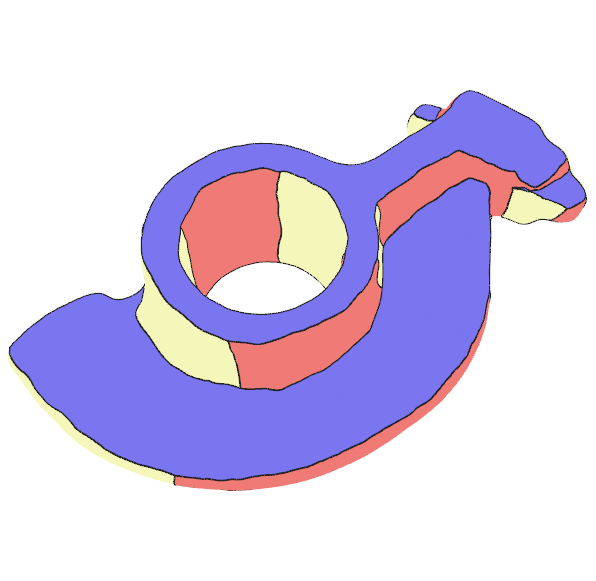}
        \includegraphics[width=0.13\linewidth]{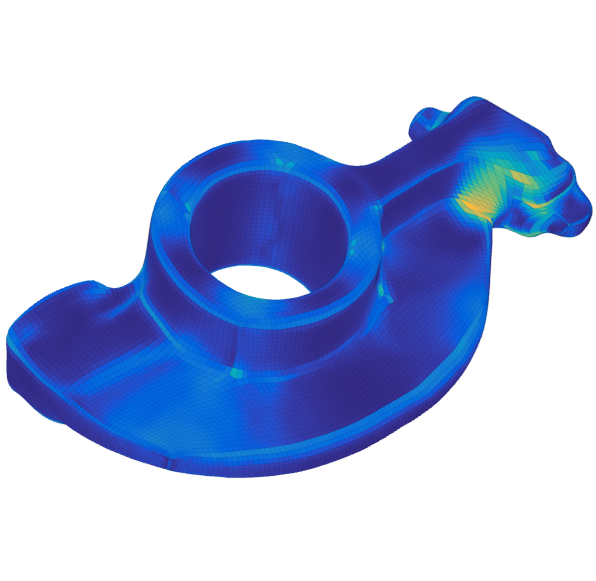}
    }

    \vspace{15pt}
    
    \subcaptionbox{\texttt{spot}}{%
        \includegraphics[width=0.13\linewidth]{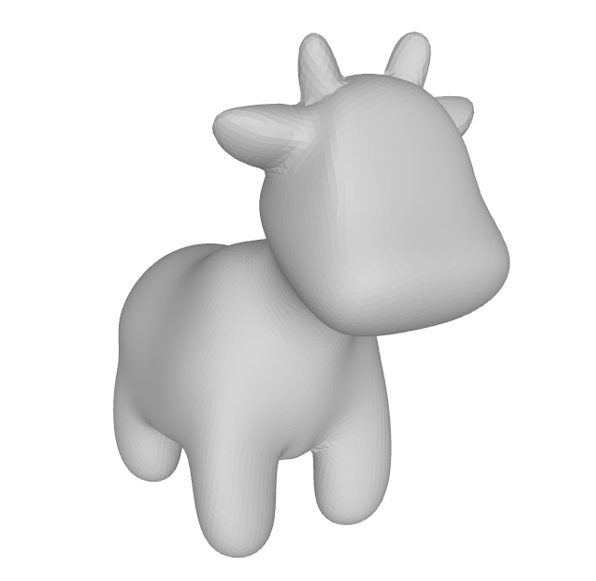}
    }
    \hfill
    \subcaptionbox*{
    $\text{SJ}_\text{min}=0,01$\enskip$\text{SJ}_\text{avg}=0,94$\enskip$\overline{\text{HD}}=1,38$
    }{%
        \includegraphics[width=0.13\linewidth]{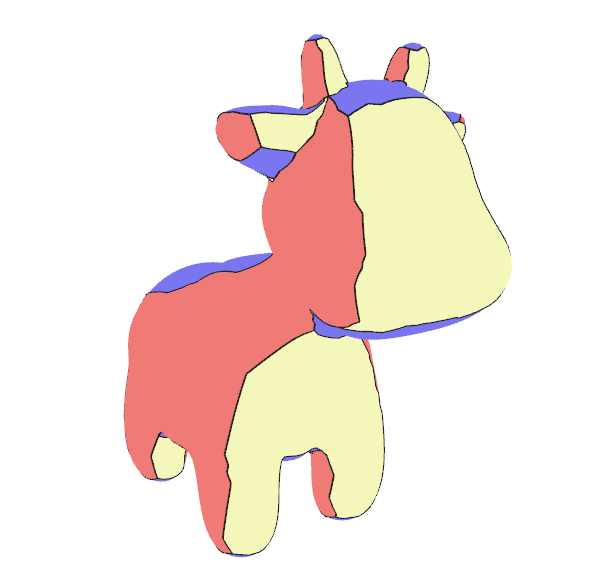}
        \includegraphics[width=0.13\linewidth]{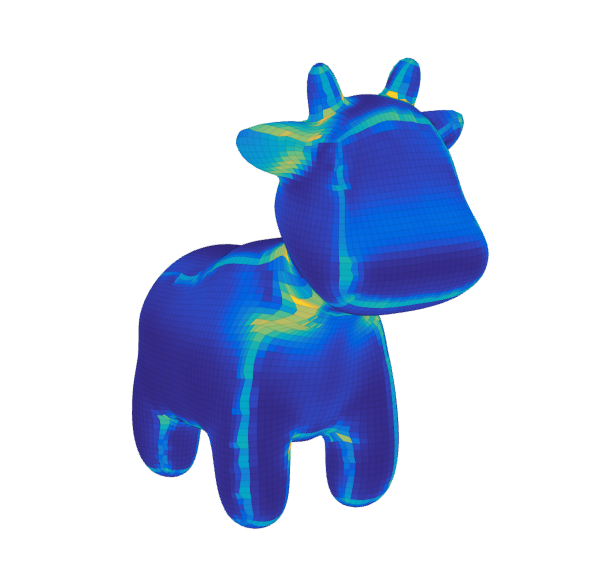}
    }
    \hfill
    \subcaptionbox*{
    $\text{SJ}_\text{min}=0,03$\enskip$\text{SJ}_\text{avg}=\mathbf{0,95}$\enskip$\overline{\text{HD}}=1,60$
    }{%
        \includegraphics[width=0.13\linewidth]{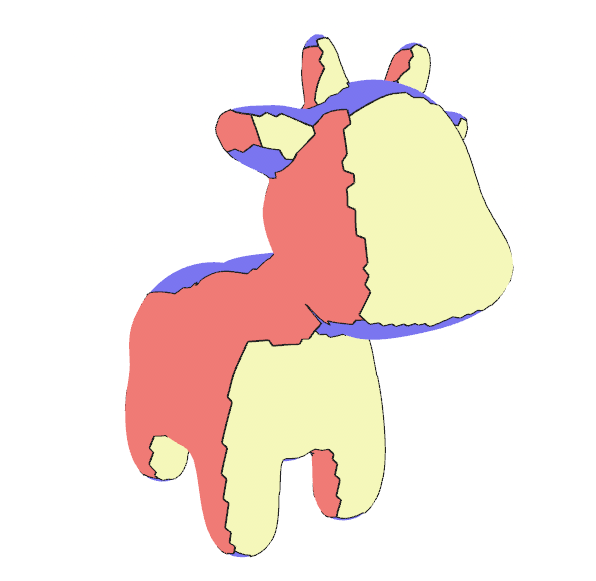}
        \includegraphics[width=0.13\linewidth]{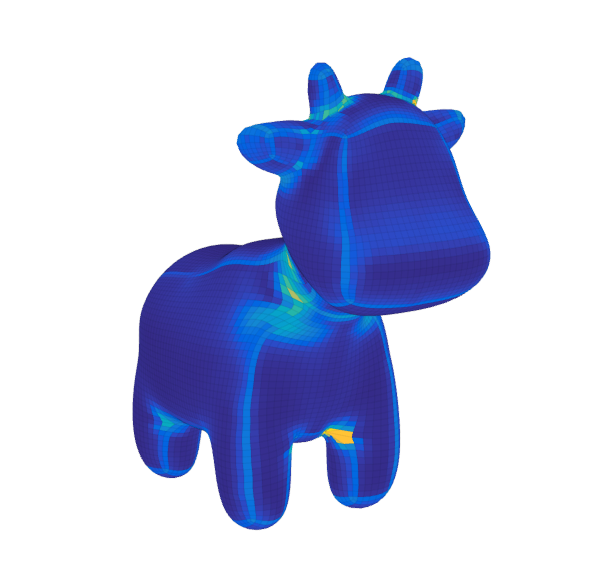}
    }
    \hfill
    \subcaptionbox*{
    $\text{SJ}_\text{min}=\mathbf{0,05}$\enskip$\text{SJ}_\text{avg}=0,94$\enskip$\overline{\text{HD}}=\mathbf{1,32}$
    }{%
        \includegraphics[width=0.13\linewidth]{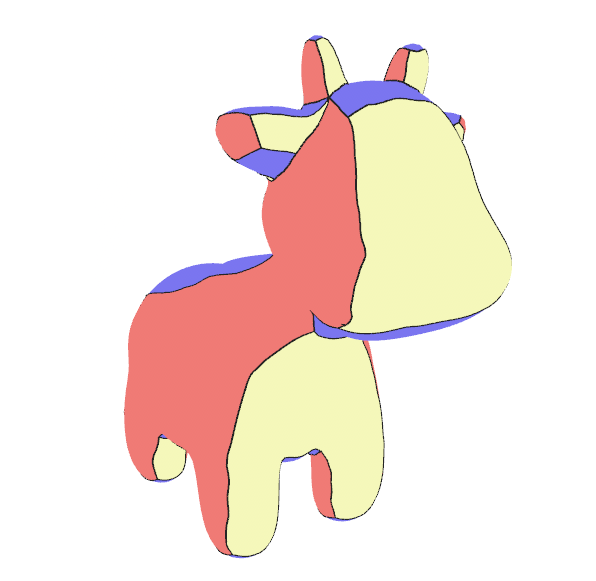}
        \includegraphics[width=0.13\linewidth]{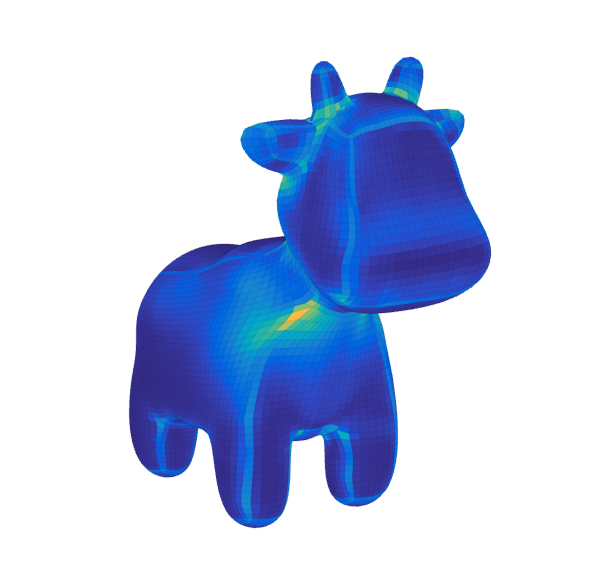}
    }

    \caption{
    Comparison of the three methods PolyCut~\cite{livesu2013polycut}, EvoCube~\cite{dumery2022evocube}, and our method DualCube. We show the input mesh, and polycube segmentation and resulting hexahedral meshes per method. We include values for $\text{SJ}_\text{min}$, $\text{SJ}_\text{avg}$, and $\overline{\text{HD}}$ per hexahedral mesh.
    }
    \label{fig:comparison}
\end{figure*}

Qualitative results can be seen in Figure~\ref{fig:showcase} and~\ref{fig:comparison}. The hexahedral meshes are visualized using HexaLab\cite{bracci2019hexalab}, with the hex quality color setting showing the Scaled Jacobian of the cells. For models that contain features well-aligned with the principal axes, all three methods result in similar polycube segmentations, and as a result, similar hexahedral meshes. On the contrary, our algorithm sometimes fails to capture features well that are not aligned with any of the principal directions, such as the horns of \texttt{goathead}. Small features, such as the hands and toes of \texttt{armdillo} and \texttt{homer}, require specific loops to be captured well. The \texttt{screw\_head} model shows a case where the solutions produced by PolyCut and EvoCube lead to poor hexahedral meshes, most likely due to invalid topology. Our method produces a polycube segmentation with a valid corresponding polycube. While the resulting hexahedral mesh is slightly improved compared to the others, it is not optimal.

\section{Conclusion}
\label{sec:conclusion}

We presented DualCube: an iterative algorithm for constructing polycube segmentations of 3D models. The algorithm is guaranteed to produce valid solutions, even for inputs with higher genus. Our method leverages a recent dual characterization of polycubes~\cite{snoep2025polycubes} to initialize a simple, valid polycube of the appropriate genus and iteratively refine it alongside its dual loop structure. Throughout the refinement process, the validity of the solution is maintained. Our experiments demonstrate that the final polycube segmentations match or exceed the quality of state-of-the-art approaches.

A key limitation of our method is the need for a \emph{manual} initialization on high-genus inputs. While we believe that a fully automated initialization is possible, but we do not yet have a solution.

Our current loop computations are based on alignment with the principal axes, which generally yields high-quality results. However, this approach has two notable limitations. First, the loop computation (and resulting polycube segmentation) is sensitive to the input mesh's orientation. Although this sensitivity could be mitigated by re-orienting the mesh using methods such as Principal Component Analysis (PCA), some downstream applications require a fixed global orientation, making such re-orientation impractical. Second, some input meshes may benefit from loops that deviate slightly from strict axis alignment. For example, features that are not well aligned with any principal axis, even after re-orientation, may be more effectively captured using loops that adapt to local rather than global directions. A natural extension of our method is to incorporate more expressive direction fields into the loop computation process.

Lastly, our work focuses solely on generating the polycube segmentation. The downstream applications may require the accompanying polycube with additional constraints, such as non-intersecting geometry, integer vertex coordinates, or fixed edge lengths. We do not have a solution for these constraints yet, though we think the dual structure gives us the additional information needed to construct polycubes with these added constraints.


\bibliographystyle{eg-alpha-doi} 
\bibliography{bib}   

\newcommand{\etalchar}[1]{$^{#1}$}
\begin{thebibliography}{\uppercase{HWFQ09}}

\bibitem[BBG{\etalchar{*}}09]{boelens2009f16}
\textsc{Boelens O., Badcock K., Görtz S., Morton S., Fritz W., Karman S., Michal T., Lamar J.}:
\newblock F-16xl geometry and computational grids used in cranked-arrow wing aerodynamics project international.
\newblock \emph{Journal of Aircraft 46} (2009).
\newblock \href {https://doi.org/10.2514/1.34852} {\path{doi:10.2514/1.34852}}.

\bibitem[BG04]{biedl2004when}
\textsc{Biedl T., Genc B.}:
\newblock When can a graph form an orthogonal polyhedron?
\newblock In \emph{Proceedings of the 16th Canadian Conference on Computational Geometry (CCCG)} (2004).
\newblock URL: \url{www.cccg.ca/proceedings/2004/15.pdf}.

\bibitem[BG09]{biedl2009cauchy}
\textsc{Biedl T., Genc B.}:
\newblock Cauchy's theorem for orthogonal polyhedra of genus 0.
\newblock In \emph{Proceedings of the 17th European Symposium on Algorithms (ESA)} (2009).
\newblock \href {https://doi.org/10.1007/978-3-642-04128-0_7} {\path{doi:10.1007/978-3-642-04128-0_7}}.

\bibitem[BK23]{baumeister2023how}
\textsc{Baumeister M., Kobbelt L.}:
\newblock How close is a quad mesh to a polycube?
\newblock \emph{Computational Geometry 111} (2023).
\newblock \href {https://doi.org/10.1016/j.comgeo.2022.101978} {\path{doi:10.1016/j.comgeo.2022.101978}}.

\bibitem[BTP{\etalchar{*}}19]{bracci2019hexalab}
\textsc{Bracci M., Tarini M., Pietroni N., Livesu M., Cignoni P.}:
\newblock Hexalab.net: An online viewer for hexahedral meshes.
\newblock \emph{Computer-Aided Design 110} (2019).
\newblock \href {https://doi.org/10.1016/j.cad.2018.12.003} {\path{doi:10.1016/j.cad.2018.12.003}}.

\bibitem[CBK12]{campen2012dual}
\textsc{Campen M., Bommes D., Kobbelt L.}:
\newblock Dual loops meshing: quality quad layouts on manifolds.
\newblock \emph{ACM Transactions on Graphics 31}, 4 (2012).
\newblock \href {https://doi.org/10.1145/2185520.2185606} {\path{doi:10.1145/2185520.2185606}}.

\bibitem[CLS16]{cherchi2016polycube}
\textsc{Cherchi G., Livesu M., Scateni R.}:
\newblock Polycube simplification for coarse layouts of surfaces and volumes.
\newblock \emph{Computer Graphics Forum 35}, 5 (2016).
\newblock \href {https://doi.org/10.1111/cgf.12959} {\path{doi:10.1111/cgf.12959}}.

\bibitem[DPM{\etalchar{*}}22]{dumery2022evocube}
\textsc{Dumery C., Protais F., Mestrallet S., Bourcier C., Ledoux F.}:
\newblock Evocube: A genetic labelling framework for polycube-maps.
\newblock \emph{Computer Graphics Forum 41}, 6 (2022).
\newblock \href {https://doi.org/10.1111/cgf.14649} {\path{doi:10.1111/cgf.14649}}.

\bibitem[EM10]{eppstein2010steinitz}
\textsc{Eppstein D., Mumford E.}:
\newblock Steinitz theorems for orthogonal polyhedra.
\newblock In \emph{Proceedings of the 26th Annual Symposium on Computational Geometry (SoCG)} (2010).
\newblock \href {https://doi.org/10.1145/1810959.1811030} {\path{doi:10.1145/1810959.1811030}}.

\bibitem[FBL16]{fu2016efficient}
\textsc{Fu X.-M., Bai C.-Y., Liu Y.}:
\newblock Efficient volumetric polycube-map construction.
\newblock \emph{Computer Graphics Forum 35}, 7 (2016).
\newblock \href {https://doi.org/10.1111/cgf.13007} {\path{doi:10.1111/cgf.13007}}.

\bibitem[FXBH16]{fang2016all}
\textsc{Fang X., Xu W., Bao H., Huang J.}:
\newblock All-hex meshing using closed-form induced polycube.
\newblock \emph{ACM Transactions on Graphics 35}, 4 (2016).
\newblock \href {https://doi.org/10.1145/2897824.2925957} {\path{doi:10.1145/2897824.2925957}}.

\bibitem[GLYL20]{guo2020cut}
\textsc{Guo H.-X., Liu X., Yan D.-M., Liu Y.}:
\newblock Cut-enhanced polycube-maps for feature-aware all-hex meshing.
\newblock \emph{ACM Transactions on Graphics 39}, 4 (2020).
\newblock \href {https://doi.org/10.1145/3386569.3392378} {\path{doi:10.1145/3386569.3392378}}.

\bibitem[GSZ11]{gregson2011all}
\textsc{Gregson J., Sheffer A., Zhang E.}:
\newblock All-hex mesh generation via volumetric polycube deformation.
\newblock \emph{Computer Graphics Forum 30}, 5 (2011).
\newblock \href {https://doi.org/10.1111/j.1467-8659.2011.02015.x} {\path{doi:10.1111/j.1467-8659.2011.02015.x}}.

\bibitem[HJS{\etalchar{*}}14]{huang2014l1}
\textsc{Huang J., Jiang T., Shi Z., Tong Y., Bao H., Desbrun M.}:
\newblock l1-based construction of polycube maps from complex shapes.
\newblock \emph{ACM Transactions on Graphics 33}, 3 (2014).
\newblock \href {https://doi.org/10.1145/2602141} {\path{doi:10.1145/2602141}}.

\bibitem[HLW{\etalchar{*}}24]{he2024expanding}
\textsc{He L., Lei N., Wang Z., Wang C., Zheng X., Luo Z.}:
\newblock Expanding the solvable space of polycube-map via validity-enhanced construction.
\newblock In \emph{Proceedings of the 2024 SIAM International Meshing Roundtable (IMR)} (2024).
\newblock \href {https://doi.org/10.1137/1.9781611978001.4} {\path{doi:10.1137/1.9781611978001.4}}.

\bibitem[HWFQ09]{he2009divide}
\textsc{He Y., Wang H., Fu C.-W., Qin H.}:
\newblock A divide-and-conquer approach for automatic polycube map construction.
\newblock \emph{Computers \& Graphics 33}, 3 (2009).
\newblock \href {https://doi.org/10.1016/j.cag.2009.03.024} {\path{doi:10.1016/j.cag.2009.03.024}}.

\bibitem[HZ16]{hu2016centroidal}
\textsc{Hu K., Zhang Y.~J.}:
\newblock Centroidal voronoi tessellation based polycube construction for adaptive all-hexahedral mesh generation.
\newblock \emph{Computer Methods in Applied Mechanics and Engineering 305} (2016).
\newblock \href {https://doi.org/10.1016/j.cma.2016.03.021} {\path{doi:10.1016/j.cma.2016.03.021}}.

\bibitem[HZL17]{hu2017surface}
\textsc{Hu K., Zhang Y.~J., Liao T.}:
\newblock Surface segmentation for polycube construction based on generalized centroidal voronoi tessellation.
\newblock \emph{Computer Methods in Applied Mechanics and Engineering 316} (2017).
\newblock \href {https://doi.org/10.1016/j.cma.2016.07.005} {\path{doi:10.1016/j.cma.2016.07.005}}.

\bibitem[LJFW08]{lin2008automatic}
\textsc{Lin J., Jin X., Fan Z., Wang C. C.~L.}:
\newblock Automatic polycube-maps.
\newblock In \emph{Proceedings of the 5th International Conference on Geometric Modeling and Processing (GMP)} (2008).
\newblock \href {https://doi.org/10.1007/978-3-540-79246-8_1} {\path{doi:10.1007/978-3-540-79246-8_1}}.

\bibitem[LVS{\etalchar{*}}13]{livesu2013polycut}
\textsc{Livesu M., Vining N., Sheffer A., Gregson J., Scateni R.}:
\newblock Polycut: monotone graph-cuts for polycube base-complex construction.
\newblock \emph{ACM Transactions on Graphics 32}, 6 (2013).
\newblock \href {https://doi.org/10.1145/2508363.2508388} {\path{doi:10.1145/2508363.2508388}}.

\bibitem[MCBC22]{mandad2022intrinsic}
\textsc{Mandad M., Chen R., Bommes D., Campen M.}:
\newblock Intrinsic mixed-integer polycubes for hexahedral meshing.
\newblock \emph{Computer Aided Geometric Design 94} (2022).
\newblock \href {https://doi.org/10.1016/j.cagd.2022.102078} {\path{doi:10.1016/j.cagd.2022.102078}}.

\bibitem[MPBL23]{mestrallet2023limits}
\textsc{Mestrallet S., Protais F., Bourcier C., Ledoux F.}:
\newblock Limits and prospects of polycube labelings.
\newblock In \emph{Research notes of the 2023 SIAM International Meshing Roundtable (IMR)} (2023).
\newblock URL: \url{cea.hal.science/cea-04169841/}.

\bibitem[PCS{\etalchar{*}}22]{pietroni2022hex}
\textsc{Pietroni N., Campen M., Sheffer A., Cherchi G., Bommes D., Gao X., Scateni R., Ledoux F., Remacle J., Livesu M.}:
\newblock Hex-mesh generation and processing: A survey.
\newblock \emph{ACM Transactions on Graphics 42}, 2 (2022).
\newblock \href {https://doi.org/10.1145/3554920} {\path{doi:10.1145/3554920}}.

\bibitem[PRR{\etalchar{*}}22]{protais2022robust}
\textsc{Protais F., Reberol M., Ray N., Corman E., Ledoux F., Sokolov D.}:
\newblock Robust quantization for polycube maps.
\newblock \emph{Computer-Aided Design 150} (2022).
\newblock \href {https://doi.org/10.1016/j.cad.2022.103321} {\path{doi:10.1016/j.cad.2022.103321}}.

\bibitem[Sok16]{sokolov2016modelisation}
\textsc{Sokolov D.}:
\newblock \emph{Mod{\'e}lisation g{\'e}om{\'e}trique}.
\newblock Habilitation {\`a} diriger des recherches, Universit{\'e} de Lorraine, 2016.
\newblock URL: \url{inria.hal.science/tel-03180395}.

\bibitem[SR15]{sokolov2015fixing}
\textsc{Sokolov D., Ray N.}:
\newblock \emph{Fixing normal constraints for generation of polycubes}.
\newblock research report, LORIA, 2015.
\newblock URL: \url{inria.hal.science/hal-01211408}.

\bibitem[SSV25]{snoep2025polycubes}
\textsc{Snoep M., Speckmann B., Verbeek K.}:
\newblock Polycubes via dual loops.
\newblock In \emph{Proceedings of the 2025 SIAM International Meshing Roundtable (IMR)} (2025).
\newblock \href {https://doi.org/10.1137/1.9781611978575.7} {\path{doi:10.1137/1.9781611978575.7}}.

\bibitem[THCM04]{tarini2004polycube}
\textsc{Tarini M., Hormann K., Cignoni P., Montani C.}:
\newblock Polycube-maps.
\newblock \emph{ACM Transactions on Graphics 23}, 3 (2004).
\newblock \href {https://doi.org/10.1145/1015706.1015810} {\path{doi:10.1145/1015706.1015810}}.

\bibitem[WHL{\etalchar{*}}07]{wang2007polycube}
\textsc{Wang H., He Y., Li X., Gu X., Qin H.}:
\newblock Polycube splines.
\newblock In \emph{Proceedings of the 2007 ACM Symposium on Solid and Physical Modeling (SPM)} (2007).
\newblock \href {https://doi.org/10.1145/1236246.1236281} {\path{doi:10.1145/1236246.1236281}}.

\bibitem[WHL{\etalchar{*}}08]{wang2008polycube}
\textsc{Wang H., He Y., Li X., Gu X., Qin H.}:
\newblock Polycube splines.
\newblock \emph{Computer-Aided Design 40}, 6 (2008).
\newblock \href {https://doi.org/10.1016/j.cad.2008.01.012} {\path{doi:10.1016/j.cad.2008.01.012}}.

\bibitem[WYZ{\etalchar{*}}11]{wan2011topology}
\textsc{Wan S., Yin Z., Zhang K., Zhang H., Li X.}:
\newblock A topology-preserving optimization algorithm for polycube mapping.
\newblock \emph{Computers \& Graphics 35}, 3 (2011).
\newblock \href {https://doi.org/10.1016/j.cag.2011.03.018} {\path{doi:10.1016/j.cag.2011.03.018}}.

\bibitem[YFL19]{yang2019computing}
\textsc{Yang Y., Fu X.-M., Liu L.}:
\newblock Computing surface polycube-maps by constrained voxelization.
\newblock \emph{Computer Graphics Forum 38}, 7 (2019).
\newblock \href {https://doi.org/10.1111/cgf.13838} {\path{doi:10.1111/cgf.13838}}.

\bibitem[YZWL14]{yu2014optimizing}
\textsc{Yu W., Zhang K., Wan S., Li X.}:
\newblock Optimizing polycube domain construction for hexahedral remeshing.
\newblock \emph{Computer-Aided Design 46} (2014).
\newblock \href {https://doi.org/10.1016/j.cad.2013.08.018} {\path{doi:10.1016/j.cad.2013.08.018}}.

\bibitem[ZLL{\etalchar{*}}17]{zhao2017robust}
\textsc{Zhao H., Lei N., Li X., Zeng P., Xu K., Gu X.}:
\newblock Robust edge-preserved surface mesh polycube deformation.
\newblock In \emph{Proceedings of the 25th Pacific Conference on Computer Graphics and Applications (PG)} (2017).
\newblock \href {https://doi.org/10.2312/pg.20171319} {\path{doi:10.2312/pg.20171319}}.

\bibitem[ZLW{\etalchar{*}}19]{zhao2019polycube}
\textsc{Zhao H., Li X., Wang W., Wang X., Wang S., Lei N., Gu X.}:
\newblock Polycube shape space.
\newblock \emph{Computer Graphics Forum 38}, 7 (2019).
\newblock \href {https://doi.org/10.1111/cgf.13839} {\path{doi:10.1111/cgf.13839}}.

\end{thebibliography}

\end{document}